\documentclass[twocolumn]{aastex62}

\newcommand{\Msun}{$M_{\odot}$}
\newcommand{\Mbh}{$M_{\rm BH}$}

\newcommand{\Lsun}{$L_\odot$}

\newcommand{\ml}{\emph{M/L}}

\newcommand{\hst}{\emph{HST}}
\newcommand{\kms}{km~s$^{-1}$}

\newcommand{\go}{G$_{\rm outer}$}

\usepackage{multirow}
\usepackage{footnote}
\graphicspath{{./}{figures/}}

\received{February 11 2018}
\revised{XXX XX 2018}
\accepted{XXX XX 2019}%{\today}
\submitjournal{ApJ}

\shorttitle{The MBHBM$_{\star}$ Project -- I: Measuring NGC 3504 Central BH using ALMA Observations}
\shortauthors{Nguyen et al.}
\begin{document}

\title{\Large{\bf The MBHBM$_{\star}$ Project -- I: Measurement of the Central Black Hole Mass in Dwarf Galaxy NGC 3504 Using Molecular Gas Kinematics}}

\correspondingauthor{Dieu D. Nguyen}\email{d.nguyen@nao.ac.jp}
\author[0000-0002-5678-1008]{Dieu D. Nguyen}
\affil{National Astronomical Observatory of Japan (NAOJ), National Institute of Natural Sciences (NINS), 2-21-1 Osawa, Mitaka, Tokyo 181-8588, Japan}
\affiliation{Department of Physics and Astronomy, University of Utah, 115 South 1400 East, Salt Lake City, UT 84112, USA}

\author{Mark den Brok}
\affiliation{Leibniz-Institut f\"ur Astrophysik Potsdam (AIP), An der Sternwarte 16, 14482 Potsdam, Germany}

\author{Anil C. Seth}
\affiliation{Department of Physics and Astronomy, University of Utah, 115 South 1400 East, Salt Lake City, UT 84112, USA}

\author{Satoru Iguchi}
\affiliation{National Astronomical Observatory of Japan (NAOJ), National Institute of Natural Sciences (NINS), 2-21-1 Osawa, Mitaka, Tokyo 181-8588, Japan}
\affiliation{Department of Astronomical Science, Graduate University for Advanced Studies (SOKENDAI), 2-21-1 Osawa, Mitaka, Tokyo 181-8588, Japan}

\author{Jenny E. Greene}
\affiliation{Department of Astrophysics, Princeton University, Princeton, NJ 08540, USA}

\author{Timothy Davis} 
\affiliation{School of Physics and Astronomy, Cardiff University, Queens Buildings, The Parade, Cardiff, CF24 3AA, UK}

\author[0000-0001-6186-8792]{Masatoshi Imanishi}
\affiliation{National Astronomical Observatory of Japan (NAOJ), National Institute of Natural Sciences (NINS), 2-21-1 Osawa, Mitaka, Tokyo 181-8588, Japan}
\affiliation{Department of Astronomical Science, Graduate University for Advanced Studies (SOKENDAI), 2-21-1 Osawa, Mitaka, Tokyo 181-8588, Japan}

\author[0000-0001-9452-0813]{Takuma Izumi} 
\altaffiliation{NAOJ fellow}
\affiliation{National Astronomical Observatory of Japan (NAOJ), National Institute of Natural Sciences (NINS), 2-21-1 Osawa, Mitaka, Tokyo 181-8588, Japan}

\author[0000-0002-1283-8420]{Michelle Cappellari}
\affiliation{Sub-department of Astrophysics, Department of Physics, University of Oxford, Denys Wilkinson Building, Keble Road, Oxford OX1 3RH, UK}

\author[0000-0002-6922-2598]{Nadine Neumayer}
\affiliation{Max Planck Institut f\"ur Astronomie (MPIA), K\"onigstuhl 17, D-69121 Heidelberg, Germany}

\author[0000-0003-1991-370X]{Kristina Nyland}
\affiliation{National Research Council, resident at the Naval Research Laboratory, Washington, DC 20375, USA}

\author{Takafumi Tsukui}
\affiliation{National Astronomical Observatory of Japan (NAOJ), National Institute of Natural Sciences (NINS), 2-21-1 Osawa, Mitaka, Tokyo 181-8588, Japan}
\affiliation{Department of Astronomical Science, Graduate University for Advanced Studies (SOKENDAI), 2-21-1 Osawa, Mitaka, Tokyo 181-8588, Japan}

\author[0000-0002-6939-0372]{Kouichiro Nakanishi}
\affiliation{National Astronomical Observatory of Japan (NAOJ), National Institute of Natural Sciences (NINS), 2-21-1 Osawa, Mitaka, Tokyo 181-8588, Japan}
\affiliation{Department of Astronomical Science, Graduate University for Advanced Studies (SOKENDAI), 2-21-1 Osawa, Mitaka, Tokyo 181-8588, Japan}

\author{Phuong M. Nguyen}
\affiliation{Department of Physics, Quy Nhon University, 170 An Duong Vuong, Quy Nhon, Vietnam}

\author{Quang L. Nguyen}
\affiliation{AA$\&$AI lab, IBM Canada, 120 Bloor Street East, Toronto, ON, M4Y 1B7, Canada}
\affiliation{Graduate School of Natural Sciences, Nagoya City University, 1 Yamanohata, Mizuho-cho, Nagoya, Aichi 467-8501, Japan}

\author[0000-0003-1820-2041]{Sabine Thater}
\affiliation{Leibniz-Institut f\"ur Astrophysik Potsdam (AIP), An der Sternwarte 16, 14482 Potsdam, Germany}

\author{Martin Bureau}
\affiliation{Sub-department of Astrophysics, Department of Physics, University of Oxford, Denys Wilkinson Building, Keble Road, Oxford OX1 3RH, UK}

\author{Kyoko Onishi}
\affiliation{Research Center for Space and Cosmic Evolution, Ehime University, 2-5 Bunkyo-cho, Matsuyama, Ehime 790-8577, Japan}

\author[0000-0001-6215-0950]{Karina T. Voggel}
\affiliation{Department of Physics and Astronomy, University of Utah, 115 South 1400 East, Salt Lake City, UT 84112, USA}

\author{Ngan M. Le}
\affiliation{Department of Space and Aeronautics, University of Science and Technology of Hanoi, 18 Hoang Quoc Viet, Hanoi, Vietnam}

\author{Trung V. Dinh}
\affiliation{Institute of Physics, Vietnamese Academy of Science and Technology, 10, Dao Tan, Ba Dinh, Hanoi, Vietnam}

\begin{abstract}

We present the first measurement of the mass of a supermassive black hole (SMBH) in the nearby double-barred spiral galaxy NGC 3504 as part of the Measuring Black Holes Below the Milky Way ($M_{\star}$) mass galaxies (MBHBM$_{\star}$) Project. Our analysis is based on Atacama Large Millimeter/sub-millimeter Array (ALMA) Cycle-5 observations of the ${\rm ^{12}CO(2-1)}$ emission line. NGC 3504 has a circumnuclear gas disk (CND), which has a relatively high velocity dispersion of  30 \kms. Our dynamical models of the CND yield a \Mbh~of $M_{\rm BH}=1.02^{+0.18}_{-0.15}\times10^7$\Msun and a mass-to-light ratio in $H$-band of \ml$_{\rm H}=0.66^{+1.44}_{-0.65}$ (\Msun/\Lsun). This black hole (BH) mass is consistent with BH--galaxy scaling relations. We also detect a central deficiency in the ${\rm ^{12}CO(2-1)}$ integrated intensity map with a diameter of 2.7 pc at the putative position of the SMBH. However, this hole is filled by a dense gas tracer ${\rm CS(5-4)}$ that peaks at the galaxy center found in one of the three low-velocity-resolution continuum spectral correlators. The ${\rm CS(5-4)}$ line has the same kinematics with the ${\rm ^{12}CO(2-1)}$ line within the CND, suggesting that it is also an alternative transition for measuring the central \Mbh~in NGC 3504 probably more accurately than the current commonly used of ${\rm ^{12}CO(2-1)}$ due to its centralization. 

\end{abstract}


%%%%%%%%%%%%%%%%%%####################
\section{Introduction}\label{sec:intro}

Supermassive black holes (SMBHs, $M_{\rm BH} \gtrsim10^6M_\odot$) are believed to reside at the centers of massive galaxies ($M_{\star}\gtrsim10^{11}$\Msun) and their masses correlate to macroscopic properties of the host (e.g., bugle velocity dispersion, bulge mass, and bulge luminosity). These remarkable discoveries are based on large efforts of observations \citep[e.g.,][]{Kormendy95, Magorrian98, Ferrarese00, Gebhardt00, Graham01, Marconi03, Haring04, Gultekin09, Beifiori12, Kormendy13, McConnell13, Saglia16, vandenBosch16} and theoretical works of self-regulated mechanisms of active galactic nuclei (AGN) feedback onto the outer gas reservoirs \citep{Silk98, DiMatteo08, Fabian12, Barai14, Netzer15}.  The underlying physics involve physically small scales but gravitationally large influences of SMBHs on their galactic environments that are recorded in the $M_{\rm BH}$--galaxy scaling relations. These correlations suggest SMBHs may play a pivotal role in the growth and evolution of galaxies \citep[e.g.,][]{Schawinski07, Kormendy13, Saglia16, vandenBosch16}.  

The observational census on the scaling relations in the regime of low-mass galaxies is very incomplete, as direct measurements of black hole (BH\footnote{In this article, we use the acronyms of SMBH and BH exchangeable.}) masses are lacking. This includes the increased scatter around the relations for the Milky Way late-type galaxies \citep[LTGs;][]{Greene10, Greene16, Lasker16} and roughly 2--3 orders of magnitude below the bulge mass relation for the lowest mass galaxies \citep[][hereafter N18, N19]{Scott13, Graham15, Nguyen17, Chilingarian18, Nguyen18, Nguyen19}.  Many hypotheses have been proposed to explain this change that may be due to (i) the formation history of the bulge \citep{Kormendy12, Krajnovic18}, (ii) the star formation history (SFH) of the galaxy \citep{Caplar15, Terrazas17}, or (iii) the bimodal population of high-$z$ BH seeds in which the light seeds ($M_{\rm BH} \lesssim10^3M_\odot$) accrete inefficiently with largely sub-Eddington rates and duty cycles, while the massive seeds ($M_{\rm BH} \gtrsim10^4M_\odot$) grow efficiently \citep{Pacucci15, Inayoshi16, Park16, Pacucci17, Pacucci18, Pacucci18b}
 
Recently, the number of known intermediate-mass black holes ($10^{3}$\Msun$ <M_{\rm BH}\lesssim10^{6}$\Msun), which are so called million/sub-million Solar masses BHs, are increasing dramatically with their masses inferred from a variety of methods including: (1) the velocity widths of their optical broad-line emissions \citep{Barth04, Greene07, Thornton08, Dong12, Reines13, Baldassare15, Reines15, Chilingarian18}, (2) the accretion signatures of  the narrow-line emissions \citep[e.g.,][]{Moran14} and the coronal emission in the mid-infrared \citep[MIR;][]{Satyapal09}, (3) tidal-disruption events \citep[TDE; e.g.,][]{Maksym13, Stone17}, (4) hard X-ray emission \citep[e.g.,][]{Gallo08, Desroches09, Gallo10, Miller15, She17}, (5) the dynamics of accretion disks containing megamasers \citep{Miyoshi95, Lo05, Kuo11, vandenBosch16}, and (6) the dynamics of nuclear stars and gas to measure \Mbh~in low-mass galaxies \citep[$5\times10^{8}$\Msun$ <M_\star\lesssim10^{10}$\Msun;][N18, N19]{Verolme02, Valluri05, Neumayer07, vandenBosch10, Seth10a, denbrok15, Nguyen17, Thater17} and ultracompact dwarfs \citep[UCDs, $1\times10^{7}$\Msun$ <M_\star\leq5\times10^{8}$\Msun;][]{Seth14, Ahn17, Afanasiev18, Ahn18, Voggel18}. The importance of the $\lesssim$$10^6M_\odot$ BH population and their masses are discussed in detail in N18 and N19. 

Atacama Large Millimeter/sub-millimeter Array (ALMA) observations of molecular gas at mm/sub-mm wavelengths offer a promising way to characterize the full spectrum of BH populations across the Hubble sequence from LTGs to early-type galaxies \citep[ETGs;][]{Davis13, Onishi15, Barth16a, Barth16b, Davis17, Onishi17, Davis18} because of the high angular resolution and sensitivity. Moreover, ALMA can improve on many problems affecting \Mbh~estimates via existing optical/infrared instruments including (1) high angular resolution capable of resolving the sphere of influence (SOI) and (2) ability to obtain dynamical measurements in dusty/obscured nuclei, which are inaccessible at optical wavelengths.  The physical idea behind the dynamical method utilizing the observations of ALMA is that the \Mbh~is derived by detecting the Keplerian turnover motion of the cold gas disk at the galactic center directly.  However, the cold gas dynamical method at mm/sub-mm works well for gas-rich galaxies those host well-defined and relaxed circumnuclear gas disks (CNDs) only.

This article is the first of a series in the Measuring Black Holes Below the Milky Way ($M_{\star}$) mass galaxies (MBHBM$_{\star}$) project. This project aims to gather a large sample of low-mass gas-rich galaxies and measure their central dark masses, which are likely BHs, using molecular gas tracer observed with ALMA. In this project, we select targets based on the presence of well-defined CNDs of gas and dust, which serve as morphological evidence for rotating dense gas about galaxy centers based on previous low-spatial-resolution surveys. Seven targets for \Mbh~measurements have already been observed by ALMA in Cycle-5. Here, we start the project with the \Mbh~measurement for NGC 3504, the first galaxy in the sample that has been observed.

We also want to test the growing power of ALMA and the capacity of the gas dynamical methods to (i) populate the sub-million Solar masses BH populations, (ii) constrain the scatters and slopes of BH--galaxy scaling relations at the low-mass regime, and (iii) precisely measure the {\it occupation fraction} of low-mass galaxies hosting central BHs. The occupation fraction is an important parameter helping to constrain the unknown formation mechanisms of BH seeds in the early Universe, which either form from the direct collapse of primitive gas clouds and produce massive seeds \citep{Lodato06, Bonoli14} or from the  remnants of the first stars (Population III) and produce lighter seeds \citep{Volonteri08, vanWassenhove10, Volonteri10, Volonteri12a, Volonteri12b, Fiacconi16, Fiacconi17}.  

The paper is organized into nine Sections. The properties of the galaxy NGC 3504 are presented in Section \ref{sec:ngc3504}.  In Section \ref{sec:data}, we present the ALMA observations of $^{12}{\rm CO(2-1)}$ nucleus gas and data reduction. We also report the evidence of a dense gas tracer CS($5-4$) at the center of the galaxy in Section \ref{sec:cs54}. The mass modeling of NGC 3504 is discussed in Section \ref{sec:massmodel}. We model the $^{12}{\rm CO(2-1)}$ gas disk and estimate the central \Mbh~and uncertainties via Kinematics Molecular Simulation \citep[KinMS;][]{Davis14} model and Tilted-ring model \citep[e.g.,][]{Neumayer07, denbrok15} in Sections \ref{sec:kinms} and \ref{sec:ring}, respectively. We discuss our results in Section \ref{sec:discussions} and conclude in Section \ref{sec:conclusions}.  Throughout the paper, unless otherwise indicated, all quantities quoted in this work have been corrected for a foreground extinction $A_V=0.306$ \citep{Schlafly11} using the interstellar extinction law of \citet{Cardelli89}.

%%%%%%%%%%%%%%%%%%%%%%%%%%%%%%%%%%%%%%%%%%%%%%%%
\section{The Galaxy NGC 3504}\label{sec:ngc3504}
  
NGC 3504 is a nearby double-barred spiral LTG (Hubble type (R)SAB(s)ab) located at the distance of $D=13.6$ Mpc estimated from Tully-Fisher relation \citep{Russell02} in the Leo Minor Group \citep{deVaucouleurs75}, giving a physical scale of 68 pc arcsec$^{-1}$. Using Sloan Digital Sky Survey (SDSS) 
photometric magnitude: $g=12.10$ mag, $r=11.39$ mag and \citet{Roediger15} $(g-r)$--$M/L_r$ relation give $M/L_r=2.5$ (\Msun/\Lsun), then the total stellar mass of NGC 3504 is $M_\star\sim6\times10^9$\Msun~($M_r=-19.3$ mag). 
\citet{Laurikainen04} decomposed the bulge mass of NGC 3504 using bugle-disk-bar decomposition model ($H$--band image of the Ohio State University Bright Galaxies Survey, OSUBGS) and found the bulge/disk ratio is 0.356. Adopting the above total stellar mass gives the bulge mass of $M_{\rm Bulge}=2.1\times10^9$\Msun. Recent work of \citet{Salo15} also used images at IR wavelengths (3.6 and 4.5 $\mu$m) from the Spitzer Survey of Stellar Structure in Galaxies (S$^4$G) imaging \citep{Sheth10} to decompose the bugle-disk-bar structure of NGC 3504. They found the total apparent magnitude of the bulge is $m_{\rm Bulge}=11.59$ mag\footnote{\url{https://www.oulu.fi/astronomy/S4G\_PIPELINE4/MAIN/\\decomp0087.html\#entry0009}}. Taking into account the distribution of the disk and bar within the region of dominant bulge ($\sim$25$\arcsec$) and assuming \ml$_{3.6\mu {\rm m}}\sim0.7$ (\Msun/\Lsun) (the \citet{Bell03} $(g-r)$--$M/L_K$ relation gives $M/L_K=0.85$ (\Msun/\Lsun)), the bulge mass is estimated as $M_{\rm Bulge}=1.7\times10^9$\Msun. These bulge mass estimates are both close to our bulge mass estimate using our a few first multiple Gaussian expansion \citep[MGE;][]{Emsellem94a, Cappellari02} components model in Section \ref{ssec:scalingrelation} distributed at the effective radius of the bulge \citep{Laurikainen04}. 

The galaxy center is determined at (R.A., Decl.) =  ($11^{\rm h}03^{\rm m}11^{\rm s}\!\!.210$, $27\degr58\arcmin21\farcs00$) in the $\alpha$(J2000) Equatorial coordinate system and has a systemic velocity of $1525.0\pm2.1$ \kms~determined from the Fabry--Perot observations in the frame of the Gassendi $H\alpha$ survey of SPirals \citep[GHASP;][]{Epinat08}. Photometry shows the inclination between the line-of-sight (LOS) and the polar axis of the galaxy is 26.4$^{\circ}$\footnote{\url{http://leda.univ-lyon1.fr/}} and oriented with a position angle (PA) of 150$^{\circ}$ \citep{Paturel00}.

Kinematic study of NGC 3504 finds its velocity field is dominated by circular motions \citep{Peterson82}. Palomar spectroscopic survey measures the stellar velocity dispersion of the bulge is $\sigma=119.3\pm10.3$ \kms~\citep{Ho09}, suggesting a central SMBH with mass of $\sim$$0.9^{+3.8}_{-0.4}\times10^7$\Msun~for this galaxy based on the $M_{\rm BH}-\sigma$ relation \citep{Kormendy13}. 

Both optical and radio data suggest that the galaxy has a composite nucleus with both AGN and emission from star formation \citep{Keel84}. $H\alpha$ obtained from 0.9 m telescope at Kitt Peak National Observatory \citep{Kenney93} shows a central starburst localized within 4$\arcsec$ (272 pc) and peaks in a ring of 1--2$\arcsec$ (68--136 pc). The star formation rate (SFR) within a diameter of 11$\arcsec$ of NGC 3504 is estimated as $2.3\pm0.4$~\Msun~yr$^{-1}$ (D. Nguyen et al. in preparation), confirming it is a well-known starburst \citep{Kenney93, Boselli15}, while the total SFR of the whole galaxy body is $0.92\pm0.37$~\Msun~yr$^{-1}$ \citep{Erroz-Ferrer15}. Moreover, Palomar optical survey defines the nucleus of NGC 3504 as a transition object \citep{Ho93b, Ho97} and suggests the central starburst component is probably powered by hot O-type stars \citep{Ho93}. NGC 3504 shows a compact nuclear unresolved source observed by very-long-baseline interferometry (VLBI) with a luminosity of $L_{\rm 1.4 GHz}=0.6\times10^{21}$ W Hz$^{-1}$ sr$^{-1}$ at 1.5 GHz \citep{Condon98, Deller14}.

%%%%%%%%%%%%%%%% HST H Band Image %%%%%%%
\begin{figure*}
    \centering\includegraphics[scale=0.207]{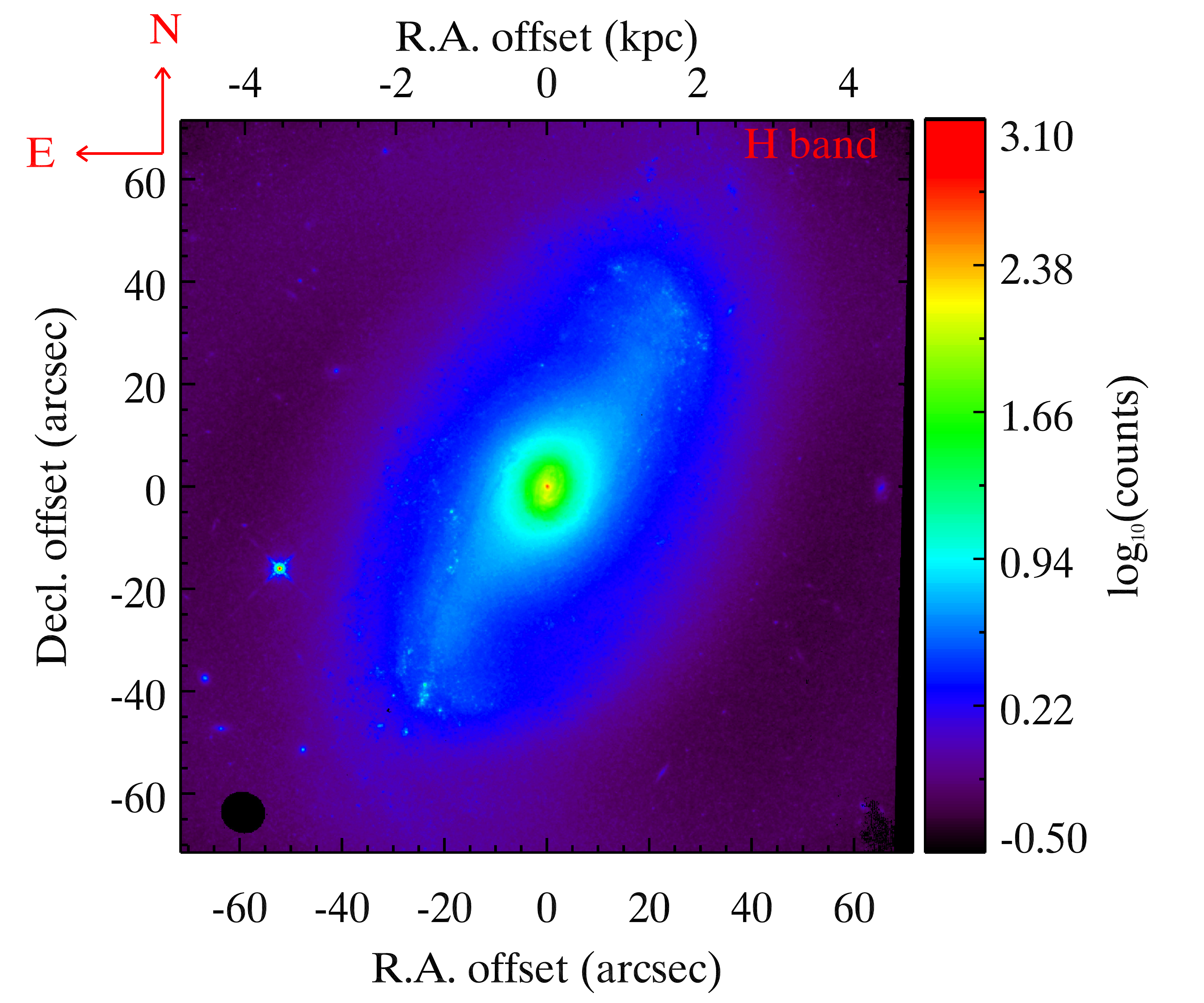}
	          \includegraphics[scale=0.222]{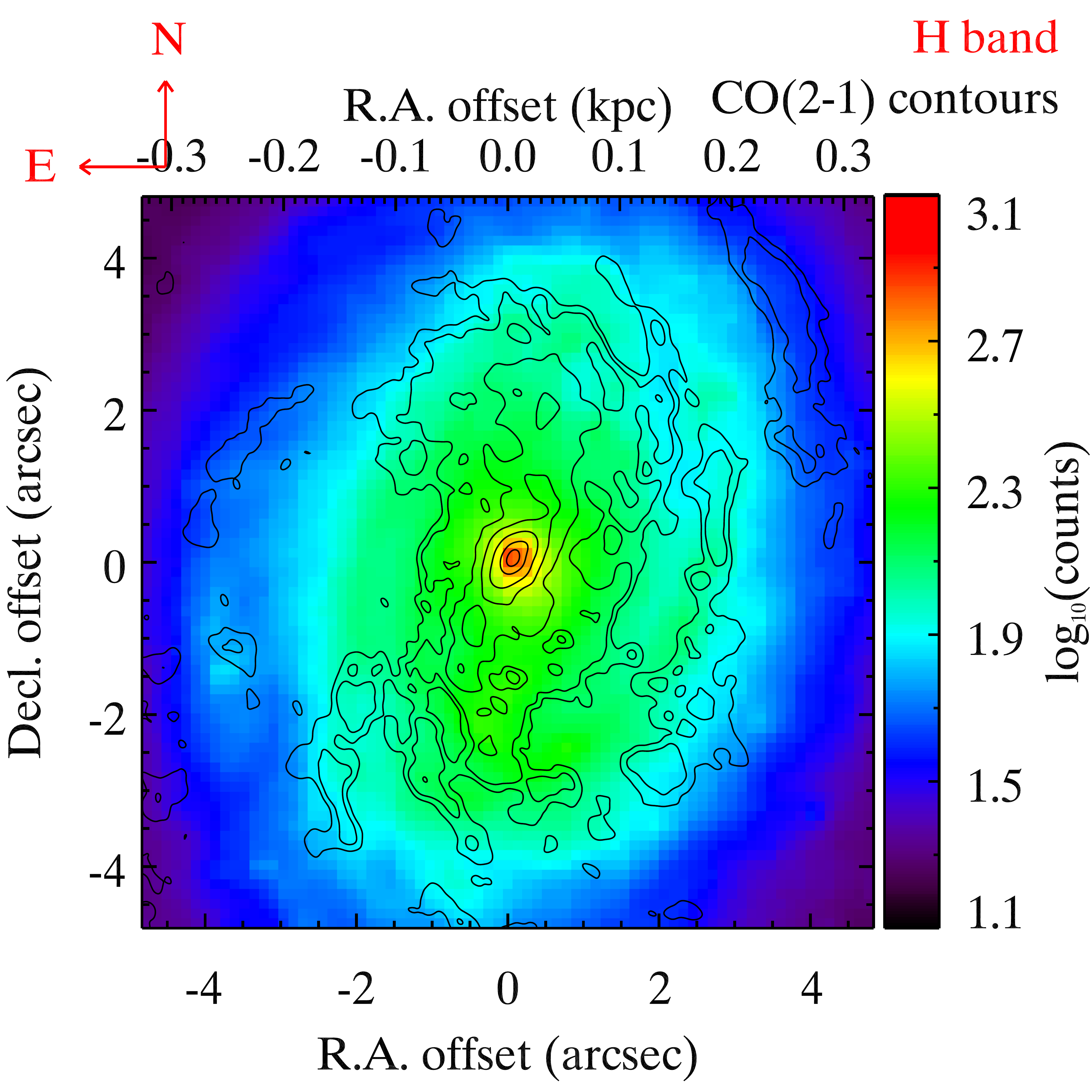}	 
\caption{Left panel: \hst/WFC3 IR F160W image of NGC 3504 within the field-of-view (FOV) of $140\arcsec\times140\arcsec$ (10 kpc\,$\times$\,10 kpc). Right panel: the zoom-in for the nucleus region within the FOV of $10\farcs0\times10\farcs0$ (0.68 kpc\,$\times$\,0.68 kpc) overlaid with black $^{12}{\rm CO(2-1)}$ integrated intensity contours from our ALMA observation from the low-resolution measurement set (MS).}
\label{hstimage}   
\end{figure*}
%%%%%%%%%%%%%%%%%%%%%%%%%%%%%%%%%

%%%%%%%%%%%%%%%%%%%%%%%%%%%%%%%%%%%%%%%%%
\begin{figure*}
    \centering\includegraphics[scale=0.7]{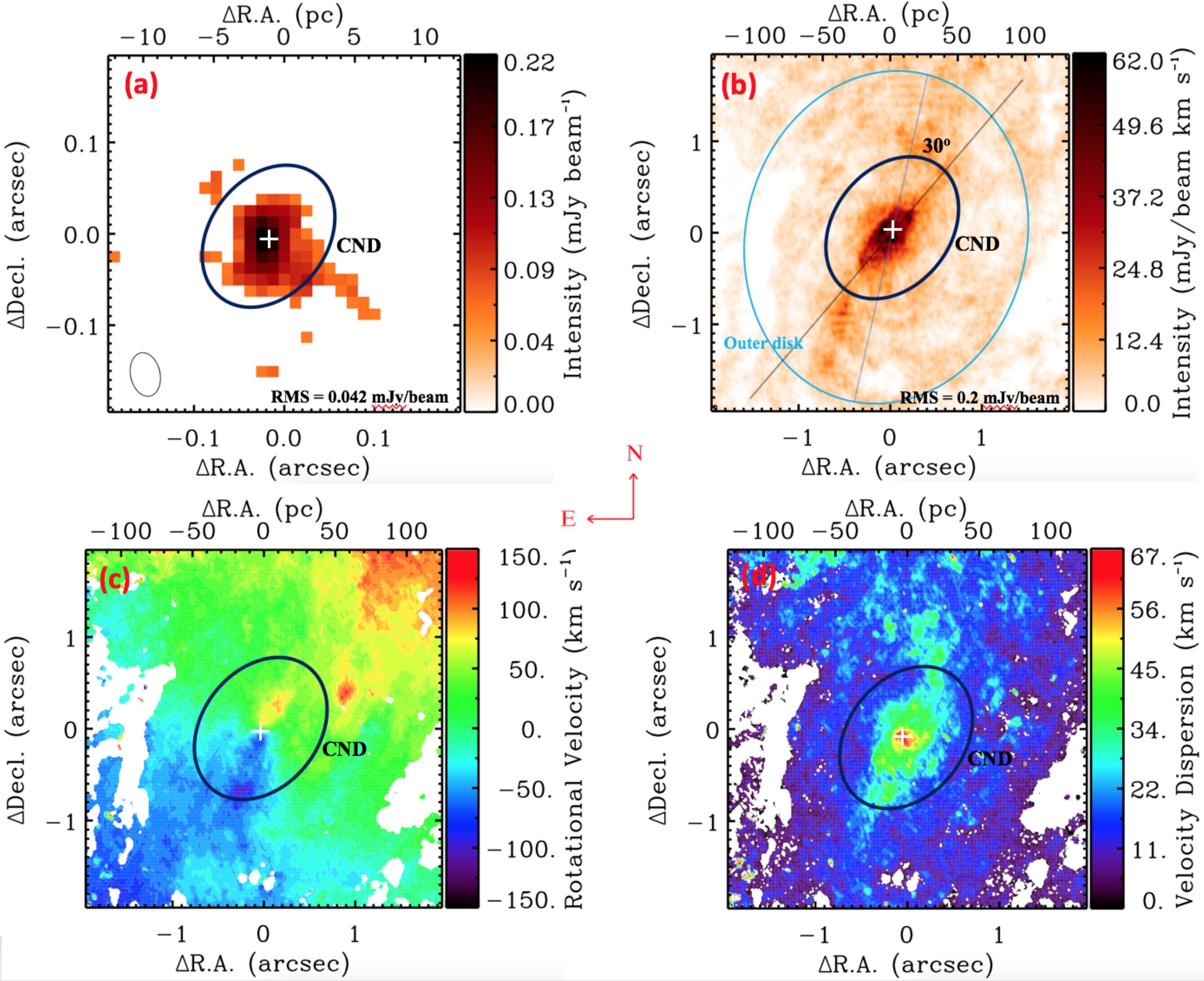}
\caption{The panel-a shows the zoom-in map of the 1.3 mm continuum emission at the center of NGC 3504. The synthesized beam of $0\farcs044\times0\farcs031$ (2.9 pc $\times$ 2.1 pc) is shown as an ellipse at the bottom left of the panel. The three rest panels are the moments of the detected $^{12}{\rm CO(2-1)}$ emission created using the masked moment technique described in Section \ref{sssec:line} including the zeroth moment map (panel-b), the first (panel-c) and the second moment (panel-d) maps. The synthesized beam cannot be seen in these large field-of-view (FOV) plots. White plus indicates the kinematic center and also the galaxy center.}   
\label{maps}  
\end{figure*}
%%%%%%%%%%%%%%%%%%%%%%%%%%%%%%%%%%%%%%%%%

%%%%%%%%%%%%%%%%%%%%%%%%%%%%%%%%%%%%%%%%%%%%%%%%
\section{Data and Data Reduction}\label{sec:data}

%%%%%%%%%%%%%%%%%%%%%%%%%%%%%%%%%%%%%%%%%%%%%%%
\subsection{Hubble Space Telescope (HST) Imaging}\label{ssec:hst}   

We use \hst~observations in wide field camera 3 (WFC3) IR band F160W to create a mass model (Section \ref{sec:massmodel}), which will be used as an input ingredient for dynamical models in Sections \ref{sec:kinms} and \ref{sec:ring}.  The \hst~data was observed in 2012 May 01 (GO-12450, PI: Kochanek) with a total exposure time of 1398 s. We downloaded this image from the Hubble Legacy Archive (HLA) directly and used them throughout our analysis. However, to test the sky background level accurately, we also downloaded the flat-fielded ({\tt flt}) images from the \hst/The Barbara A. Mikulski Archive for Space Telescopes (MAST), combined these images in the same filter using \texttt{drizzlepac/Astrodrizzle} \citep{Avila12}, and compared the combined image with the HLA image. We choose F160W filter as the default image to model the mass-follows-light map, while the F110W image is used as an alternative mass model to estimate the uncertainty caused by different wavelengths and extinction (Section \ref{sssec:stellarmasserror}).

Figure~\ref{hstimage} shows the large structure of NGC 3504 in the F160W image with prominent bar connecting the bulge and galactic disk on the left and its zoom-in at the field-of-view (FOV) of $10\arcsec\times10\arcsec$ on the right. There is a few regular dust lanes circling the galaxy center, suggesting a circumnuclear gas disk (CND), which extends to a radius of at least $5\arcsec$ from the center. 

The photocenter of the HLA images is at (R.A., Decl.) = ($11^{\rm h}03^{\rm m}11^{\rm s}.210,  +27^{\circ}58^{\prime}21\farcs00$) in the $\alpha$(J2000) system as presented in HLA images. This is offset $\sim$$0\farcs21$ compare to the peak of the compact $^{12}{\rm CO(2-1)}$ continuum emission and the optical center of the galaxy derive from optical images (Section \ref{sssec:cont}); those are consistent to each other. We therefore align the \hst~images to the $^{12}{\rm CO(2-1)}$ continuum emission map to correct for this astrometric mismatch. 

We used {\tt Tiny Tim} point spread functions \citep[PSFs;][]{Jedrzejewski87a, Jedrzejewski87b} for the WFC3/IF F160W and F110W images to decompose the MGE model \citep{Emsellem94a, Cappellari02} in Section \ref{sec:massmodel}. 

%%%%%%%%%%%%%%%%%%%%%%%%%%%%%%%%%%%%%%%%%
\begin{figure}[!ht]  
\centering\hspace{-5mm}
	\includegraphics[scale=0.5]{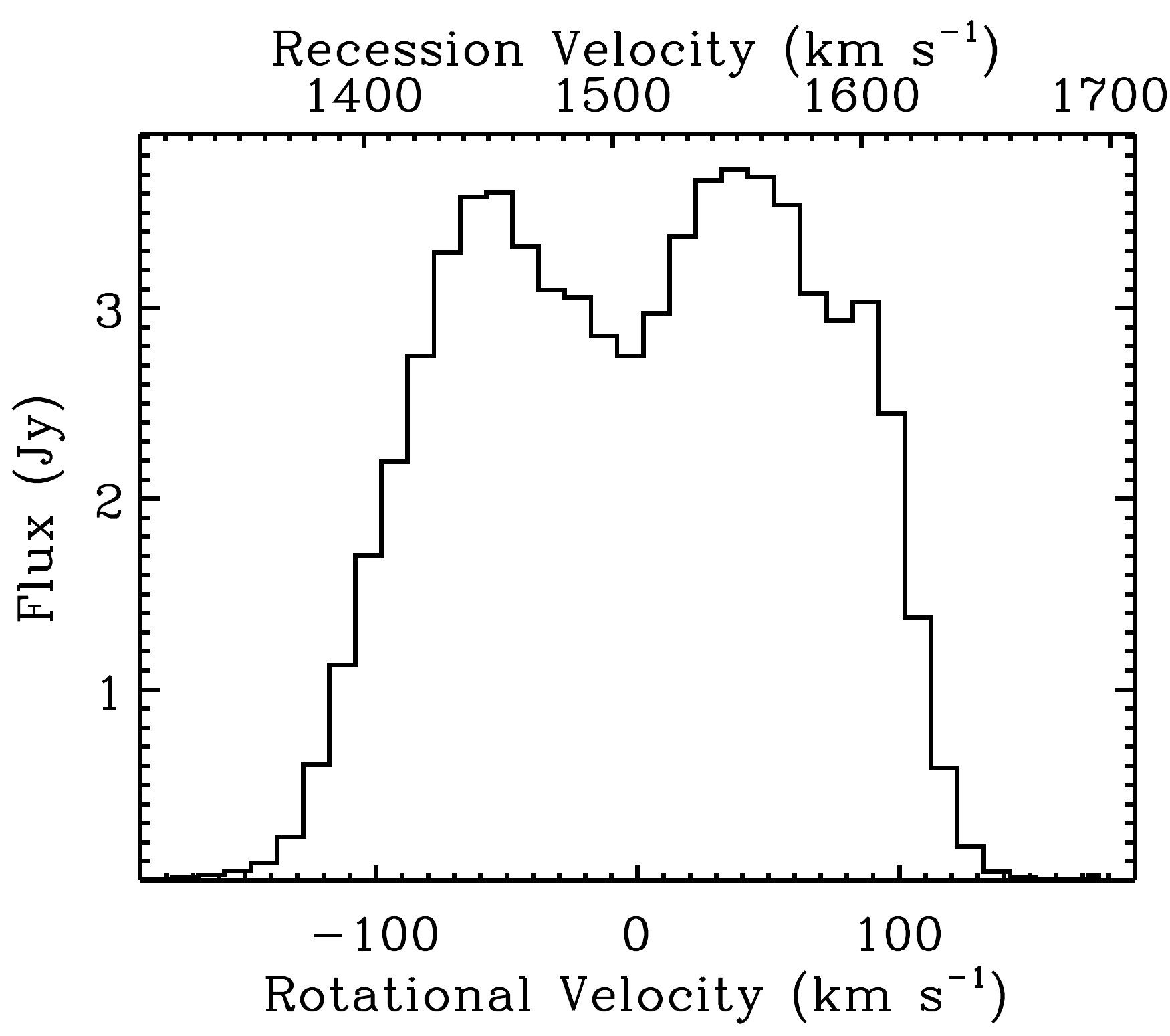} 
\caption{The integrated $^{12}{\rm CO(2-1)}$ spectrum extracted within the nuclear region of $10\arcsec\times10\arcsec$ (680 pc $\times$ 680 pc), where includes all the detected emission.  We observe the classic symmetric double horn shape of a rotating disk.} 
\label{spec}   
\end{figure}
%%%%%%%%%%%%%%%%%%%%%%%%%%%%%%%%%%%%%%%%%

%%%%%%%%%%%%%%%%%%%%%%%%%%%%%%%%%%%%%%%%%
\begin{figure*}[!ht]  
\centering
        \includegraphics[scale=0.45]{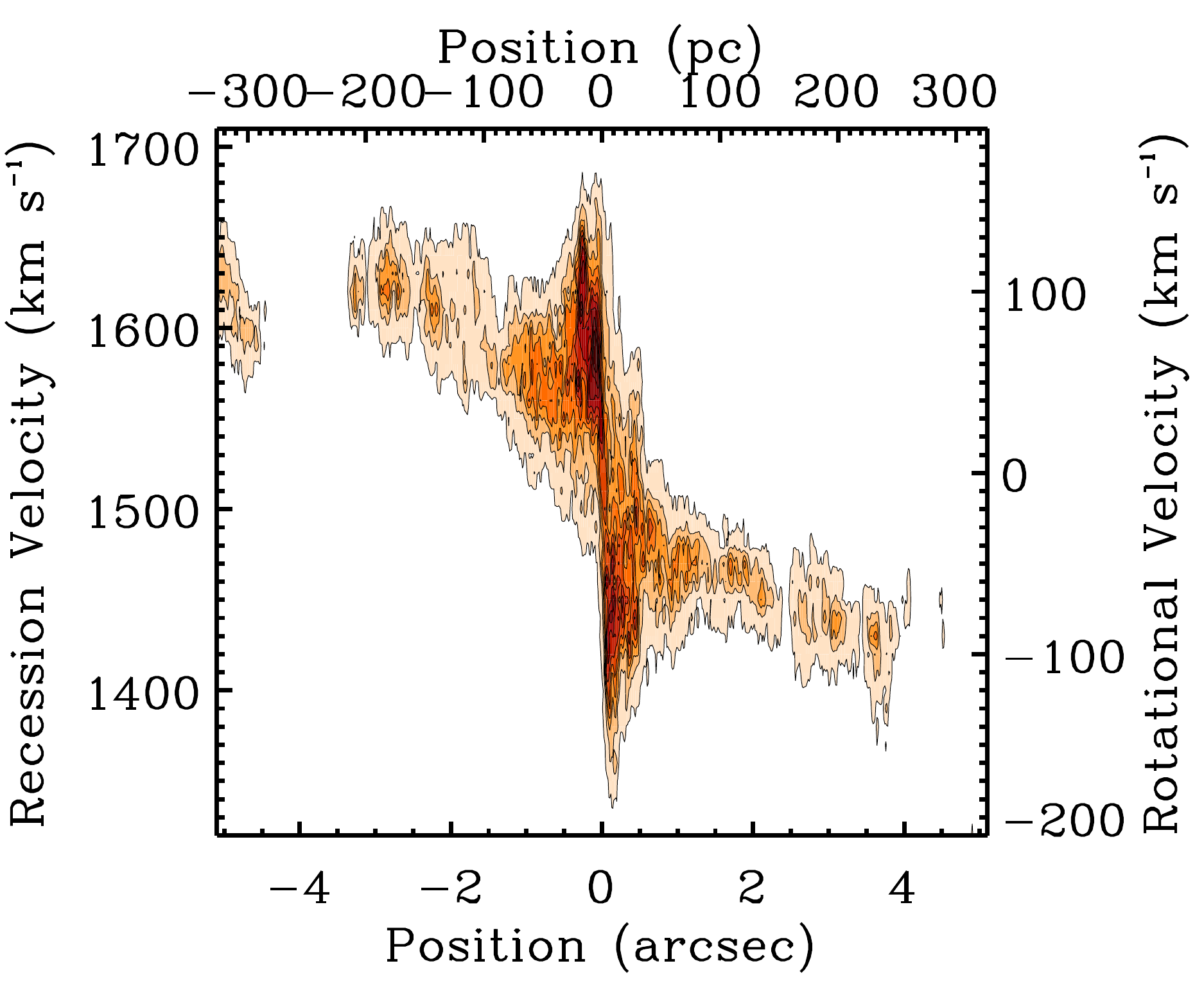}
\hspace{5mm}\includegraphics[scale=0.45]{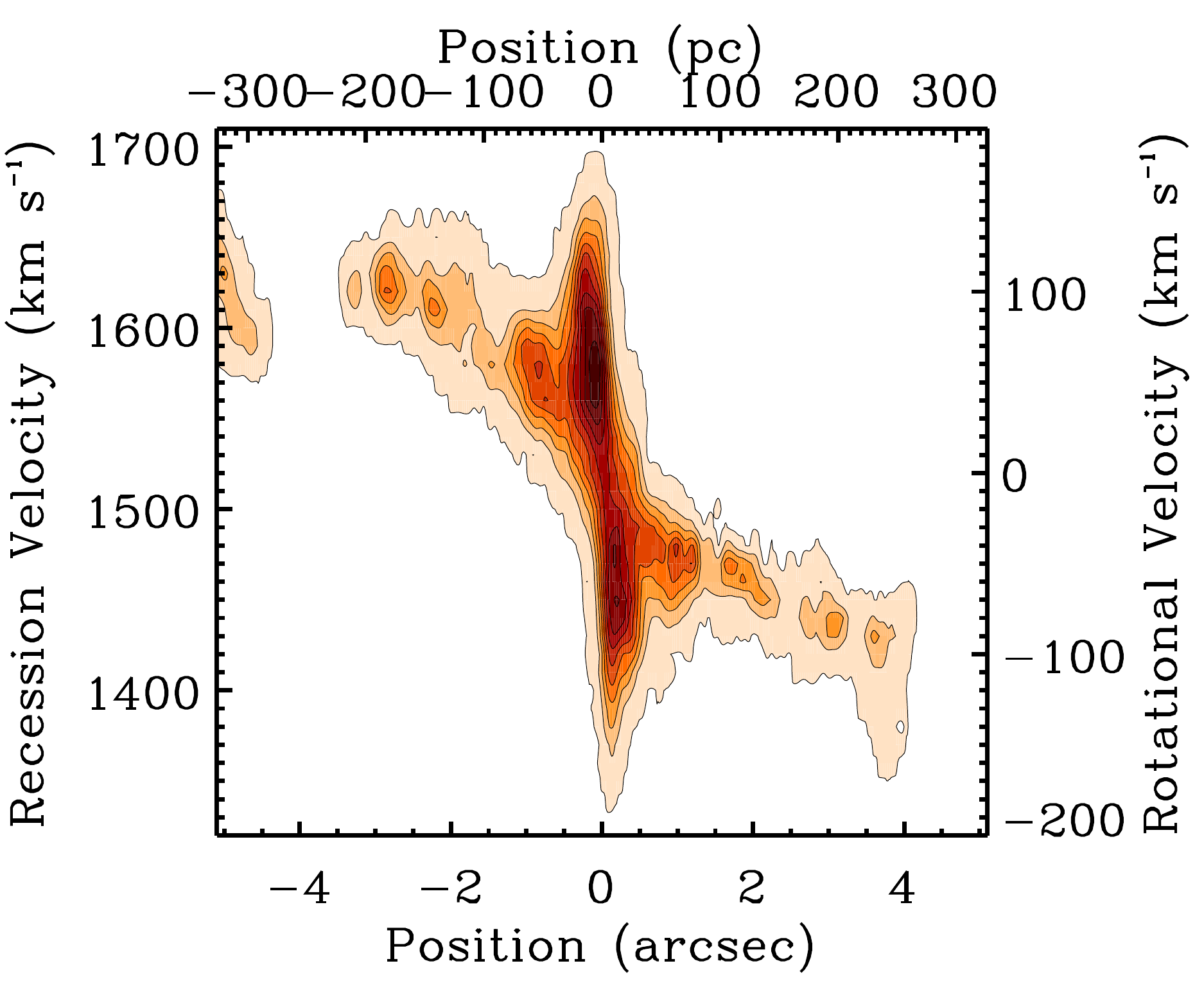} 	
\caption{Left panel: The position-velocity diagram (PVD) of the  $^{12}{\rm CO(2-1)}$ emission in the same field-of-view (FOV) as presented in moment maps, extracted along the kinematic major axis with a slit of five pixels in width ($0\farcs065$ or 4.4 pc). We do not show the beamsize of $0\farcs044\times0\farcs031$ (3.0 pc $\times$ 2.1 pc) and velocity channel width 10 \kms\,explicitly in this plot, as they are very small compared to the plotted ranges. Clear channelization is present in the PVD, suggesting the velocity dispersion in the gas is as small as compared to our channel width.  Right panel:  The same PVD for the low-resolution cube with the beamsize of $0\farcs221\times0\farcs164$ (15.0 pc $\times$ 11.2 pc).} 
\label{pvd}   
\end{figure*}
%%%%%%%%%%%%%%%%%%%%%%%%%%%%%%%%%%%%%%%%%

%%%%%%%%%%%%%%%%%%%%%%%%%%%%%%%%%%%%%%%%%%%%%%%%%%%%%%
\subsection{\emph{ALMA} Observations}\label{ssec:alma}

The observations of the $^{12}{\rm CO(2-1)}$ line in the nucleus of NGC 3504 are carried out with ALMA as a part of the, ``Weighing Black Hole Masses in Low-Mass Galaxies'' project (Program 2017.1.00964.S, PI: Nguyen, Dieu). The correlators cover the data in four spectral windows including one 1875 MHz FDM spectral window covers over the $^{12}{\rm CO(2-1)}$ line and three 2 GHz TDM spectral windows added simultaneously to detect continuum emission. The observations use 46 ALMA's 12m antennas with the C43-9 and C43-6 configurations so that the data has a maximum recoverable scale (MRS) of 10$\arcsec$ in diameter. The raw ALMA data are calibrated by the ALMA regional center staff using the standard ALMA pipeline. Flux and bandpass calibration are conducted using the quasar J1058+0133, while the atmospheric phase offsets of the data are determined as a phase calibrator using J1102+2757.  More details of these observations are summarized in Table \ref{almaobs}.

%%%%%%%%%%%%%%%%% Table1: ALMA observing parameters %%%%%%%%%%%%%%
\begin{table}
\caption{ALMA observation parameters.}
\centering
\begin{tabular}{lcc} \hline \hline
Phase center    & R.A. & Decl. \\
 & $11^{\rm h}03^{\rm m}11^{\rm s}.205$&$+27^{\circ}58^{\prime}20\farcs80$\\
\hline
                &High Res.& Low Res.\\ 
Configurations  &  C43-9  & C43-6\\ 
\hline
Obs. Date       &2017 Oct. 24&2018 Jan. 01\\
Exposure time   & 46.6 min.  & 25.7 min.  \\
Beamsize        &$0\farcs042\times0\farcs030$&$0\farcs221\times0\farcs164$\\  
(or FWHM)&2.9 pc $\times$ 2.0 pc&15.0 pc $\times$ 11.2 pc  \\  
Beam PA         &$-$2$^{\circ}$.0   &   32$^{\circ}$.3   \\
Velocity resolution&  1.5 \kms &  39.5 \kms   \\
Frequency resolution&1.13 MHz&31.25 MHz\\
\hline
\end{tabular}
\tablenotemark{}  
\tablecomments{FWHM$^\star$: Full width at half of maximum.}
\label{almaobs}
\end{table}
%%%%%%%%%%%%%%%%%%%%%%%%%%%%%%%%%%%%%%%%%%%%%%%%%

Continuum emission is detected at the galactic center of NGC 3504 only (panel-a, Figure \ref{maps}) and measured over the full line-free bandwidth, fitted by a power law function, and then subtracted from the data in the $uv$-plane using the task {\tt uvcontsub} of the \texttt{Common Astronomy Software Applications} ({\tt CASA}) package version 5.1.1. 

We first combine the visibility files of these two measurement sets into a final continuum-free calibrated data using the {\tt CASA} task {\tt concat} with the optimal combination ratios of 1/3 and 2/3 for the high- and low-spatial-resolution measurement set during the {\tt visweightscale} mode, respectively. Second, we create a three-dimensional (3D) (R.A., Decl., velocity)-combined cube from the continuum-free calibrated file using the {\tt clean} task. To dynamically model the CND and estimate the \Mbh, we image the combined cube using the channel width of 10\,\kms~\citep{Davis14}, Briggs weighting with a robust parameter of 0.5, and pixel size of $0\farcs$013 in order to optimize the sensitivity of diffuse gas and resolution. The sidelobes on the image are reduced using a mask during interactive mode. We estimate the root-mean square noise, RMS = 42 $\mu$Jy beam$^{-1}$, in a few blank signal channels, then set 3 $\times$ RMS as the clean threshold in regions of source emission for dirty channels. We also do primary beam correction during the imaging cube. The final calibrated $^{12}{\rm CO(2-1)}$ cube has a synthesized beam of $0\farcs044\times0\farcs031$ (3.0 pc $\times$2.1 pc), PA $\sim$ 9.7$^{\circ}$, and RMS $\sim$ 0.187 mJy beam$^{-1}$ \kms. The nucleus molecular gas emission of NGC 3504 is detected from 1340 to 1680~\kms.

%%%%%%%%%%%%%%%%%%%%%%%%%%%%%%%%%%%%%%%%%%%%%%%%%%%%
\subsubsection{Continuum Emission}\label{sssec:cont}

The continuum emission is clearly resolved and centrally peaked as seen in the panel-a of Figure \ref{maps}. We identify the continuum peak as the galaxy/kinematic center. To determine the position, size, and total integrated intensity of this source, we fit this continuum profile with a Gaussian using the {\tt CASA} task {\tt imfit}. The emission is centered at (R.A., Decl.) = ($11^{\rm h}03^{\rm m}11^{\rm s}.115\pm0\farcs040$, $+27^{\circ}58^{\prime}20\farcs80\pm0\farcs18$) with the error bars include both of the {\tt imfit} fit and ALMA astrometric uncertainties. The SDSS data release 14 \citep{Abolfathi18} also report an optical center at ($11^{\rm h}03^{\rm m}11^{\rm s}.112$, $+27^{\circ}58^{\prime}20\farcs77$) that is consistent to what we found here.  {\tt imfit} estimates the size of the emission source of $30^{+10}_{-10}\;10^{-3}$ arcsec $\times$ $10^{-3}$ arcsec (or $2.0^{+0.7}_{-0.7}$ pc $\times$ pc), with a PA of $\sim$30$^{\circ}$ and the total integrated intensity of $4.70\pm0.16\pm0.24$ mJy over the emissions free of USB (442--446 GHz) and LSB (227.5--231.5 GHz) frequency windows. Note that the former/latter errors are the systematic error/ALMA flux calibration error, respectively.
 
%%%%%%%%%%%%%%%%%%%%%%%%%%%%%%%%%%%%%%%%%%%%%%%%%%%%%%%%%%%%%%%%%%%%
\subsubsection{$^{12}${\rm CO(2--1)} Line emission}\label{sssec:line}

We create the integrated intensity, intensity-weighted mean velocity field, and intensity-weighted velocity dispersion maps for the nuclear $^{12}{\rm CO(2-1)}$ gas disk from the combined cube directly in Figure \ref{maps}. The emission is significant within the radius of 5$\arcsec$. We enhance the quality of these maps using the moments masking technique; details of this technique were described in \citet{Davis17, Onishi17, Davis18}. 

The integrated intensity map reveals the presence of a nuclear rotating disk. However, as seen in the panel-b and c of Figure~\ref{maps}, this disk seems separating into two distinct rotating disks including a dense compact CND distributed within 1$\arcsec$ and a faint extended disk of spiral arms and voids. \texttt{Kinemetry}\footnote{\url{http://davor.krajnovic.org/idl/\#kinemetry}}  \citep{Krajnovic06} examination reveals these two disks are clearly misalignment by an angle of $\sim$30$^{\circ}$, this is also consistent to what found in \citet{Knapen02}. 

The velocity field (panel-c) shows a regular disk-like rotation with a total velocity width of $\sim$$300$ \kms. The velocity dispersion map (panel-d) are quite flat with constant values of $\sim$30 \kms~and $\sim$10 \kms~in the inner and outer gas disks, respectively. However, there is a suddenly increasing peak $\sim$65 \kms~at the galactic center. This is not an intrinsic velocity dispersion but beam smearing over the velocity gradient at the galaxy center and the LOS integration through a nearly edge-on orientation of the inner gas disk ($i\gtrsim60^{\circ}$). 

Figure \ref{spec} shows the integrated $^{12}{\rm CO(2-1)}$ spectrum of NGC 3504 that has a classical double-horn shape of a rotating disk. While in Figure \ref{pvd}, we plot the position-velocity diagram (PVD) extracting from a cut through the major axis of the gas disk (PA $\sim332^{\circ}$). We show both PVDs of the combined cube (high resolution) and the low-resolution cube side by side in this figure to demonstrate the increased rotation towards the center even at different spatial scales. We interpret this motion as a Keplerian curve caused either by an SMBH existing at the heart of NGC 3504 or by non-circular motions of an inner ring/spiral within the innermost regions, which is typically seen in barred galaxies. 

\begin{figure*}[!ht]
\centering
	\includegraphics[scale=0.71]{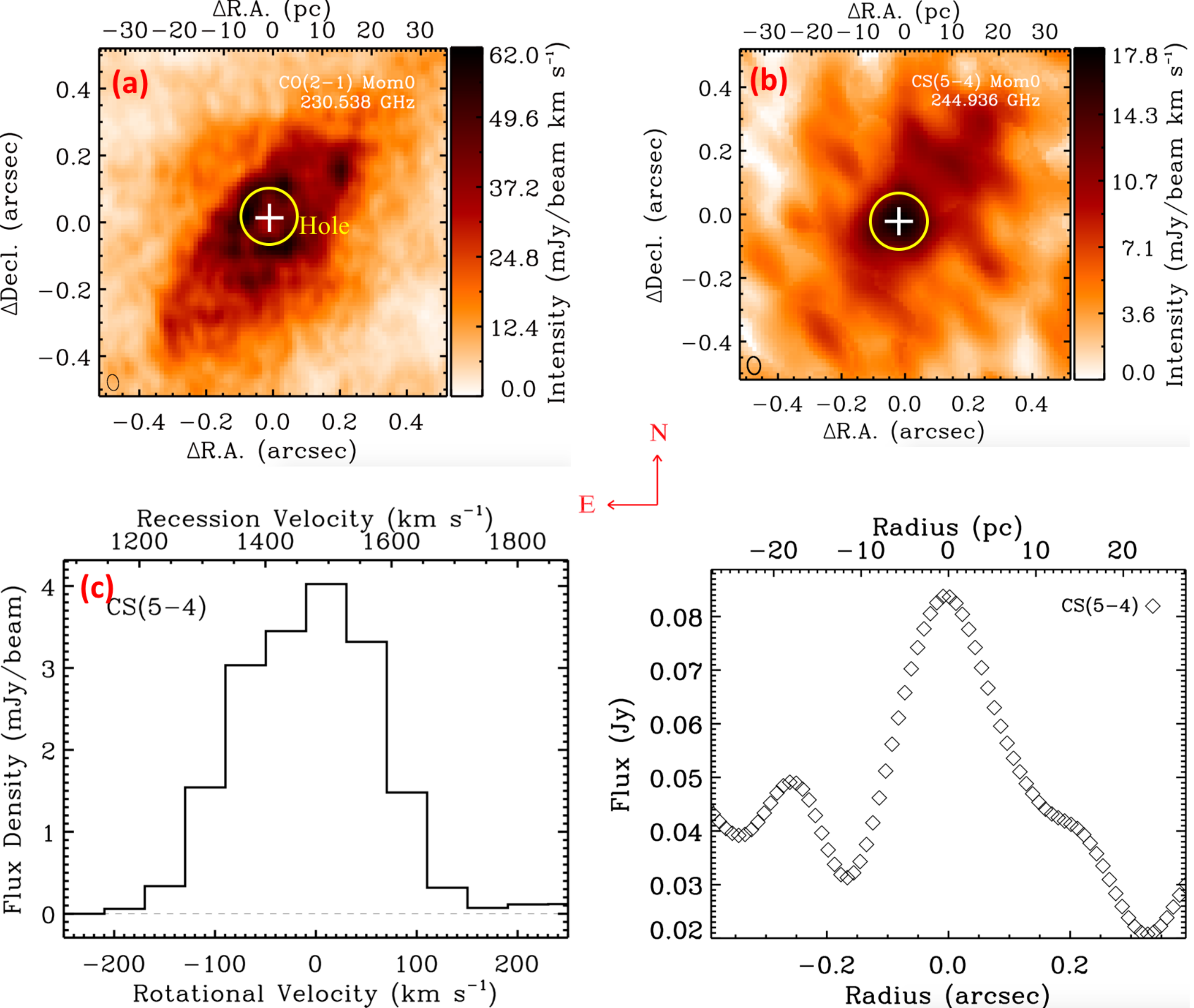} 
\caption{(a): The nuclear integrated intensity map of the $^{12}{\rm CO(2-1)}$ gas emission zoom-in to the field-of-view (FOV) of $1\arcsec\times1\arcsec$. The map shows a CND attenuation hole, which has a diameter of $\sim$2.7 pc and is centered at the location of the putative SMBH (white crosses). (b): The same FOV for our detected dense gas tracer ${\rm CS(5-4)}$ emission line in one of the continuum spectral windows. The ${\rm CS(5-4)}$ line is clearly centrally concentrated and peaks at the  position of the central SMBH where we observe the $^{12}{\rm CO(2-1)}$ attenuated hole. The integrated spectrum (c) and radial profile (d) of ${\rm CS(5-4)}$, both show the line is centrally filled rather than a hole. White pluses indicate the kinematic center.}  
\label{densegas}   
\end{figure*}

The intensity map also shows a central hole, which is marginally resolved on pc scales as seen in the zoom-in $^{12}{\rm CO(2-1)}$ intensity map (panel-a, Figure \ref{densegas}). The intensity remains above zero inside the hole is due to either the angular extent of the synthesized beam or some galactic foreground gas distribute further above the torus and align along the LOS. Another possibility of the non-zero intensity in the hole is a genuine existence of $^{12}{\rm CO(2-1)}$ in the hole but faint due to high density, making CO gas is excited to higher-J transitions. Based on its physical radius of $\sim0\farcs04$ (or 2.7 pc, estimated in Section \ref{sec:kinms}) and centralization, we believe that we are marginally observing a dusty molecular torus surrounding an accreting SMBH, a fuel source of the unified AGN model \citep{Antonucci93a}. We estimate the size of this dusty molecular torus using MIR flux observations \citep{Rieke72} and the torus size-luminosity relation from \citet{Tristram11}. The radius of the hole is $\sim$2 pc, roughly corresponding to the hole in our $^{12}{\rm CO(2-1)}$ observations. This radius is bigger than the inner dust sublimation radius of the torus, which is typically of $<$1 pc \citep{Barvainis87a}. The reason for this mismatch is the $^{12}{\rm CO(2-1)}$ emission is sometimes deficient in the nuclear regions, and therefore may not be the best tracer for the torus \citep{Imanishi18, Izumi18}. 

 %%%%%%%%%%%%%%%%%%%%%%%%%%%%%%%%%%%%%%%%%
%%%%%%%%%%%%%%%%%%%%%%%%%%%%%%%%%%%%%%%%%

%%%%%%%%%%%%%%%%%%%%%%%%%%%%%%%%%%%%%%%%%
\begin{figure*}[!ht]
\centering\includegraphics[scale=0.3]{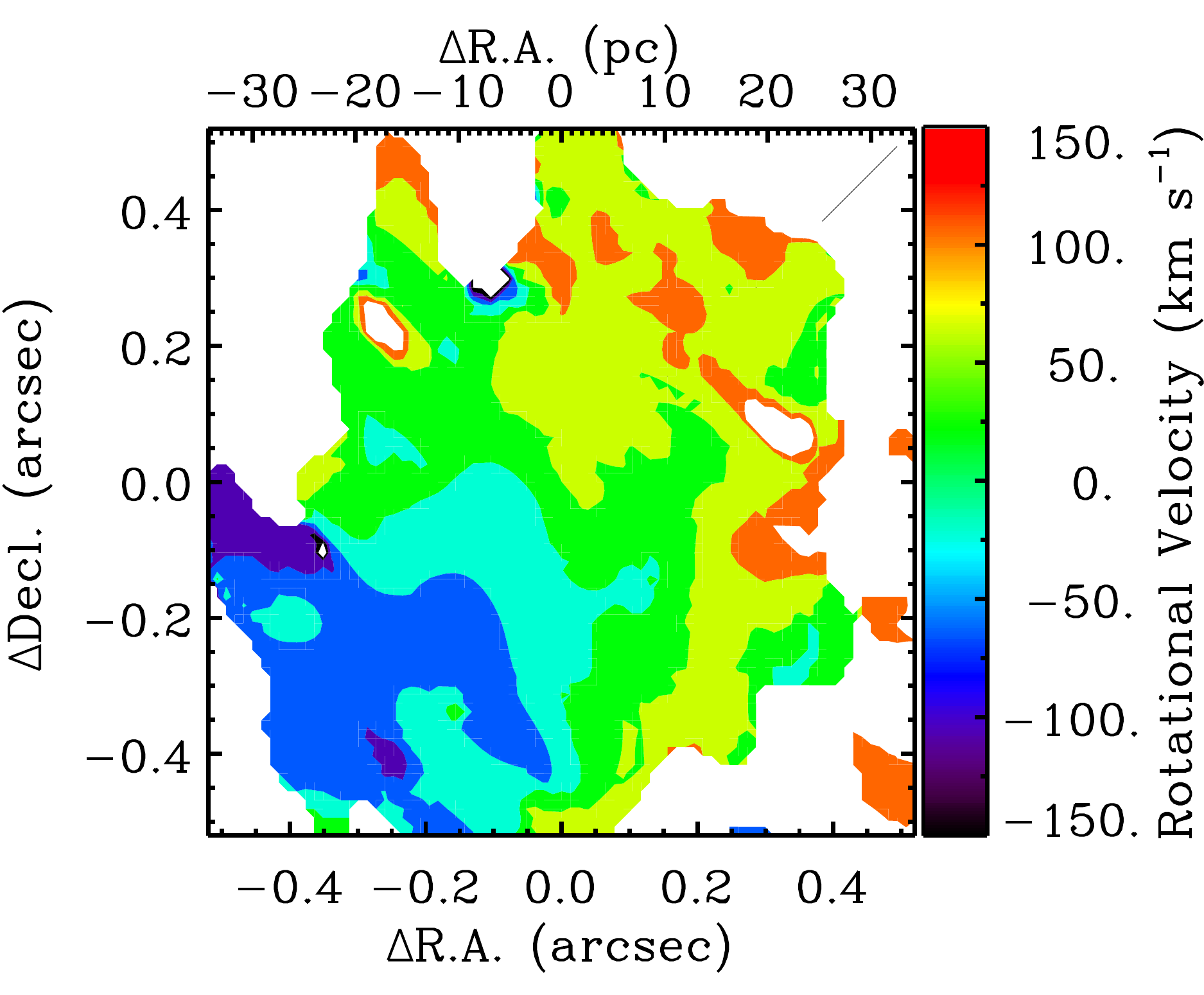} 
\hspace{3mm}\includegraphics[scale=0.3]{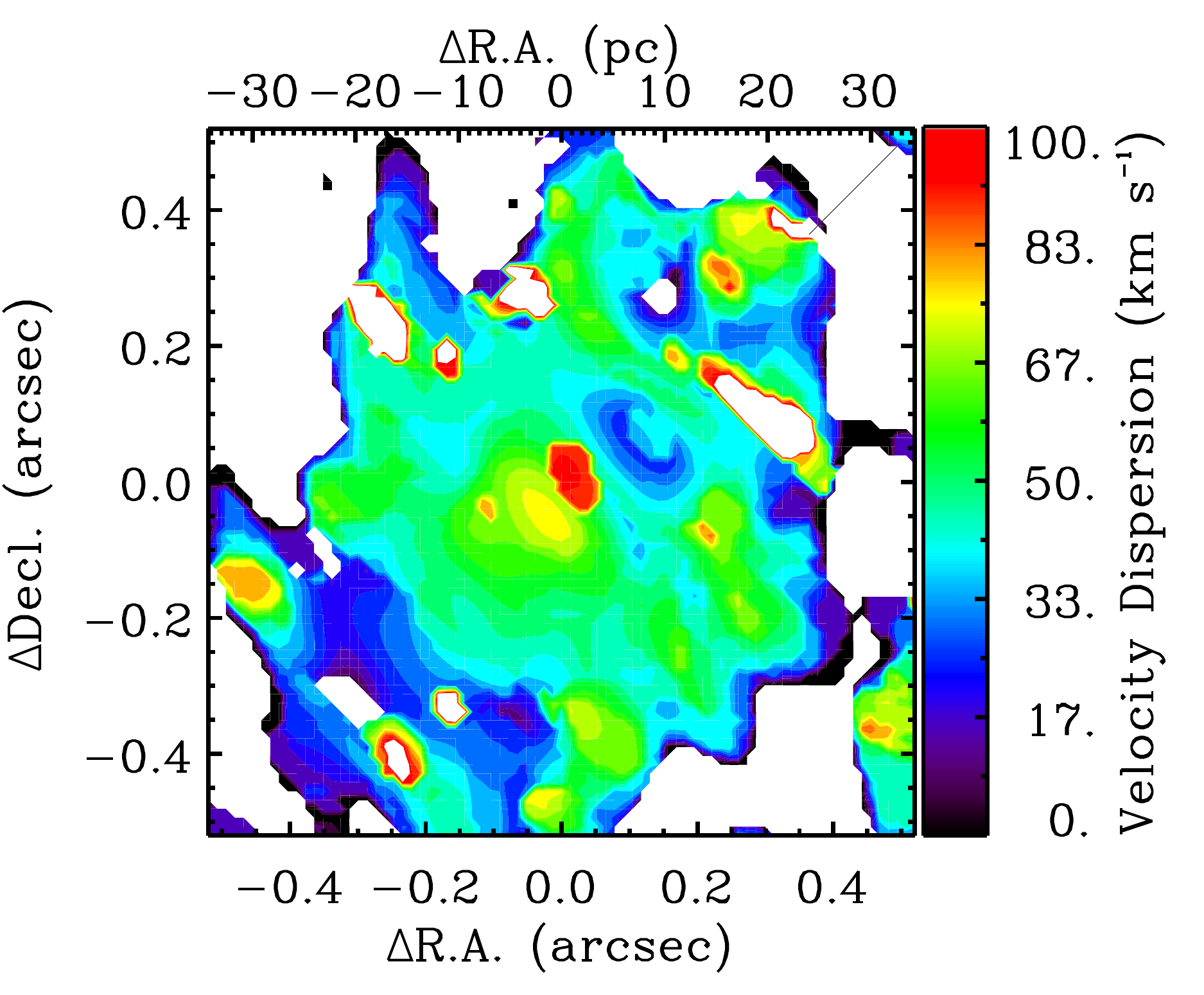} 
\hspace{1mm}\includegraphics[scale=0.3]{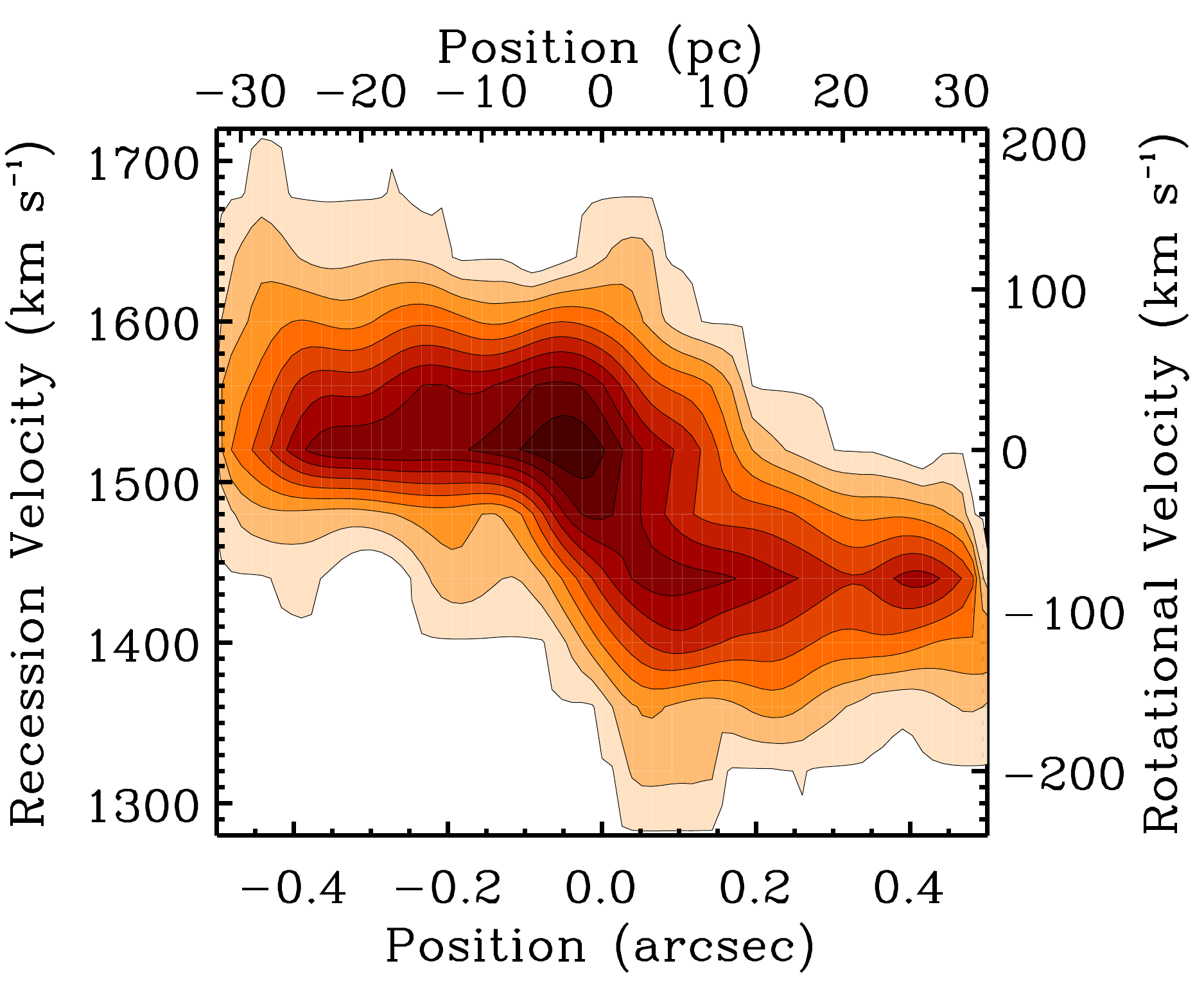} 
\caption{The moment maps of the detected ${\rm CS(5-4)}$ emission created using the masked moment technique described in Section \ref{sssec:line} including the first (rotational velocity--left panel) and the second moment (velocity dispersion--middle).  Right panel: the position-velocity diagram of ${\rm CS(5-4)}$, which has the same recession velocity and velocity width of $^{12}{\rm CO(2-1)}$, as well as the rotational signature around the galactic center. We note that these maps have low-velocity resolution $\sim$40~\kms.}  
\label{densegas1}   
\end{figure*}
%%%%%%%%%%%%%%%%%%%%%%%%%%%%%%%%%%%%%%%%%

%%%%%%%%%%%%%%%%%%%%%%%%%%%%%%%%%%%%%%%%%%%%%%%%%%%%%%%%%%%%%%%%
\section{Detection of Dense Gas CS(5--4)}\label{sec:cs54}

We detect a dense gas tracer ${\rm CS(5-4)}$ in one of the continuum spectral windows. This ${\rm CS(5-4)}$ is centrally concentrated, filling in the hole of the $^{12}{\rm CO(2-1)}$ map. The integrated intensity map, spectrum, and radial profile of ${\rm CS(5-4)}$ are shown in the panel-b, c, and d of Figure \ref{densegas}. The detection of ${\rm CS(5-4)}$ is significant above 20$\sigma$ ($\sigma\sim0.3$ mJy beam$^{-1}$ \kms) but in a very low-velocity-resolution spectral window with only 128 channels over 2 GHz band width ($\sim$40~\kms). We estimate the total flux is $1.76\pm0.42$ Jy \kms~with 10\% of the error budget comes from the flux calibration uncertainty of ALMA data.  The ${\rm CS(5-4)}$ line is more centrally concentrated than the $^{12}{\rm CO(2-1)}$ emission as shown in the moment 1 and moment 2 maps and the PVD in Figure \ref{densegas1}. These features suggest the ${\rm CS(5-4)}$ line is an alternative transition that could provide a better constraint on the \Mbh~than the $^{12}{\rm CO(2-1)}$ line. This is because the central concentration of ${\rm CS(5-4)}$ would recover the high-velocity upturn in the data at the very center that is missing in the current $^{12}{\rm CO(2-1)}$ map due to the central hole. However, the velocity resolution of ${\rm CS(5-4)}$ is not good enough to perform such dynamical models. A {\it higher-velocity-resolution} observation of ${\rm CS(5-4)}$ with $\sim$30 times improvement in velocity resolution is required to reduce the uncertainty on \Mbh~determination.

%%%%%%%%%%%%%%%%%%%%%%%%%%%%%%%%%%%%%%%%%
\begin{figure*} 
\centering\includegraphics[scale=0.580]{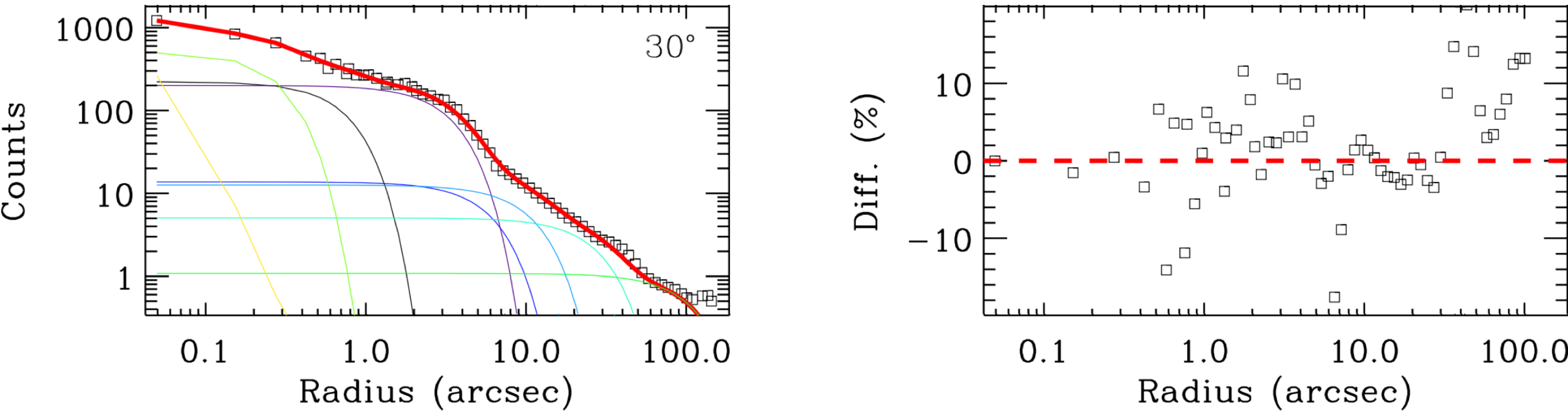}
	      \includegraphics[scale=0.605]{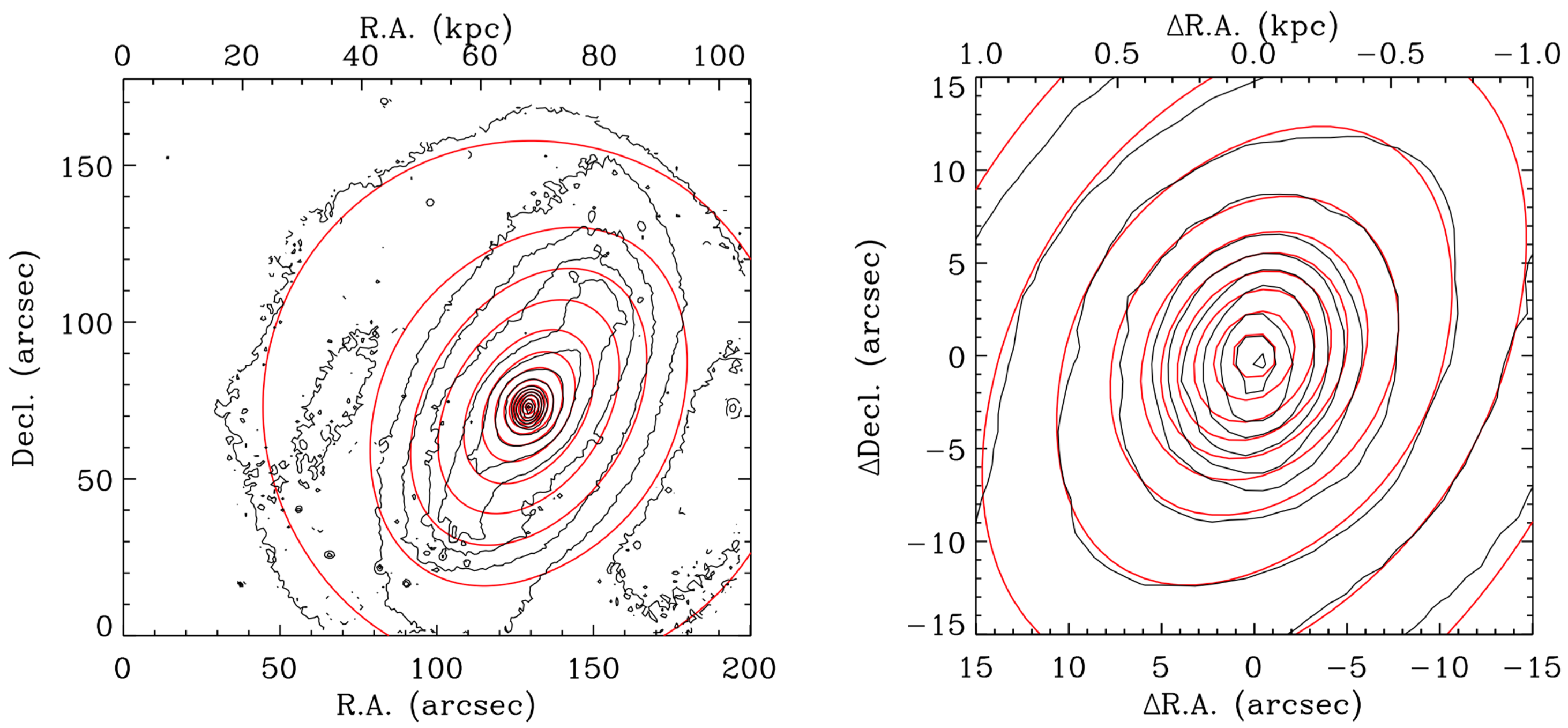}
\caption{Top plots: the comparison between the \hst/WFC3 F160W photometry of NGC 3504 (open squares) and its corresponding best-fit light MGE models (red solid line), which are summed up from multiple Gaussians (color thin lines), is shown in the left panel. We show the best-fit model that is projected along 5$^{\circ}$-wide sector and has an angle of 30$^{\circ}$ between the major and the minor axis. The fractional residuals {\tt (Data-Model)/Data} is shown in the right panel with the agreement between data and model within 15\% across 100$\arcsec$. Bottom plots: the comparison between the F160W light distribution and its best-fit MGE model in the form of 2D light surface density contours for NGC 3504 at the whole galactic scale (left panel) and the central zoom-in of $30\arcsec\times30\arcsec$ field-of-view (FOV; right panel). Black contours show the data, while the red contours show the model to highlight the agreement between data and model at the same radii and contour levels. The mismatch between data and model on the south side of the nucleus is caused by the small jet due to AGN activity visually in Figure \ref{hstimage}, which has been masked out during the MGE fit.}  
\label{massmodel}   
\end{figure*}
%%%%%%%%%%%%%%%%%%%%%%%%%%%%%%%%%%%%%%%%%

%%%%%%%%%%%%%%%%%%%%%%%%%%%%%%%%%%%%%%%%%%%%%%%%%%%%
\section{Creating a Mass Model}\label{sec:massmodel}

In galaxies the stellar \ml~can vary due to the presence of complex nuclear stellar populations \citep[][N18; N19]{Seth10a, McConnell13, Ahn17, Nguyen17, Ahn18}. This variation causes large uncertainties on dynamical \Mbh~estimates \citep[][henceforth N17; N19]{Nguyen17}. Unlike our work in N17 and N19, we are lacking the nuclear stellar spectroscopic information for NGC 3504. However, we examine the nuclear $J-H$ color map of NGC 3504. Here, we assume \hst/WFC3 F110W $\approx$ $J$ and F160W $\approx$ $H$.  We find a constant color of $J-H\sim0.85$ mag across the FOV where we measure the $^{12}{\rm CO(2-1)}$ kinematics, although there is variability in few central dust lances and at larger scale of the galaxy; those will be masked out. Photometry examination with Jacobus Kapteyn Telescope (JKT) images also found a constant $B-I\sim1.3$ across the FOV of 5$\arcsec$ \citep{Knapen02}. The effect of uniform $J-H$ and $B-I$ colors in the region of interest suggests we can use a constant \ml~to create the mass model for NGC 3504. We utilize the MGE code\footnote{\url{https://www-astro.physics.ox.ac.uk/~mxc/software/\#mge}} \citep{Emsellem94a, Cappellari02} to decompose the light surface density into individuals Gaussian components that can then be deprojected. We use the \texttt{mge\_fit\_sectors IDL} version 4.14 \citep{Cappellari02} to fit the F160W image and deconvolve the effects of a PSF. 

We first parameterize the PSF using Gaussian functions in the first MGE fit, then use them as an input during the second MGE fit to obtain a deconvolved MGE model of the galaxy. This PSF MGE model is tabulated in Table~\ref{tab_psfmges}. During the second MGE fit, we supply a mask map that masks out pixels contaminated by prominent dust lanes, a small-jet like feature, and bright stars near the center, as well as the bar that connects the central bulge and the larger structure along the major axis seen in the left panel of Figure \ref{hstimage}. This bar causes a twist on the MGE model if we allow the PA changing as a free parameter, which cannot be modeled in an axisymmetric dynamical models. In this work, we fix the PA~$=-28^{\circ}$ during the MGE fit based on the orientation of the whole galactic disk for an axisymmetric model.

%%%%%%%%%%%%%%%%%%%%%%%%%%%%%%%%%%%%%%%%%%%%%%
\begin{table}
\caption{MGE parameters of the \hst/WFC3 IR F160W PSF}
\centering
\begin{tabular}{ccccc}
\hline\hline   
$j$  &Total Count&$\sigma$& $a/b$ \\
     &of Gaussian$_j$&(arcsec)&   \\
 (1) &   (2)         &   (3)  &(4)\\
\hline
1 &  0.343   &   0.049  &   0.999 \\
2 &  0.538   &   0.130  &   0.999 \\
3 &  0.060   &   0.403  &   0.999 \\
4 &  0.033   &   0.897  &   0.996 \\
5 &  0.033   &   1.716  &   0.993 \\
\hline
\end{tabular}
\tablenotemark{}  
\tablecomments{Column 1: Gaussian component number. Column 2: the MGE model which represented for the total light of each Gaussian.  Column 3: the Gaussian width (FWHM or dispersion) along the major axis.  Column 4: the axial ratios.}
\label{tab_psfmges}
\end{table}
%%%%%%%%%%%%%%%%%%%%%%%%%%%%%%%%%%%%%%%%%%%%%%

Each Gaussian can be deprojected analytically with a specific axis ratio (or inclination) to reconstruct a 3D light distribution model. To optimize data vs. MGE model, we set the axis ratio in the range of $q=0.65-1.0$. This MGE parameters are listed in Table \ref{tab_f160wmges}. Note that we assume \hst/WFC3 F160W solar Vega absolute magnitude system\footnote{\url{http://mips.as.arizona.edu/~cnaw/sun.html}}.% of $\sim$3.37 mag. 

We show the radial light surface brightness density and its best-fitting MGE model in the upper-left panel of Figure \ref{massmodel}, while the fractional residual indicates the agreement between the best-fit model and the data are shown in the upper right panel.  The 2D light surface density is also plotted with their MGE model in the bottom panels of Figure \ref{massmodel} with the whole image on the left and the zoom-in $30\arcsec\times30\arcsec$ FOV on the right.  The figure shows the agreement of the data and its MGE model at the same radii and contours levels. The difference between the best-fit model and the data are $<$15\% across the FOV ($>$100$\arcsec$). %, although in N17 we found that our dynamical models were unchanged when the MGE model was larger than 6$\arcsec$. 

%%%%%%%%%%%%%%%%%% Table3: HST H band  %%%%%%%%%%%%%%%% 
\begin{table}
\caption{The \hst/WFC3 IR 160W MGE Model of NGC 3504}
\centering
\begin{tabular}{cccc}
\hline\hline   
$j$ &$\log$(Luminosity Density)&$\sigma^{\prime}$&$a/b$\\
    &($L_{\odot}/{\rm pc}^2$)&  (arcsec)         &     \\
 (1)&       (2)                &  (3)            &  (4)\\
\hline
1$^\star$ & 4.641 & 0.049& 0.889 \\ 
2    &  4.639  &     0.210   & 1.000 \\
3    &  4.268  &     0.546   & 1.000 \\
4    &  4.217  &     2.649   & 0.780 \\
5    &  3.051  &     4.325   & 1.000 \\
6    &  3.016  &     9.147   & 0.650 \\
7    &  2.620  &    23.588   & 0.650 \\
8    &  1.949  &    78.853   & 1.000 \\
\hline
\end{tabular}
\tablenotemark{}  
\tablecomments{MGE models using in KinMS and Title-ring model fits in Sections~\ref{sec:kinms} and \ref{sec:ring}.  Column 1: Gaussian component number. Column 2:    the MGE model represents for the galaxy luminosity model.  Column 3: the Gaussian width (FWHM or dispersion) along the major axis.  Column 4: the axial ratios. $^\star$ means the leave out marginalized resolved Gaussian component during the dynamical models.}
\label{tab_f160wmges}
\end{table}
%%%%%%%%%%%%%%%%%%%%%%%%%%%%%%%%%%%%%%%%%%%%%%

%%%%%%%%%%%%%%%%%%%%%%%%%%%%%%%%%%%%%%%%%%%%%%%%%%
\section{KinMS Dynamical Modeling}\label{sec:kinms}

In this section, we describe the KINematic Molecular Simulation dynamical model \citep[KinMS\footnote{\url{https://github.com/TimothyADavis/KinMS}};][]{Davis14} that we employ to measure the \Mbh~in NGC 3504 and state our results. 

The KinMS model is a mm-wave observational simulation tool developed by \citet{Davis13} using the Markov Chain Monte Carlo (MCMC) method to simulate dynamical motion of molecular/atomic cold gas distributions under the influence of galaxy and central dark massive object's gravity. In practice, the MCMC technique allows to plug in initial guesses for the true gas distribution and kinematics based on the assumption that the gas is rotating in circular orbits. The model creates a simulated data cube, which can be compared to the observed data via $\chi^2$-minimized likelihood function \citep{Davis17, Onishi17, Davis18}. The model explores the parameter space using Bayesian analysis technique with a set of walkers using the {\tt emcee} algorithm \citep{Foreman-Mackey13} and affine-invariant ensemble sampler \citep{Goodman10}. The relative likelihood for each walker at each step will determine for their next move through parameter space. The best-fit model parameters are then obtained from the posterior distribution of the full pool of model parameters. In this work, we use the Python code {\tt KINMSpy\_MCMC}\footnote{\url{https://github.com/TimothyADavis/KinMSpy\_MCMC}} to find the best set of model parameters.

%%%%%%%%%%%%%%%%%%%%%%%%%%%%%%%%%%%%%%%%%%%%%%%%%%%%%%%%
\subsection{Nuclear Gas Morphology}\label{ssec:gasmorph}

The KinMS model requires surface brightness distribution function of gas, $\Sigma(r)$, that describes the gas morphology radially. We assume an axisymmetric morphology for the nuclear $^{12}{\rm CO(2-1)}$ gas in NGC 3504 as a thin disk distributed continuously out to $\sim$$5\arcsec$.

%%%%%%%%%%%%%%%%%%%%%%%%%%%%%%%%%%%%%%%%%
\begin{figure} 
\centering
	\includegraphics[scale=0.48]{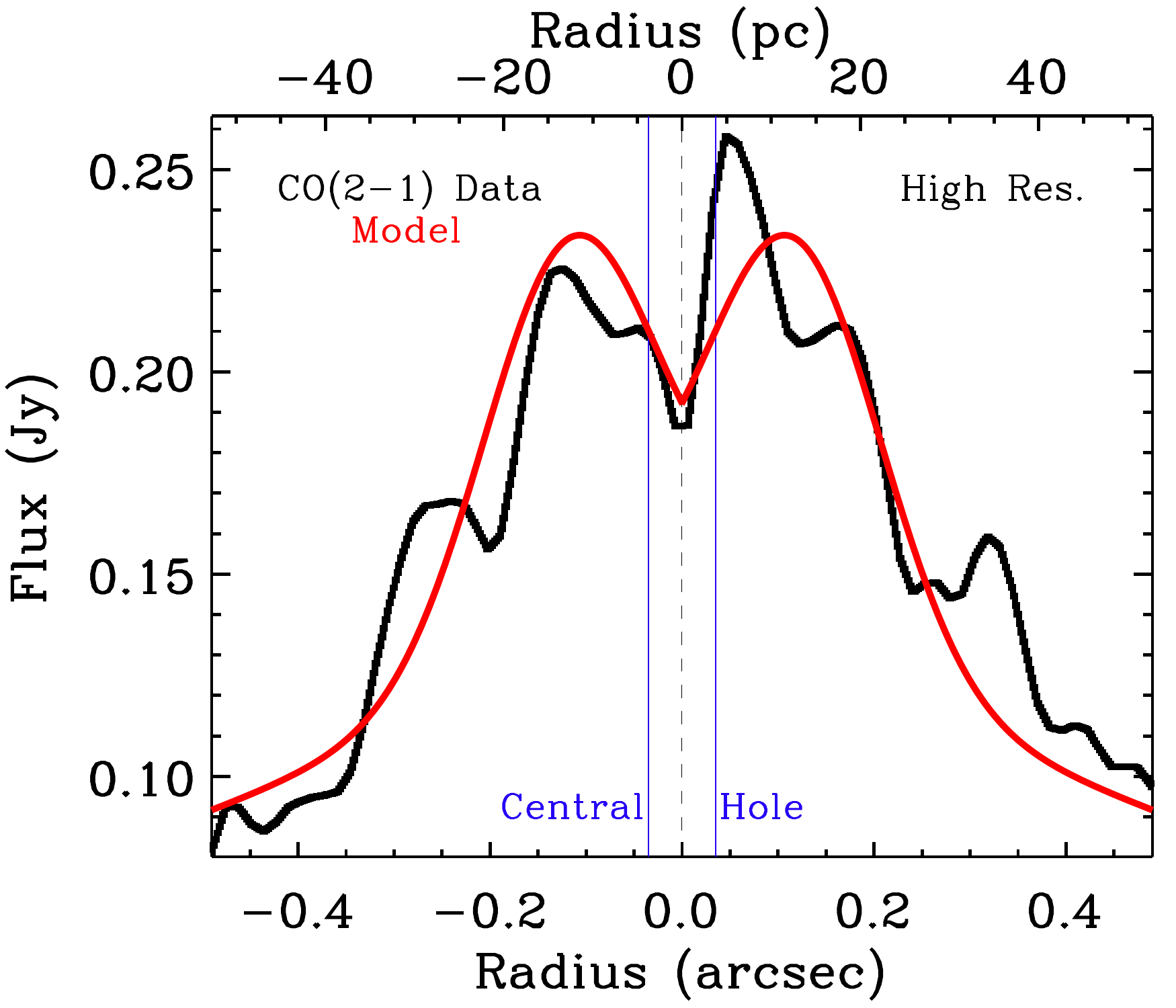} 
\caption{The nuclear gas morphological distribution of the combined cube (high spatial resolution) shows with a cut along major axis through the center of NGC 3504 integrated intensity map $(1\arcsec\times1\arcsec)$. The data are plotted in black diamond, while and our best-fit KinMS model for our chosen surface brightness profile is plotted in red solid line.} 
\label{gasmorph}   
\end{figure}
%%%%%%%%%%%%%%%%%%%%%%%%%%%%%%%%%%%%%%%%%

%%%%%%%%%%%%%%%%%%%%%%%%%%%%%%%%%%%%%%%%%
\begin{figure}[!ht]  
\centering\includegraphics[scale=0.165]{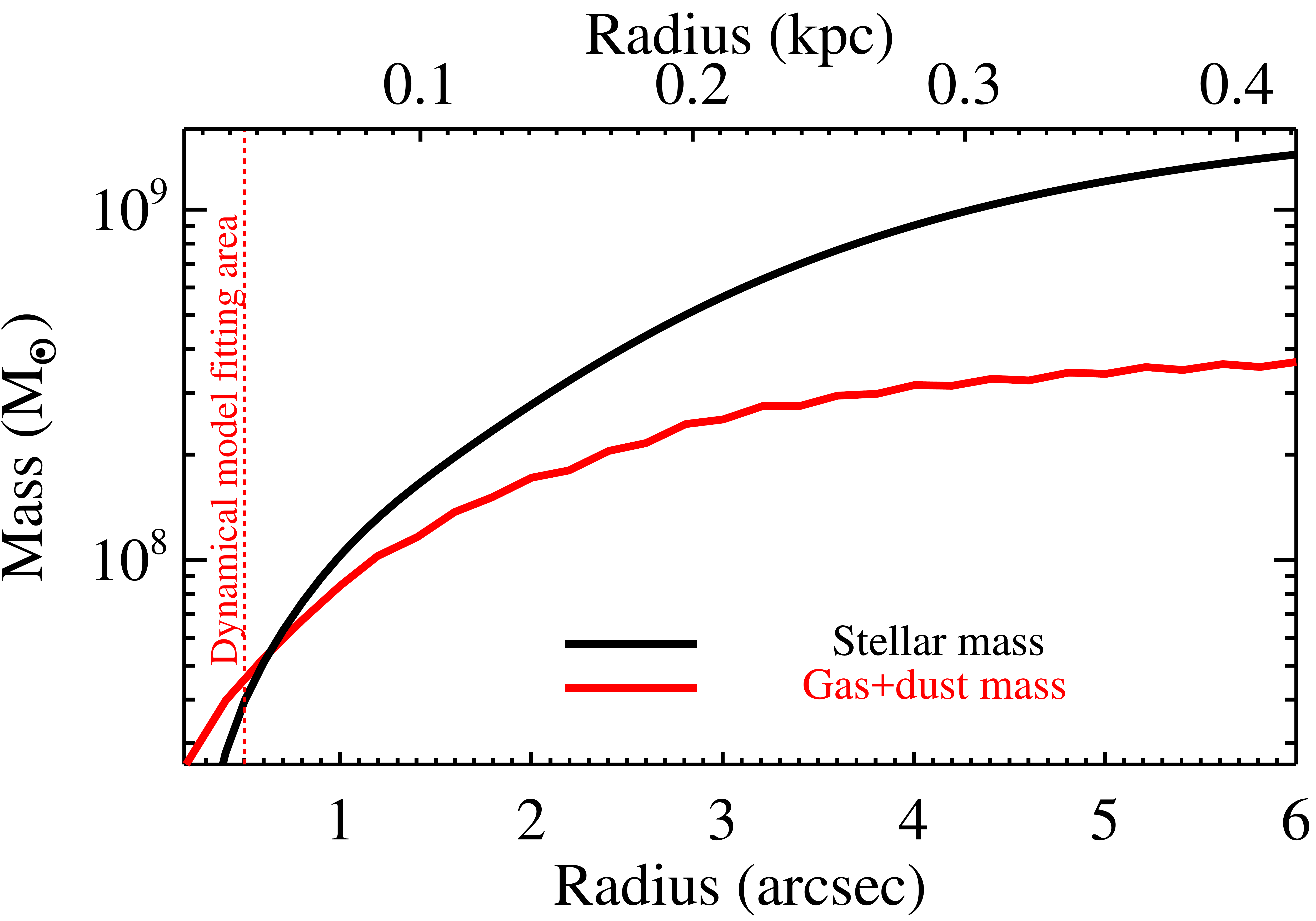} 
\caption{One dimensional (1D) cumulative mass profile of stellar mass (black) and gas + dust (red) components plotted within the radius of 6$\arcsec$ from the center.} 
\label{mass1d}   
\end{figure}
%%%%%%%%%%%%%%%%%%%%%%%%%%%%%%%%%%%%%%%%%

%%%%%%%%%%%%%%%%%%%%%%%%%%%%%%%%%%%%%%%%%
\begin{figure*}[!th]  
\hspace{-10mm}
	\includegraphics[scale=0.27]{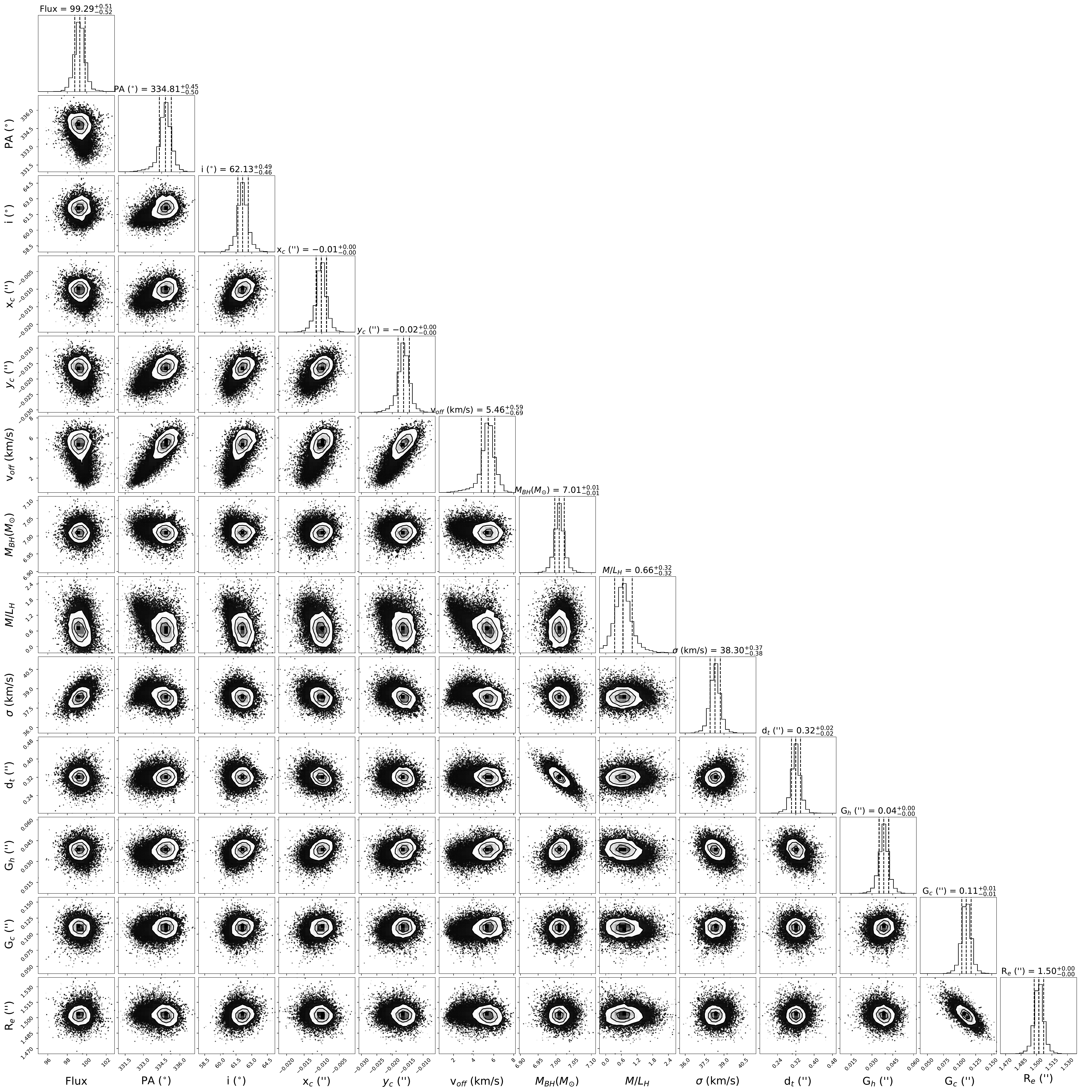}
\caption{The multidimensional parameter space posterior distributions that are explored by the KinMS dynamical model to the combined cube (high resolution) from the central $1\arcsec\times1\arcsec$ field-of-view (FOV) of NGC 3504. The top panel of each column is an one-dimensional (1D) histogram shows the marginalized posterior distribution of that parameter, with 68\% (1$\sigma$) confidence interval, which is corresponding to the innermost contour showed in the two-dimensional (2D) marginalization of those fitted parameters in the panels below. We also show the distribution of the parameter space within 2$\sigma$ (97.0\%) and 3$\sigma$ (99.7\%) confidence intervals within the second and the outermost contours, respectively. See Table \ref{fittable} for a quantitative description of the likelihoods of all fitting parameters. Note that \Mbh~is flat in log scale, others are in linear scales.}
\label{posterior}   
\end{figure*}
%%%%%%%%%%%%%%%%%%%%%%%%%%%%%%%%%%%%%%%%%

%%%%%%%%%%%%%%%%%%%%%%%%%%%%%%%%%%%%%%%%%
\begin{figure*}[!th]  
\centering
    \includegraphics[scale=0.61]{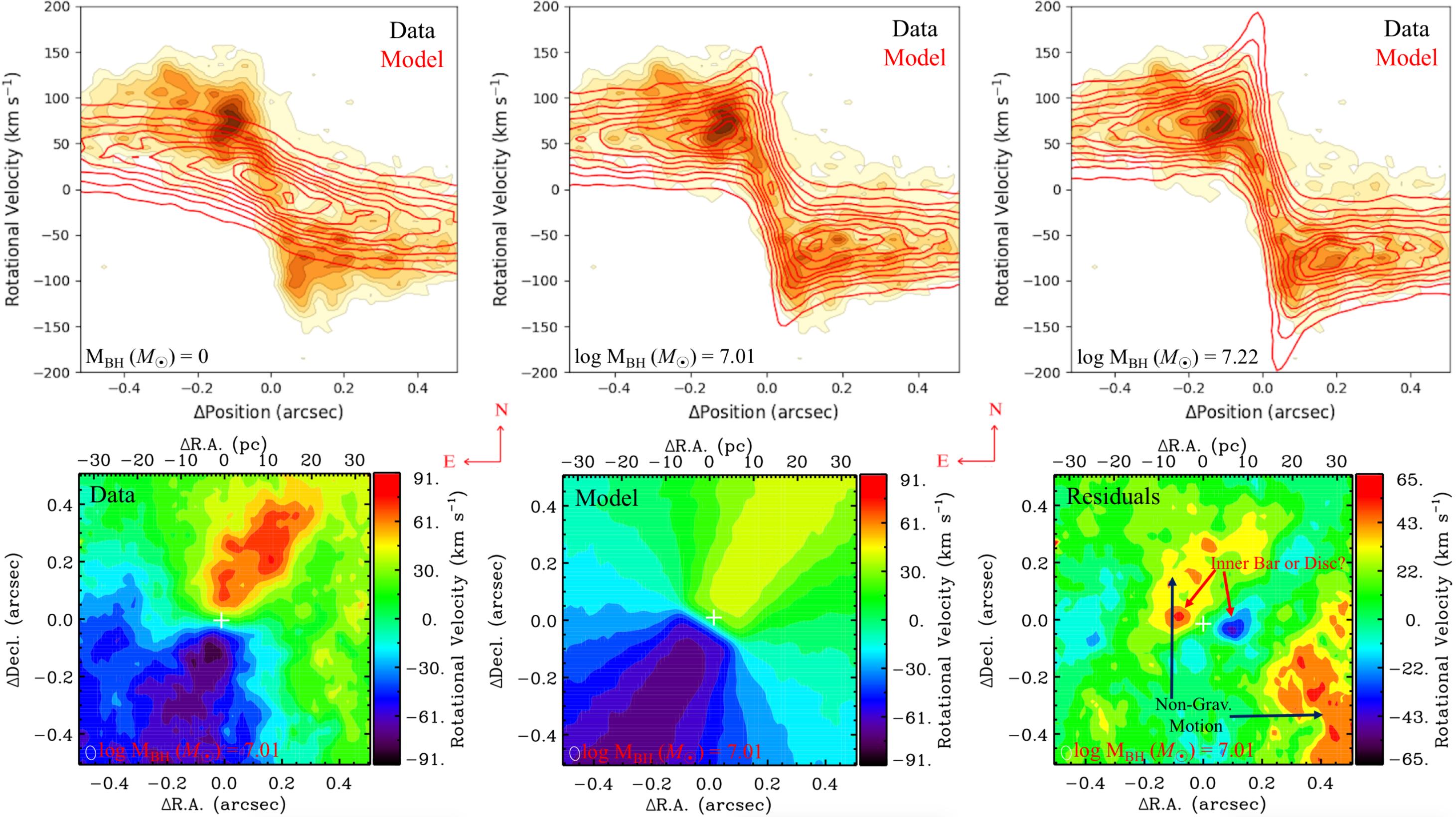} 
    \caption{Top panels: comparisons between the combined cube (high resolution) with a few specific KinMS dynamical model including the best fit. The position-velocity diagram (PVD) of the $^{12}{\rm CO(2-1)}$ emission in NGC 3504 extracted along the major-axis (orange scale and grey contours) in the same manner we did in Figure \ref{pvd}. The model PVDs are extracted in an identical fashion from models that are different by the central \Mbh~(red contours) only. The left panel shows the case with no SMBH, the central panel shows our best-fitting \Mbh, and the right panel has an overly massive SMBH. The models with no/overly large SMBHs are clearly not the good fits to the data in the central parts.  Bottom panels: The central $1\arcsec\times1\arcsec$ two-dimensional (2D) velocity map of the $^{12}{\rm CO(2-1)}$ emission in NGC 3504 extracted from the bottom-left panel of Figure \ref{maps} (left), the velocity map derived from the KinMS model with the best-fit \Mbh~(middle), and the residual velocity map between the data and the model, {\tt (Data - Model)} (right). White plus indicates the kinematic center and also the galaxy center. The synthesized beam of the observation is shown as an ellipse at the bottom left of the panel.}
\label{bestfit_pvd}   
\end{figure*}
%%%%%%%%%%%%%%%%%%%%%%%%%%%%%%%%%%%%%%%%%

We examine the morphological distribution of the nuclear $^{12}{\rm CO(2-1)}$ gas by extracting the flux of the low-resolution cube in a central box with seven pixels in width along the major axis and plotting the slit within the area of $1\arcsec\times1\arcsec$ where we actually fit for the KinMS model in Figure \ref{gasmorph}. Note that we limit our fitting region within $1\arcsec\times1\arcsec$ to avoid the contamination from the outer rotating disk. We model the inner gas disk morphology using a (1) a central offset single-Gaussian function to describe the ring morphology and central attenuated hole of the $^{12}{\rm CO(2-1)}$ integrated intensity map and (2) an exponential disk, which is fitted quite well to the outer disk of the nuclear gas reservoirs. Overall, we describe the nuclear gas distribution by three parameters including the exponential disk length scale ($R_e$), the Gaussian FWHM ($G_h$), and center peaks ($G_{\rm c}$). 

The best-fit profile of the nuclear gas distribution produced by the KinMS model is plotted as a red solid line in Figure \ref{gasmorph}. We should note that the KinMS model matches the observations by fitting a set of free parameters including the above gas morphology and the total flux (scaling factor), PA, and $i$ of the gas disk, as well as its kinematic center in R.A. ($x_c$), Decl. ($y_c$), and velocity offset ($v_{\rm off.}$), \ml, $M_{\rm BH}$, gas disk thickness ($d_{\rm t}$), and internal velocity dispersion of the gas ($\sigma$). This dispersion is assumed as a spatial-constant parameter. Due to the axisymmetric model, we constraint $i$ and PA in each fit as a single value throughout the disk to avoid warp. So, our initial KinMS model thus has 13 free parameters listed in Table \ref{fittable}. 

%%%%%%%%%%%%%%%%%%%%%%%%%%%%%%%%%%%%%%%%%%%%%%
\subsection{Velocity Model}\label{ssec:vmodel}     

We assume the gas rotates around the galaxy center in circular orbits and that vary radially. The KinMS model takes in the circular velocity curve in the equatorial plane of the CND to produce a simulated cube. This circular motion of the gas is controlled by the gravitational potential of the galaxy calculated from the stellar mass model (Section \ref{sec:massmodel}) and the point source potential of the SMBH described in \citet{Cappellari02}.  We calculate this circular velocity profile using the {\tt mge\_vcirc} procedure within the Python Jeans Axisymmetric Modelling \citep[{\tt JamPy}\footnote{\url{http://www-astro.physics.ox.ac.uk/~mxc/software/\#jam}};][]{Cappellari08} package.

%%%%%%%%%%%%%%%%%%%%%%%%%%%%%%%%%%%%%%%%%%%%%%%%
\subsection{Fitting Process}\label{ssec:fitting} 

We first run the KinMS model with an area of 300 pixels $\times$ 300 pixels ($4\arcsec\times4\arcsec$), where roughly covers most of important features of the molecular gas disk, and a number of iterations of $10^5$. To avoid the possible effect of AGN on our dynamical \Mbh~and \ml~estimates simultaneously, we leave out the marginalized resolved Gaussian component during the model fitting; the first MGE component denoted by an upper star in Table \ref{tab_f160wmges}. From this run, we obtain rough estimates of the model parameters. The second fit then fits in the central 80 pixels $\times$ 80 pixels ($1\arcsec\times1\arcsec$ or 68 pc $\times$ 68 pc) area only, and starts with reasonable flat priors for all free parameters to ensure our kinematic fitting process converge. These priors are the best fit from the first fit. We note that the prior on the \Mbh~was flat in log-space, while the inclination of the gas disk was allowed to vary over the full physical range allowed by the MGE model. The search ranges of these parameters are shown in column 2 of Table \ref{fittable}, and good fits were always found well within these ranges. To account the covariance between pixels, which describes the neighbouring spaxels strongly correlated by the synthesized beam due to the Nyquist spatial sample of ALMA data \citep{ Davis17, Onishi17, Davis18}, we increase the RMS by a factor $(2N\times M)^{1/4}$, where $N$ and $M$ are the numbers of spaxels and velocity channels included in the fit \citep{Mitzkus17}.  

We take in to account the distributions of any interstellar material including gas and dust within the fitting region into our mass model. The total flux in this region is $6.9_{\pm0.7}^{\pm0.2}$ Jy \kms, where the upper and lower uncertainty are systematic and $\sim$10\% flux calibration uncertainty in the measurement and ALMA data, respectively.  This total flux provides an estimate on the total mass of H$_2$ gas of $M_{\rm H_2}=(4.2\pm0.5)\times10^7$\Msun~by assuming the line ratio $^{12}{\rm CO(2-1)}$/$^{12}{\rm CO(1-0)}=0.8$ \citep{Bigiel08} and H$_2$-to-CO conversion factor for starburst galaxies: $N({\rm H_2})/I_{(\rm 1-0)}=X_{\rm CO}=0.5\times X_{\rm CO}^{\rm Milky Way}=1\times10^{20}$ cm$^{-1}$ (K \kms)$^{-1}$ for converting $^{12}{\rm CO(1-0)}$ to H$_2$ mass \citep{Kuno00, Kuno07, Bolatto13}. The dust mass of $(5.1\pm0.7)\times10^5$\Msun~is calculated in D. Nguyen et al. in preparation. In Figure \ref{mass1d}, we plot the 1D stellar mass and gas + dust mass profiles as the radial functions simultaneously within the radius of $6\arcsec$. In our fitting area, the gas and dust mass distributions are dominant over the stellar mass, suggesting the gravitational effect of these components plays an important role in determining the \Mbh~in NGC 3504 accurately.  We add these masses in the KinMS \texttt{gasGrav} mechanism, assuming the dust and gas is co-spatial distribution. 

We run the model fits with velocity channels are from $-$200 to 200 \kms, w.r.t. the systemic velocity of 1525 \kms. The total number of iterations are set to be $3\times10^6$ and the first 20\% of iterations are considered as the burn-in phases to produce our final posterior probability distributions of these 13 model parameters. 

%%%%%%%%%%%%%%%%%%%%%%%%%%%%%%%%%%%%%%%%%%%%%
\subsection{Model Results}\label{ssec:result}

We clearly detect an SMBH that causes the increasing Keplerian curve towards the center. The best-fit model parameters are identified directly from our Bayesian analysis, relying on the likelihood probability distribution functions (PDFs) generated via MCMC. We choose the best-fit parameters that are the medians of the parameter posterior PDFs. Particularly, the probability is marginalized over to produce a best-fit value by taking the median of the marginalized posterior samples for each parameter. The 1$\sigma$, 2$\sigma$, and 3$\sigma$ confidence levels (CLs) or uncertainties are estimated from all models within (16\% and 84\%), (3\% and 97\%), and (0.2\% and 99.8\%) of the PDFs, respectively.  At $3\sigma$ CL, the central \Mbh~is measured at $M_{\rm BH}=1.02^{+0.18}_{-0.15}\times10^7$\Msun, while the stellar \ml$_{\rm H}=0.66^{+1.44}_{-0.65}$ (\Msun/\Lsun) and $i=61.01^{\circ}$$^{+3.11}_{-3.51}$. The reduced chi-square ($\chi^2_{\rm red}$) for the best fit is 1.16, indicating a good fit. We list all the best-fitting parameters and their likelihoods in Table \ref{fittable}. 

Figure \ref{posterior} shows the 1D and 2D marginalization of the physical parameters included in the fit. All correlated parameters are well constrained by the data. We show the observed PVD overlaid with the PVD of the best-fit model in Figure \ref{bestfit_pvd}. To enhance the illustration of our result and how good the best-fit model fit to the data, we add the models with no SMBH and with an overly large \Mbh. The best-fit model with an appropriate \Mbh~clearly reproduces the kinematics of the molecular gas better than the other two. The velocity map of the data, the velocity map of the KinMS best-fit model, and the velocity-residual map ${\tt (Data-Model)}$ are shown in the bottom-row plots of Figure~\ref{bestfit_pvd}. 

We find evidence of (1) some non-gravitational motions resulted in high residual on the southwest side of the nucleus as a signature of either outflows or caused by a warped inner disk/bar and (2) signature of a nuclear small rotating structure at the scale of $r\sim0\farcs15$. This central structure is oriented along the minor axis. However, its morphology is very ambiguous. The detailed investigations of these nuclear rotating structure and non-circular motions will be discussed in D. Nguyen et al. in preparation.  

In Figure \ref{bestfit_velchan} we show our best-fit model overlaid on the channel maps of our ALMA observations. Also, we demonstrate the same results for the low-resolution data cube in Figure \ref{bestfit_pvd_lowRes} of the Appendix. The full list of the best-fit parameters and their likelihoods are presented in Table \ref{fittable_lowres}. We note that the central $^{12}{\rm CO(1-0)}$ hole does not appear in the low-resolution cube because its size is smaller than the cube's beamsize. We model the central $^{12}{\rm CO(1-0)}$ surface brightness profile within the fitting area by a single Gaussian with the peak tighten to the kinematic center ($G_c=0$), leaving the KinMS model with 11 free parameters. From the result of this fit, we do not find any signatures of the inner rotating structure in the velocity residual map as seen in the case of high-resolution cube fit. This is because the spatial scale of this structure is smaller than the spatial angular resolution of this cube.

With the stellar velocity dispersion of the bulge of $\sigma=119.3\pm10.3$ \kms~\citep{Ho09}, the BH in NGC 3504 has an intrinsic SOI radius $R_{\rm SOI}=G M_{\rm BH}/\sigma^2=4.6$ pc ($0\farcs067$), where $G$ is the gravitational constant, $M_{\rm BH}$ is the \Mbh, and $\sigma$ is the bugle stellar velocity dispersion. However, \citet{Davis14} argue the spatially angular resolution (in pc scale) that is required to detect a BH with specific \Mbh~and $\alpha$ CL in an object with a circular velocity caused by luminous matter in their equation (9). Using this formula for our NGC 3504 observations and expecting to detect the given \Mbh~at $3\sigma$ CL, we find the essential angular resolution of $\sim$3.1 pc ($0\farcs045$). This scale corresponds to the FWHM of the synthesized beamsize of the combined cube. All of these scales mean that our ALMA observations are able to constrain \Mbh~and kinematic properties of the CND in NGC 3504. 

%%%%%%%%%%%%%%%%%%%%%%%%%%%%%%%%%%%
\begin{table*}
\caption{Best-fit KinMS Model Parameters and Statistical Uncertainties for the Combined ALMA Data Cube (High Resolution)}
\centering  
\begin{tabular}{lccccccc} 
\hline\hline       
 Parameter Names (Notations; Units)&\multicolumn{3}{c}{Search Range}&Best Fit&$1\sigma$ Error&$3\sigma$ Error\\
                &              &               &                    &        &  (68\% conf.) &(99.7\% conf.) \\  
        (1)     &              &     (2)       &                    &  (3)   &     (4)       &    (5)        \\  
\hline 
\underline{{\bf Black Hole:}}  &               &                    &        &               &          &    \\  
\Mbh~in log scale ($\log_{10} M_{\rm BH}$; \Msun)& 4.00 &$\longrightarrow$&9.00& 7.01 &$-$0.01, +0.01&$-$0.07, +0.07\\
Mass-to-light ratio in $H$-band (\ml$_H$; \Msun/\Lsun)  & 0.01 &$\longrightarrow$&3.00& 0.66 &$-$0.32, +0.32&$-$0.65, +1.44\\
\underline{{\bf Molecular Gas Disc:}}                   &      &                 &    &      &              &              \\  
Position angle (PA; $^{\circ}$)                         &150.0 &$\longrightarrow$&360.0&334.81&$-$0.50, +0.45&$-$3.04, +2.03\\
Inclination angle ($i$; $^{\circ}$)                     & 45.0 &$\longrightarrow$&90.0&62.13 &$-$0.46, +0.49&$-$3.51, +3.11\\
Gas velocity dispersion ($\sigma$; \kms)                & 1.0  &$\longrightarrow$&80.0& 38.30&$-$0.38, +0.37&$-$0.06, +0.06\\   
Disk thickness ($d_t$; arcsec)                          & 0.01 &$\longrightarrow$&1.50& 0.32 &$-$0.02, +0.02&$-$0.05, +0.05\\
Gaussian peak (G$_c$; arcsec)                           & 0.01 &$\longrightarrow$&1.00&0.04  &$-$0.00, +0.00&$-$0.02, +0.02\\
Gaussian HWHM (G$_h$; arcsec)                           & 0.01 &$\longrightarrow$&1.00&0.11  &$-$0.01, +0.01&$-$0.03, +0.03\\
Exponential disk length scale ($R_e$; arcsec)           & 0.01 &$\longrightarrow$&10.0&1.50  &$-$0.00, +0.00&$-$0.02, +0.02\\     
\underline{{\bf Nuisance Parameters:}}                  &      &   &           &              &           &        \\  
CO surface brightness scaling factor (Flux)             &  10.0 &$\longrightarrow$&500.0& 97.26 &$-$0.52, +0.51&$-$2.21, +2.75\\
R.A. offset ($x_c$; arcsec)                             &$-$0.10&$\longrightarrow$&+0.10&$-$0.01&$-$0.00, +0.00&$-$0.01, +0.01\\
Decl. offset ($y_c$; arcsec)                            &$-$0.10&$\longrightarrow$&+0.10&$-$0.02&$-$0.00, +0.00&$-$0.01, +0.01\\
Systemic velocity offset ($v_{\rm off.}$; \kms)         &$-$10.0&$\longrightarrow$&+10.0&$-$5.46&$-$0.69, +0.59&$-$2.61, +1.83\\
\hline
\end{tabular}
\tablenotemark{}
\tablecomments{Column 1: A list of the fitted model parameters. Column 2: A list of the priors of the fitted model parameters and their search ranges. The prior are constructed in the uniform linear space with only SMBH is in logarithmic space. Columns 3--5: The best-fit value of each parameter and their uncertainties at 1$\sigma$ and 3$\sigma$ confident levels. The R.A., Decl., and velocity offset nuisance parameters are defined relative to the ALMA data phase center position ($11^{\rm h}03^{\rm m}11^{\rm s}.205$, $+27^{\circ}58^{\prime}20\farcs80$, $V_{\rm sys.}=1525$ \kms).}
\label{fittable}
\end{table*}
%%%%%%%%%%%%%%%%%%%%%%%%%%%%%%%%%%%

%%%%%%%%%%%%%%%%%%%%%%%%%%%%%%%%%%%%%%%%%%%%%%%%%%%%%%%%%%%%%%%%%%%%%%%%%%%%%%%%
\subsection{Uncertainty Sources on the \Mbh~Estimate}\label{ssec:masserror}

%%%%%%%%%%%%%%%%%%%%%%%%%%%%%%%%%%%%%%%%%%%%%%%%%%%%%%%%%%%%%%%%
\subsubsection{Stellar Mass Models}\label{sssec:stellarmasserror}

We test the robustness of the mass model by create a new MGE mass model using the \hst/WFC3 F110W image (approximate to $J$-band), which gives the best-fit \Mbh~of $8.05_{-0.21}^{+0.27}\times10^6$\Msun~and \ml$_J=0.76_{-0.61}^{+1.17}$ (\Msun/\Lsun). Moreover, the way we fit the light MGE model from \hst~images also causes some error on the mass model. Since large structure of NGC 3504 hosts an outer Lindblard resonance \citep[OLR;][]{Buta92, Buta96}, while its smaller scale contains central elongated bar \citep{Kuno00}.  We rerun two MGE fits as follows: (1) there is no constrain on $q$ to get a MGE with elongated barred Gaussian dominant, and (2) set $q=0.9-1$  to get a MGE with OLR Gaussian dominant. We have double checked these MGEs within $16\arcsec\times16\arcsec$ central region do not change significantly. The KinMS model with the former MGE gives the best fit ($M_{\rm BH}$, $\ml_{H}$, $i$) = ($8.74_{-0.08}^{+0.09}\times10^6$\Msun, $0.72_{-0.60}^{+1.26}$ (\Msun/\Lsun), 58.34$^\circ$$_{-3.16}^{+3.21}$), while that of the latter MGE gives the best fit ($M_{\rm BH}$, $\ml_{H}$, $i$) = ($1.16_{-0.21}^{+0.26}\times10^7$\Msun, $0.61_{-0.58}^{+1.36}$ (\Msun/\Lsun), 60.45$^\circ$$_{-3.76}^{+3.52}$); other nuisance parameters vary within 15\% compare to the best fit of the default model. We thus conclude that our \Mbh~estimate is robust to the systematic errors of our mass models.

%%%%%%%%%%%%%%%%%%%%%%%%%%%%%%%%%%%%%%%%%%%%%%%%%%%%%%
\subsubsection{AGN and Distance}\label{sssec:agnerror}

The contribution of AGN at the center of NGC 3504 as seen in optical \citep{Ho93} and radio \citep{Deller14} may attribute to the F160W waveband flux at the very center, which is represented by the first MGE component in Table \ref{tab_f160wmges}. We rerun the KinMS model included this component to test for the possible effect of AGN distribution, providing  \Mbh~$=1.30_{-0.05}^{+0.09}\times10^7$\Msun~and \ml$_H$~$=0.58_{-0.51}^{+1.26}$ (\Msun/\Lsun). This systemic uncertainty of the \Mbh~is within our 3$\sigma$ CL, suggesting the impact of AGN on our mass model is insignificant.   

In addition, the \Mbh~estimate is systematically affected by the distance to the galaxy based on the relation $M_{\rm BH}\propto D$. In this work, we assumed the Tully-Fisher distance to NGC 3504 is $D=13.6\pm1.4$ Mpc \citep{Russell02}. Therefore, the systematic distance uncertainties on the \Mbh~are of 10\% as similar as the random uncertainties. We should note that the six existing distance estimates\footnote{\url{https://ned.ipac.caltech.edu/}} to NGC 3504 are in a wide range of 8.7--26.5 Mpc, suggesting the systemic errors due to different distances adopted are much larger than that of the distance estimate which we are using \citep{Russell02} up to $\sim$95\%.

%%%%%%%%%%%%%%%%%%%%%%%%%%%%%%%%%%%%%%%%%%%%%%%%%%%%%%%%%%%%
\subsubsection{Non-circular Motions}\label{sssec:noncircle}

In this work, we assume the gas in NGC 3504 is in purely circular motion. However, non-circular motions (e.g., inflow, outflow, streaming) are usually seen in barred galaxies like NGC 3504, where the non-circular motion is primarily caused by gas streaming along the bar and could affect our analysis significantly. The large scale bar ($>$20$\arcsec$) does not strongly affect to our dynamical results because our interested in region is deeply small within the bulge and the nuclear molecular gas disk resides within the radius of $\lesssim$5$\arcsec$. 

However, from the residual map of Figure \ref{bestfit_pvd} we find the orientation of the nuclear small rotating structure, which is perpendicular to the projected major axis. \citet{Randriamampandry15} shows the non-circular motions derived from mass profiles of strongly barred galaxies vary dramatically if the bar is orientated at specific angles w.r.t. the LOS. Particularly, there are under/overestimates the circular motions when the bar is parallel/perpendicular to the projected major axis. As a SAB(s)ab type galaxy like NGC 3504, the expected overestimate rotation is minimum as suggested by Figure 2 of \citet[][top-row plots]{Randriamampandry15}. This is because the contribution of gas from the bulge is compensated for gas streaming motion along the bars.    

To calculate the effect of this non-circular motion on our dynamical results, we run the KinMS model with a new velocity model extracted along the major axis of the moment 1 map (Figure \ref{maps}) using the \texttt{IDL Kinemetry} code \citep{Krajnovic06} instead of the velocity model built up from the mass MGE model in Section \ref{ssec:vmodel}.  The best-fit model gives \Mbh~of $1.12_{-0.08}^{+0.07}\times10^7$\Msun~and \ml$_H=0.64_{-0.56}^{+1.38}$ (\Msun/\Lsun), while all other parameters are variable within 9\% compare to the best-fit value listed in Table \ref{fittable}.
 
%%%%%%%%%%%%%%%%%%%%%%%%%%%%%%%%%%%%%%%%%%%%%%%%%%%%%%%%%%%%%
\subsubsection{Gas Velocity Dispersion}\label{sssec:gasdisp}

During the above analysis, we assumed a constant gas velocity dispersion. In reality, the velocity dispersion could vary with radius and azimuth within the gas disk. In the central part of the galaxy, where beam smearing is important, an increase of velocity dispersion could lead to overestimated \Mbh.  To quantify this effect, we allow a variable velocity dispersion as a function of radius. We test the turbulent velocity dispersion profile using the following prescriptions for $\sigma(r)_{\rm gas}$:

 {\it (a) Linear gradient:} $\sigma(r)_{\rm gas}=a\times r + b$, where $a$ and $b$ are free parameters. We find $a\approx0$ and $b=37.57$ \kms, and KinMS results are consistent with the default model of constant velocity dispersion. 

 {\it (b) Exponential:} $\sigma(r)_{\rm gas}=\sigma_0\exp{(-r/r_0)}+\sigma_1$, where $\sigma_0$, $r_0$, and $\sigma_1$ are free parameters. As discussed in \citet{Barth16a}, we set the lower boundary for $\sigma(r)_{\rm gas,\;min} =1$ \kms~during the fit to prevent the line-profile widths becoming arbitrarily small. The best-fit KinMS model provides $M_{\rm BH}=9.12_{-0.08}^{+0.10}\times10^6$\Msun~and \ml$_H=0.70_{+1.37}^{-0.62}$ (\Msun/\Lsun) with an exponential dispersion model with $\sigma_0=36.95_{-2.55}^{+1.40}$ \kms, $r_0=1\farcs41_{-1.13}^{+0.30}$, and $\sigma_1=9.89_{-1.27}^{+1.31}$ \kms.  Other best-fit parameter values are consistent with the default model.

{\it (c) Gaussian:} $\sigma(r)_{\rm gas}=\sigma_0\exp{(-(r-r_0)^2/2\mu^2)}+\sigma_1$, where $\sigma_0$, $r_0$, $\mu$, and $\sigma_1$ are free parameters. We allow the parameter $r_0$ to vary over positive and negative values because the line width is sometimes offset from the center and also set the lower boundary for $\sigma(r)_{\rm gas,\;min} =1$ \kms~during the fit. The best-fit KinMS model provides $M_{\rm BH}=8.73_{-0.05}^{+0.09}\times10^6$\Msun~and \ml$_H=0.71_{-0.56}^{+1.28}$ (\Msun/\Lsun) with a Gaussian dispersion model with $\sigma_0=36.96_{-3.10}^{+2.11}$ \kms, $r_0=0\farcs1_{-0.15}^{+0.12}$, $\mu=0\farcs74_{-0.10}^{+0.12}$, and $\sigma_1=10.01_{-1.32}^{+1.28}$ \kms. Other best-fit parameter values are similar to the default model.
	
The minimum $\chi^2_{\rm red}$ are determined at 0.91, 0.87, and 0.83 for the linear, exponential, and Gaussian dispersion profile, respectively.

%%%%%%%%%%%%%%%%%%%%%%%%%%%%%%%%%%%%%%%%%%%%%%%%%%%%%%%%%%%%%%%%%%%%%%%%%%%%%%%%%%%%%%%%%%%%%%%
\subsubsection{Different Observational Scales $\&$ Synthesis Beamsizes}\label{ssec:co21deficit} 

The dynamical model of \Mbh/\ml$_{H}$ estimated from the combined cube is higher/lower than that from the low-resolution cube by 32\%/16\%, respectively. The mass estimate within the synthesis beamsize is also another systemic uncertainty on the \Mbh. We convert the total flux in the beam of the combined cube into molecular mass of $2.5\times10^5$\Msun. This mass is only 10\% of the 3$\sigma$ \Mbh~uncertainty provided by the KinMS model (Table~\ref{fittable}), suggesting the mass uncertainty due to the synthesis beamsize of our observation is minimal. 

%%%%%%%%%%%%%%%%%%%%%%%%%%%%%%%%%%%%%%%%%
\begin{figure*}[!th]
	\centering\includegraphics[scale=0.605]{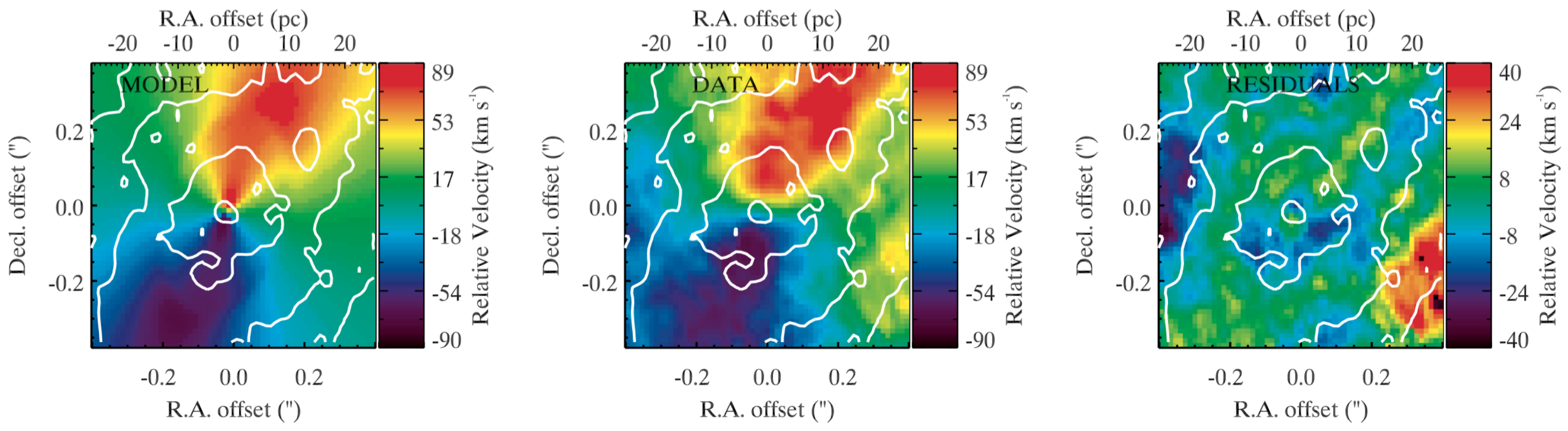} 
\caption{Comparison of the velocity fields of our best-fit title-ring model (left) with the $^{12}{\rm CO(2-1)}$ velocity field (center) and velocity residuals (right) with the area of $0\farcs8\times0\farcs8$ (55 pc $\times$ 55 pc). The while contours denote the integrated intensity of $^{12}{\rm CO(2-1)}$ emission in the the fitted area.}
\label{titlering_map}   
\end{figure*}
%%%%%%%%%%%%%%%%%%%%%%%%%%%%%%%%%%%%%%%%%

%%%%%%%%%%%%%%%%%%%%%%%%%%%%%%%%%%%%%%%%%
\begin{figure*}  
	\centering\includegraphics[scale=0.14]{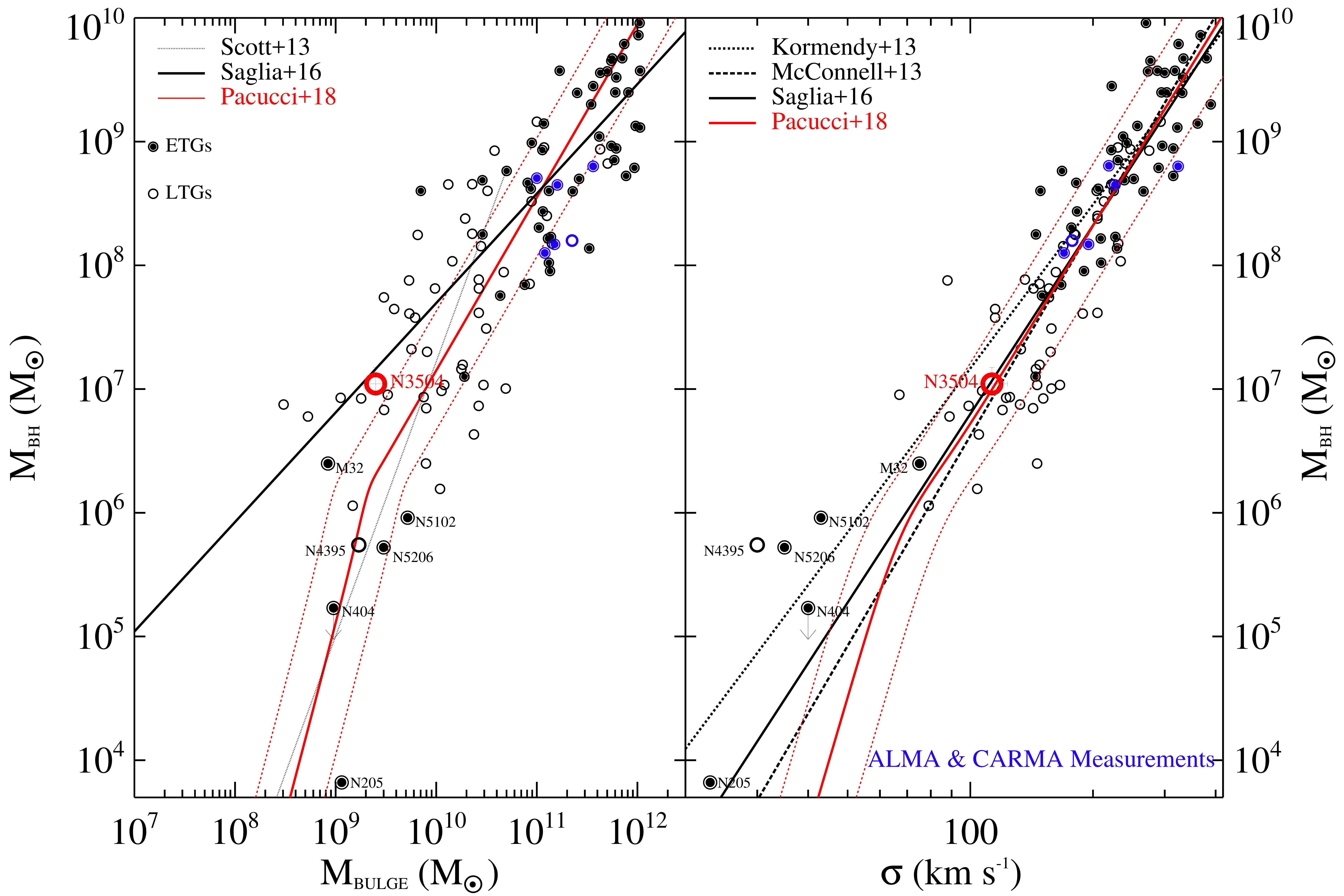}
\caption{Our NGC 3504 \Mbh~(red encircle) in the context of the $M_{\rm BH}-M_{\rm Bulge}$ (left) and $M_{\rm BH}-\sigma$ (right) scaling relations. Six measurements using ALMA \citep{Onishi15, Barth16a, Barth16b, Davis17, Davis18} and Combined Array for Research in Millimeter-wave Astronomy \citep[CARMA;][]{Davis13, Onishi17} observations are plotted in blue, while early-type galaxies (ETGs, black dots within open circles) and late-type galaxies (LTGs, black open circles) are taken from \citet{Saglia16}. The scaling relations of \citet{Scott13, Kormendy13, McConnell13, Saglia16} for ETGs and LTGs are plotted in the dotted, dashed, long-dashed lines, respectively. We also include the theoretical prediction of a bimodality from \citet[][red solid lines]{Pacucci18} and its 1$\sigma$ (red dashed line) uncertainty.}
\label{scaling}   
\end{figure*}
%%%%%%%%%%%%%%%%%%%%%%%%%%%%%%%%%%%%%%%%%

%%%%%%%%%%%%%%%%%%%%%%%%%%%%%%%%%%%%%%%%%%%%%%%%%%%%%%%%%%%%%%%
\section{Thin Disk Tilted-Ring Dynamical Model}\label{sec:ring}

We also constrain the \ml, $i$, and $M_{\rm BH}$ of the central SMBH in NGC 3504 independently using a different dynamical model, which is the so called thin disk tilted-ring model \citep{Begeman87, Quillen92, Nicholson92, Neumayer07, denbrok15}. We use the same kinematic measurements from the nuclear $^{12}{\rm CO(2-1)}$ CND and stellar mass model that are conducted in Sections \ref{sssec:line} and \ref{sec:massmodel}, respectively. The purpose of this test is to examine the robustness of the measured \Mbh~in various assumptions and dynamical models.

%%%%%%%%%%%%%%%%%%%%%%%%%%%%%%%%%%%%%%%%%%%%%%%%%%%%%%%%%
\subsection{The Tilted-Ring Models}\label{ssec:ringmodel}

We model the kinematics of $^{12}{\rm CO(2-1)}$ CND with tilted ring models in a similar approach as for the H$_2$1-0 S(1) transition kinematics in \citet{Seth10a, denbrok15}; N17. The basic idea of these models is that we assume the emitting gas (e.g., CO or H$_2$) is rotating in thin rings on concentric circular orbits around the center of NGC 3504 with a velocity that can be interpolated between the discrete points on the model grid linearly. Each ring of gas is described by three parameters: radius $R$, inclination angle $i$, and azimuthal angle $\theta$ (relative to the projected major axis), which is projected along the LOS allowing the rings become ellipses.  

To test for systemics, we fit three different models to the data as follow:
\begin{enumerate}
    \item The first model mimics the presence of two rotating disks around the center. Both the inner disk and the outer disk have their own PA and $i$, which are free parameters in the fit, as well as the radius at which the model transitions from the inner to the outer disk. 
    
    \item The second model, we assume that the ellipses change their geometry smoothly with radius. We determine the radially varying PA for this model with the {\tt Kinemetry} routine \citep[see footnote 6;][]{Krajnovic06} but allow $i$ to change linearly with radius. 
    
    \item The third model assumes the potential is axially symmetric around the $z$ axis and that the gas is rotating in the $z=0$ plane of the potential.
\end{enumerate}

We assume for the flattening of the mass distribution the standard MGE deprojection.  On each ring at a certain radius, the velocity of the gas is determined by the derivative of the potential with respect to radius \citep[][and N17]{Seth10a, denbrok15} as the following equation: 
\begin{equation}
v^2_c(r)=r\frac{d\Phi(r)}{dr}
\end{equation}
which for the spherically symmetric potential is equal to
\begin{equation}
v^2_c(r)=G(M_\star(<r)+M_{\rm BH})r.
\end{equation}

The gravitational potential of each model is based on the \Mbh~as a point source and the stellar-mass-component distribution, $M_\star(r)$, which is modeled as the deconvolved and deprojected stellar light profile, $L_\star(r)$, multiplied with an additional scaling in \ml.

During the fit, our model generates a spectral cube with the same spectral sampling as the ALMA data cube. The model distributes the flux over the cube based on its spatial distribution and the velocity and velocity dispersion for each ring across the coplanar disk to replicate the observations.  We calculate the $\chi^2$ for both the predicted rotational velocity and velocity dispersion fields, but use only the $\chi^2$ of the rotational velocity field to determine our best-fit model. In order to compare to the KinMS model, we optimize the tilted-ring model in the same fitting area. We ignore gas and dust distribution during the fit in this work; however, in the future work, we will develop the code to add in these masses in another form of MGE separately. 

%%%%%%%%%%%%%%%%%%%%%%%%%%%%%%%%%%%
\begin{table*}
\caption{Best-fit Tilted-Ring Model Parameters and Statistical Uncertainties for the Combined ALMA Data Cube (High Resolution)}
\centering  
\begin{tabular}{lccccccc} 
\hline\hline       
 Parameter Names (Notations; Units)&\multicolumn{3}{c}{Search Range}& Best Fit& $1\sigma$ Error  &$3\sigma$ Error \\
                                   &     &          &               &         &   (68\% conf.)   &(99.7\% conf.)  \\  
  (1)                   &            (2)           &   (3)   &      (4)         &    (5) &    (6)  &    (7)      \\  
\hline 
\underline{{\bf Model 1:}}         &                            &         &                  &       \\               
\Mbh~in log scale ($\log M_{\rm BH}$; \Msun)            & 6.0&$\longrightarrow$&8.0 &6.91 &$-$0.01, +0.01&$-$0.05, +0.04\\
Mass-to-light ratio in $H$-band ($M/L_{H}$; \Msun/\Lsun)&$10^{-2}$&$\longrightarrow$&5.2&0.62 &$-$0.01 +0.01 &$-$0.04, +0.04\\
Position angle of the inner disk (PA$_1$; $^{\circ}$)   & 65 &$\longrightarrow$& 90&76.20&$-$0.17, +0.13 &$-$0.53, +0.30\\
Position angle of the outer disk (PA$_1$; $^{\circ}$)   & 65 &$\longrightarrow$& 90&65.02&$-$0.01, +0.02 &$-$0.07, +0.07\\
Inclination angle of the inner disk ($i_1$; $^{\circ}$) & 25 &$\longrightarrow$& 90&54.75&$-$0.17, +0.18 &$-$0.60, +0.52\\
Inclination angle of the outer disk ($i_2$; $^{\circ}$) & 25 &$\longrightarrow$& 90&61.84&$-$0.20, +0.20 &$-$0.60, +0.62\\
Transition radius between two disks (R$_{\rm break}$; arcsec)&$10^{-2}$&$\longrightarrow$&1.0&0.29&$-$0.0, +0.0 &$-$0.04, +0.04\\
\underline{{\bf Model 2:}}        &                     &    &                    &     \\  
\Mbh~in log scale ($\log M_{\rm BH}$; \Msun)            &6.0 &$\longrightarrow$&8.0 &6.88 &$-$0.02, +0.02&$-$0.06, +0.07\\
Mass-to-light ratio in $H$-band ($M/L_{H}$; \Msun/\Lsun)&$10^{-2}$&$\longrightarrow$& 5.2 & 0.63 &$-$0.03, +0.02&$-$0.08, +0.07\\
Inclination angle of the inner ellipse ($i_1$; $^{\circ}$)&30&$\longrightarrow$& 90 & 47.54&$-$1.20, +1.34&$-$3.72, +3.92\\
Inclination angle of the outer ellipse ($i_2$; $^{\circ}$)&30&$\longrightarrow$& 90 & 60.42&$-$1.06, +0.99&$-$1.21, +1.82\\  
\underline{{\bf Model 3:}}        &                     &    &                    &    \\
\Mbh~in log scale ($\log M_{\rm BH}$; \Msun)            &6.0 &$\longrightarrow$&8.0 & 6.94 &$-$0.00, +0.04&$-$0.06, +0.13\\
Mass-to-light ratio in $H$-band ($M/L_{H}$; \Msun/\Lsun)&$10^{-2}$&$\longrightarrow$&5.2 & 0.61 &$-$0.00  +0.03 &$-$0.03, +0.09\\
Position angle of the inner disk (PA; $^{\circ}$)       & 65 &$\longrightarrow$& 90 & 65.51&$-$0.83, +0.94 &$-$2.35, +2.90\\
Inclination angle of the inner disk ($i$; $^{\circ}$)   & 50 &$\longrightarrow$& 90 & 57.77&$-$0.96, +1.99 &$-$3.23, +5.65\\
Central hole radius  (r$_{\rm inner}$; arcsec)          &$10^{-2}$&$\longrightarrow$&0.1 & 0.04 &$-$0.01, +0.01 &$-$0.03, +0.03\\
\hline
\end{tabular}
\tablenotemark{}
\tablecomments{Column 1: A list of the fitted model parameters. Columns 2--4: A list of the priors of the fitted model parameters and their search ranges. The prior are constructed in the uniform linear space with only SMBH is in logarithmic space. Columns 5--7: The best-fit value of each parameter and their uncertainties at 1$\sigma$ and 3$\sigma$ confidence levels.}
\label{ringfittable}
\end{table*}
%%%%%%%%%%%%%%%%%%%%%%%%%%%%%%%%%%%

%%%%%%%%%%%%%%%%%%%%%%%%%%%%%%%%%%%%%%%%%%%%
\subsection{Results}\label{ssec:ringresults}

We summarize the best-fit results of these three tilted-ring models in Table \ref{ringfittable}. All uncertainties of the best-fit parameters we quote in this section are determined within 3$\sigma$ CL. The tilted-ring  model derives a best-fit \Mbh~of $M_{\rm BH}=7.7_{-1.2}^{+1.5}\times10^6$\Msun. This mass is 23\% lower than the best-fit \Mbh~found by the KinMS model, while the outer disk best-fit inclination is determined at $i\sim60^{\circ}$. However, the best-fit \ml$_H$ is found at $0.63^{+0.08}_{-0.02}$ (\Msun/\Lsun); less than $\sim$5\% of the KinMS's predictions. In these models, we assume uniform errors for the velocities, the confidence intervals are therefore likely underestimated.
 
In Figure \ref {titlering_map} we plot the best-fit velocity map of of $^{12}{\rm CO(2-1)}$ gas disk, its data, and the residual ${\tt (Data-Model)}$ of the Model 2 only. As similar as the KinMS model, the tilted-ring model finds evidence of some non-circular motions on the velocity residual map as well. However, the signature of the nuclear small rotating structure is weak due to the presence of other residuals at the same scale of the structure. 
 
%%%%%%%%%%%%%%%%%%%%%%%%%%%%%%%%%%%%%%%%%%%%
\section{Discussions}\label{sec:discussions}

%%%%%%%%%%%%%%%%%%%%%%%%%%%%%%%%%%%%%%%%%%%%%%%%%%%%%%%%%%%%%%%%%%%%%%%%%%%%%%%%%%%%%%%%%%%%%
\subsection{\Mbh~Measurements in Nearby Low-mass Galaxies with ALMA}\label{ssec:posBHmass}

ALMA observations provide a new promising path for gas-dynamical \Mbh~measurements and exploration of BH demographics in a variety types of Hubble sequence in nearby low-mass galaxies; especially for LTGs where host a large fraction of gas and dust in their nuclear regions. As discuss in \citet{Barth16a}, to measure \Mbh~in these galaxies accurately, cold gas observations are required to (i) have a simple disk-like rotation and kinematically dominated, (ii) the beamsize should be at least twice of the BH SOI along the minor axis, and (iii) the gas kinematics have to mapped well supported by high surface brightness molecular line emissions. In addition, accurate modeling of host galaxy luminosity/mass profiles using NIR observations to avoid dust contamination and extinction, with angular resolutions that are as high as the ALMA data are also greatly important. %This would be achieved by next-generation telescopes equipped with adaptive optics.

In the near future, a number of galaxies with detection of gas high-velocity rotation within SOI should increase rapidly. Measuring \Mbh~in such galaxies accurately are important to anchor the local BH demographics and BH--host galaxy correlations \citep{Maoz98} and (2) constraint the BH seed formations in the early Universe \citep[e.g.,][]{Volonteri08, Bonoli14, Volonteri15, Fiacconi16, Fiacconi17}. So, the future growing number of high-resolution observations of ALMA of nearby targets will make it possible to examine molecular gas kinematics in galaxy nuclei in far greater details than any previous surveys.

%%%%%%%%%%%%%%%%%%%%%%%%%%%%%%%%%%%%%%%%%%%%%%%%%%%%%%%%%%%%%%%%%%
\subsection{\Mbh~Scaling Relations}\label{ssec:scalingrelation}

We examine our best-fit \Mbh~of NGC 3504 in the context of scaling relations of $M_{\rm BH}-M_{\rm Bulge}$ and $M_{\rm BH}-\sigma$ including the empirical compilations \citep{Kormendy13, McConnell13, Saglia16} and theoretical prediction that assumes a bimodality in BH accretion efficiency \citep{Pacucci17, Pacucci18, Pacucci18b} in Figure \ref{scaling}.  The stellar velocity dispersion of the bulge of NGC 3504 is determined from \citet{Ho09}. Here, we estimate the bulge mass of NGC 3504 using our $H$-band MGE model (Section \ref{sec:massmodel}) and  adapt the bulge effective radius from the bugle-disk-bar decomposition model \citep{Laurikainen04}. After calibrating for the distance to the galaxy and accounting for the dynamical \ml$_H$ (Section \ref{ssec:result}), we obtain $M_{\rm Bulge}=(2.3\pm0.4)\times10^9$\Msun. Our result shows that the best-fit \Mbh~of NGC 3504 is fully consistent with both empirical and theoretical  $M_{\rm BH}-\sigma$ and $M_{\rm BH}-M_{\rm Bulge}$ scaling relations of \citet{Kormendy13, Saglia16, vandenBosch16}, but outside $+1\sigma$ uncertainty of the theoretical $M_{\rm BH}-M_{\rm Bulge}$ relation of \citet{Pacucci18}. At the mass of $\sim$$10^7$\Msun~the central BH of NGC 3504 lies in the middle with one order of magnitude above/below the recent dynamical measurements in low-mass systems \citep[][N17; N18; N19]{denbrok15} and in more massive targets using molecular gas \citep{Davis13, Onishi15, Barth16a, Barth16b, Davis17, Onishi17, Davis18} 

%%%%%%%%%%%%%%%%%%%%%%%%%%%%%%%%%%%%%%%%%%%%%%%%%%%%%%%%%%%%%%%%%%%%%%%%%%%%%%%
\subsection{Complex Structure of the Molecular Gas}\label{ssec:complexstructure} 

Our both low- and high-resolution (combined) kinematics show a larger rotational disk as seen in the residual maps of velocity fields after subtracting the data to the models when we fit the models to the data with a larger FOV ($1\arcsec\times1\arcsec$). This larger disk is co-rotating but misaligned by 30$^\circ$ with the CND as mentioned in Section \ref{sssec:line}. So, both morphology and kinematics suggest a nuclear complex structure of the $^{12}{\rm CO(2-1)}$ gas in NGC 3504 with (i) a large faint disk with a size of $\pi\times5\arcsec^2$ consists of spiral arms and voids, which may be associated with bars, (ii) a dominant integrated flux CND within $0\farcs5$, and (iii) a nuclear rotating structure with a size of $r\sim0\farcs15$ that is revealed by the high-resolution kinematic measurement only. 

The higher rotational velocities of the outermost disk and CND compare to that of the nuclear rotating structure suggest that this structure is not only oriented along the minor axis but also along the LOS direction, while the larger discs are oriented perpendicular to the LOS direction. Detailed investigations of morphology, kinematics, and role of this nuclear structure on the growth of SMBH will be discussed in D. Nguyen et al. in preparation.

%%%%%%%%%%%%%%%%%%%%%%%%%%%%%%%%%%%%%%%%%%%%
\section{Conclusions}\label{sec:conclusions}

We present a dynamical mass measurement for the SMBH in NGC 3504 using $^{12}{\rm CO(2-1)}$ emission observed by ALMA and \hst~imaging.  Our main results are highlighted as follow:

\begin{enumerate}
	 \item NGC 3504 hosts a CND, which is co-spatial distribution with a obscuring dust lanes visible in the \hst~imaging within $5\arcsec$. 
	 	 
	 \item Both KinMS and tilted-ring models suggest the detection of a central SMBH with $M_{\rm BH} \sim1\times10^7$\Msun, which is consistent with both $M_{\rm BH}-M_{\rm Bulge}$ and $M_{\rm BH}-\sigma$ relations. The agreement between two models suggests they are both powerful tools to probe the BH--galaxy scaling relations.
	 	 
	 \item Our dynamical models also give \ml$_H=0.66$ (\Msun/\Lsun) for the nucleus of NGC 3504, suggesting its nucleus is in an early phase of its evolution. 
		 
	 \item Both combined (high-) and low-resolution observations give consistent constraints on the \Mbh~and \ml$_H$. These prove our observational strategy for measuring \Mbh~in nearby low-mass galaxies to explore the BH demographics in a large sample, which cannot achieve with the stellar dynamical method.  
	 	 
	 \item The CND of NGC 3504 has a relatively high velocity dispersion with $\sim$30 \kms~in the region $<$1$\arcsec$ and $\sim$10 \kms~at larger radii. The central high value ($>$60 \kms) of dispersion is not an intrinsic but a beam smearing effect along the LOS integration through the nearly edge-on orientation of the CND.
	
	\item Our multiple scale observations reveal the complexity of the nuclear molecular gas disk in NGC 3504 comprised of a small nuclear rotating structure, a CND, and a larger disk with different orientations along the LOS and galactic axes.

	\item We find a central hole that has a radius of 2.7 pc in the $^{12}{\rm CO(2-1)}$ integrated intensity map.  However, this hole is filled by a dense gas tracer ${\rm CS(5-4)}$, which is centrally peaked, coexisting, and has a similar kinematic feature to the $^{12}{\rm CO(2-1)}$ line within the CND. 
\end{enumerate}

%%%%%%%%%%%%%%%%%%%%%%%%%%
%
% Section --- Acknowledgements
%
%\Acknowledgements
\section*{Acknowledgements}

The authors would like to thank National Astronomical Observatory of Japan (NAOJ), National Institute of Natural Sciences (NINS) for supporting this work. D.D.N. delivers his gratitude to the supports from Willard L. and Ruth P. Eccles Foundation for their Eccles Fellowship during the 2017-2018 academic year at the Department of Physics and Astronomy of The University of Utah and the TAIZAI Visiting Fellowship during the Spring 2018 at NAOJ. A.C.S. acknowledges financial support from NSF grant AST-1350389 and M.C. acknowledges the support from a Royal Society University Research Fellowship. T.I. sincerely thanks supports from the JSPS KAKENHI grant number 17K14247. M.I. is supported by JSPS KAKENHI grant number 15K05030.

%and the ALMA Japan Research Grant of NAOJ Chile Observatory, NAOJ-ALMA-170. 
%We also thank the anonymous referee for careful reading and useful comments which greatly improved this paper. 

The authors would like to thank Drs Naoteru Gouda, Makoto Miyoshi, Yano Taihei, and Daisuke Iono at NAOJ for enlightenment discussions. This paper makes use of the following ALMA data: ADS/JAO.ALMA \#2017.1.00964.S. ALMA is a partnership of ESO (representing its member states), NSF (USA) and NINS (Japan), together with NRC (Canada) and NSC and ASIAA (Taiwan) and KASI (Republic of Korea), in cooperation with the Republic of Chile. The Joint ALMA Observatory is operated by ESO, AUI/NRAO and NAOJ. The National Radio Astronomy Observatory is a facility of the National Science Foundation operated under cooperative agreement by Associated Universities, Inc. We thank the ALMA operators and staff and the ALMA help desk as well for diligent feedback and invaluable assistance in processing these data.

{\it Facilities:} ALMA and \hst~WFC3.

{\it Software:} \texttt{Astropy}, \texttt{IDL}, and \texttt{emcee}

\appendix

%%%%%%%%%%%%%%%%%%%%%%%%%%%%%%%%%%%%%%%%%%%%%%%%%%%%%%%%%%%%%%%%%%%%%%%%%%%%%
\section{Supplementary Figures and Tables}\label{sec:fittable_lowres}

%%%%%%%%%%%%%%%%%%%%%%%%%%%%%%%%%%%%%%%%%
\begin{figure*}[!ht]
\centering\includegraphics[scale=0.86]{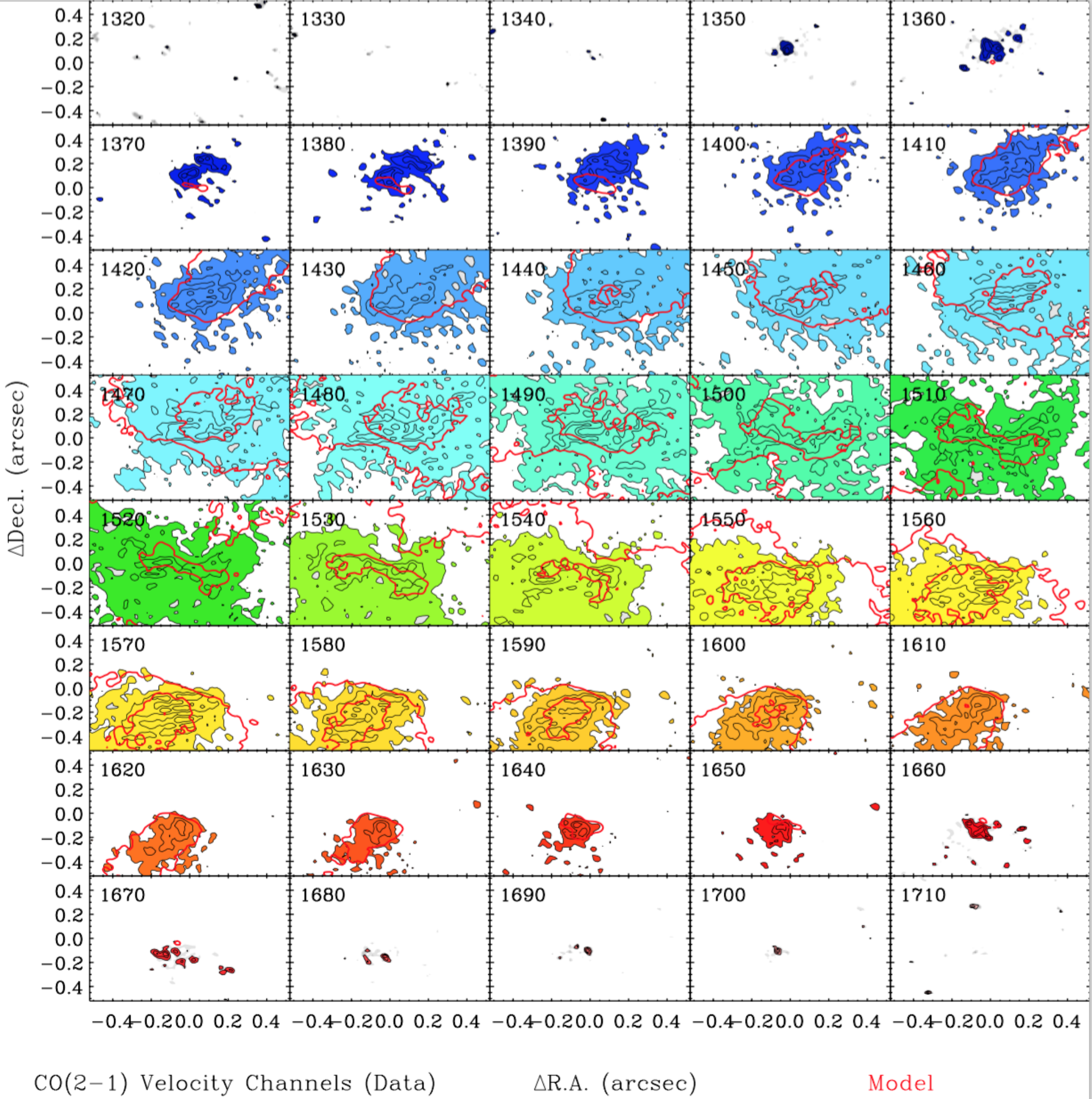} 
\caption{The inner fitted region channel maps of our combined (high resolution) ALMA $^{12}{\rm CO(2-1)}$ data cube in the velocity range where emission is detected. The colored regions with thin black contours show the areas detected with more than 2.5$\sigma$ significance.  The thick red contours are plotted with the same contour levels from our best-fitting KinMS model.}
\label{bestfit_velchan}   
\end{figure*}
%%%%%%%%%%%%%%%%%%%%%%%%%%%%%%%%%%%%%%%%%

%%%%%%%%%%%%%%%%%%%%%%%%%%%%%%%%%%%%%%%%%
\begin{figure*}[!th]  
\centering
	\includegraphics[scale=0.60]{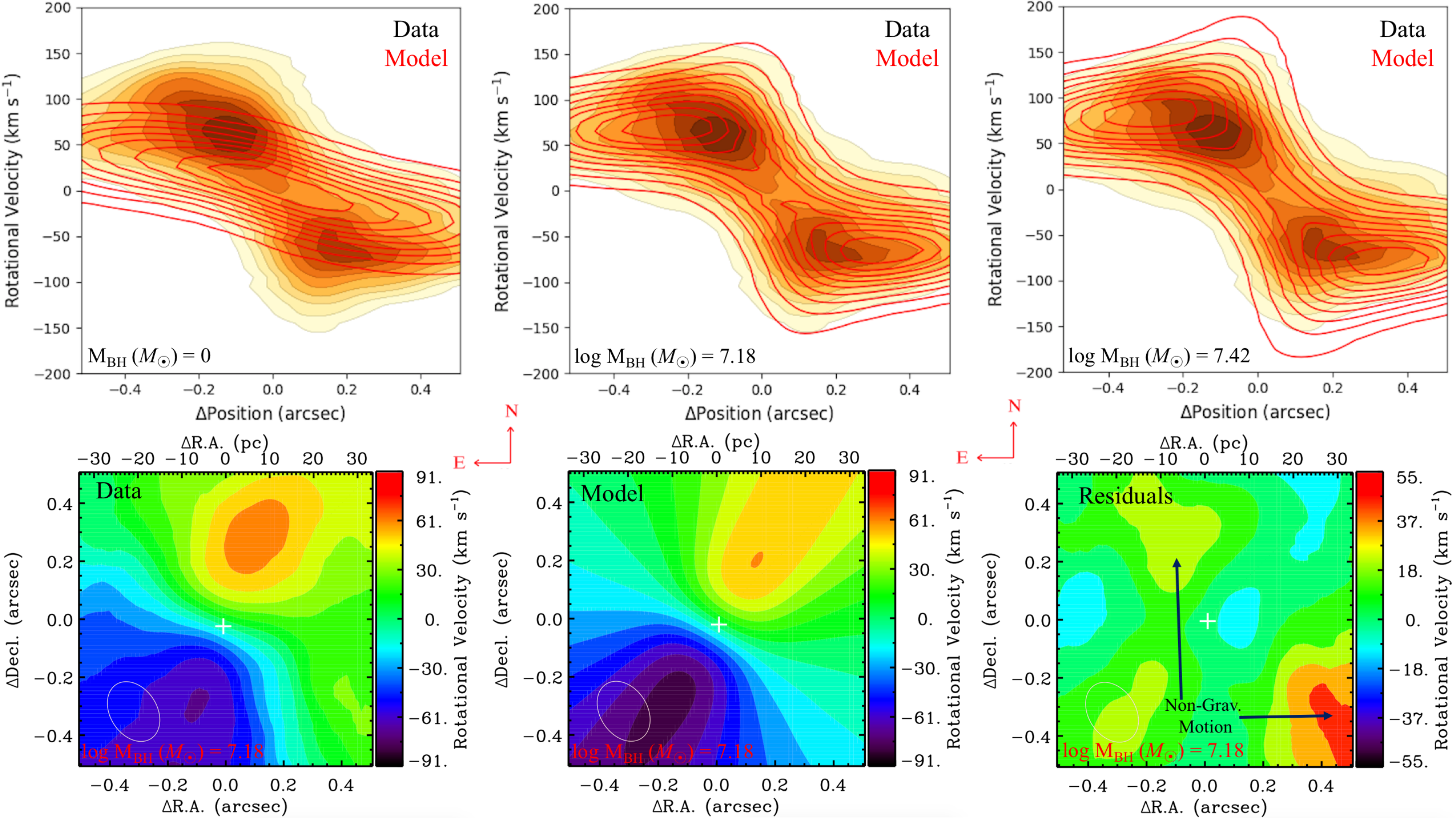} 
\caption{Comparisons between the low-resolution cube with a few specific KinMS dynamical model including the best fit. All notations and illustrations in this figure are similar to Figure \ref{bestfit_pvd}.} 
\label{bestfit_pvd_lowRes}   
\end{figure*}
%%%%%%%%%%%%%%%%%%%%%%%%%%%%%%%%%%%%%%%%%

%%%%%%%%%%%%%%%%%%%%%%%%%%%%%%%%%%%
\begin{table*}
\caption{Best-fit KinMS Model Parameters and Statistical Uncertainties for the Low-Resolution ALMA Cube}
\centering  
\begin{tabular}{lccccccc} 
\hline\hline       
 Parameter Names (Notations; Units)&\multicolumn{3}{c}{Search Range}&Best Fit&$1\sigma$ Error&$3\sigma$ Error\\
                &              &               &                   &        &  (68\% conf.) &(99.7\% conf.) \\  
        (1)     &           (2)       &      (3)   &     (4)       &    (5) &    (6)&    (7)        \\  
\hline 
\underline{{\bf Black Hole:}}  &               &                    &        &               &          &    \\  
\Mbh~in log scale ($\log_{10} M_{\rm BH}$; \Msun)& 4.00 &$\longrightarrow$&9.00& 7.18 &$-$0.04, +0.04&$-$0.10, +0.13\\
Mass-to-light ratio in $H$-band (\ml$_H$; \Msun/\Lsun)  & 0.01 &$\longrightarrow$&3.00& 0.67 &$-$0.02, +0.03&$-$0.62, +1.01\\
\underline{{\bf Molecular Gas Disc:}}                   &      &                 &    &      &              &              \\  
Position angle (PA; $^{\circ}$)                         &150.0 &$\longrightarrow$&360.0&330.17&$-$1.42, +0.90&$-$4.24, +3.25\\
Inclination angle ($i$; $^{\circ}$)                     & 45.0 &$\longrightarrow$&90.0&60.31 &$-$1.43, +1.34&$-$4.51, +4.01\\
Gas velocity dispersion ($\sigma$; \kms)                & 1.0  &$\longrightarrow$&80.0& 35.50&$-$0.53, +0.45&$-$1.56, +1.42\\   
Disk thickness ($d_t$; arcsec)                          & 0.01 &$\longrightarrow$&1.50& 0.41 &$-$0.03, +0.03&$-$0.09, +0.12\\
Gaussian HWHM (G$_h$; arcsec)                           & 0.01 &$\longrightarrow$&1.00&0.35  &$-$0.03, +0.03&$-$0.10, +0.13\\
\underline{{\bf Nuisance Parameters:}}                  &       &           &        &            &           &        \\  
CO surface brightness scaling factor (Flux)             &  10.0 &$\longrightarrow$&500.0& 149.01 &$-$0.02, +0.02&$-$0.07, +0.09\\
R.A. offset ($x_c$; arcsec)                             &$-$0.10&$\longrightarrow$&+0.10&$-$0.02&$-$0.00, +0.00&$-$0.01, +0.01\\
Decl. offset ($y_c$; arcsec)                            &$-$0.10&$\longrightarrow$&+0.10&$-$0.03&$-$0.00, +0.00&$-$0.01, +0.01\\
Systemic velocity offset ($v_{\rm off.}$; \kms)         &$-$10.0&$\longrightarrow$&+10.0&$-$0.02&$-$0.22, +0.02&$-$0.06, +0.07\\
\hline
\end{tabular}
\tablenotemark{}
\tablecomments{All notations and parameters in thí table are keeping silimar to Table \ref{fittable}.}
\label{fittable_lowres}
\end{table*}
%%%%%%%%%%%%%%%%%%%%%%%%%%%%%%%%%%%

%###########################


\begin{thebibliography}{}

\expandafter\ifx\csname natexlab\endcsname\relax\def\natexlab#1{#1}\fi

\bibitem[{{Abolfathi} {et~al.}(2018){Abolfathi}, {Aguado}, {Aguilar}, {Allende
  Prieto}, {Almeida}, {Ananna}, {Anders}, {Anderson}, {Andrews}, {Anguiano}, \&
  et~al.}]{Abolfathi18}
{Abolfathi}, B., {Aguado}, D.~S., {Aguilar}, G., {et~al.} 2018, \apjs, 235, 42

\bibitem[{{Afanasiev} {et~al.}(2018){Afanasiev}, {Chilingarian}, {Mieske},
  {Voggel}, {Picotti}, {Hilker}, {Seth}, {Neumayer}, {Frank}, {Romanowsky},
  {Hau}, {Baumgardt}, {Ahn}, {Strader}, {den{\^A} Brok}, {McDermid}, {Spitler},
  {Brodie}, \& {Walsh}}]{Afanasiev18}
{Afanasiev}, A.~V., {Chilingarian}, I.~V., {Mieske}, S., {et~al.} 2018, \mnras,
  477, 4856

\bibitem[{{Ahn} {et~al.}(2017){Ahn}, {Seth}, {den Brok}, {Strader},
  {Baumgardt}, {van den Bosch}, {Chilingarian}, {Frank}, {Hilker}, {McDermid},
  {Mieske}, {Romanowsky}, {Spitler}, {Brodie}, {Neumayer}, \& {Walsh}}]{Ahn17}
{Ahn}, C.~P., {Seth}, A.~C., {den Brok}, M., {et~al.} 2017, \apj, 839, 72

\bibitem[{{Ahn} {et~al.}(2018){Ahn}, {Seth}, {Cappellari}, {Krajnovi{\'c}},
  {Strader}, {Voggel}, {Walsh}, {Bahramian}, {Baumgardt}, {Brodie},
  {Chilingarian}, {Chomiuk}, {den Brok}, {Frank}, {Hilker}, {McDermid},
  {Mieske}, {Neumayer}, {Nguyen}, {Pechetti}, {Romanowsky}, \&
  {Spitler}}]{Ahn18}
{Ahn}, C.~P., {Seth}, A.~C., {Cappellari}, M., {et~al.} 2018, \apj, 858, 102

\bibitem[{{Antonucci}(1993)}]{Antonucci93a}
{Antonucci}, R. 1993, \araa, 31, 473

\bibitem[{{Avila} {et~al.}(2012){Avila}, {Hack}, \& {STScI AstroDrizzle
  Team}}]{Avila12}
{Avila}, R.~J., {Hack}, W.~J., \& {STScI AstroDrizzle Team}. 2012, in American
  Astronomical Society Meeting Abstracts, Vol. 220, American Astronomical
  Society Meeting Abstracts 220, \#135.13

\bibitem[{{Baldassare} {et~al.}(2015){Baldassare}, {Reines}, {Gallo}, \&
  {Greene}}]{Baldassare15}
{Baldassare}, V.~F., {Reines}, A.~E., {Gallo}, E., \& {Greene}, J.~E. 2015,
  \apjl, 809, L14

\bibitem[{{Barai} {et~al.}(2014){Barai}, {Viel}, {Murante}, {Gaspari}, \&
  {Borgani}}]{Barai14}
{Barai}, P., {Viel}, M., {Murante}, G., {Gaspari}, M., \& {Borgani}, S. 2014,
  \mnras, 437, 1456

\bibitem[{{Barth} {et~al.}(2016{\natexlab{a}}){Barth}, {Boizelle}, {Darling},
  {Baker}, {Buote}, {Ho}, \& {Walsh}}]{Barth16a}
{Barth}, A.~J., {Boizelle}, B.~D., {Darling}, J., {et~al.} 2016{\natexlab{a}},
  \apjl, 822, L28

\bibitem[{{Barth} {et~al.}(2016{\natexlab{b}}){Barth}, {Darling}, {Baker},
  {Boizelle}, {Buote}, {Ho}, \& {Walsh}}]{Barth16b}
{Barth}, A.~J., {Darling}, J., {Baker}, A.~J., {et~al.} 2016{\natexlab{b}},
  \apj, 823, 51

\bibitem[{{Barth} {et~al.}(2004){Barth}, {Ho}, {Rutledge}, \&
  {Sargent}}]{Barth04}
{Barth}, A.~J., {Ho}, L.~C., {Rutledge}, R.~E., \& {Sargent}, W.~L.~W. 2004,
  \apj, 607, 90

\bibitem[{{Barvainis}(1987)}]{Barvainis87a}
{Barvainis}, R. 1987, \apj, 320, 537

\bibitem[{{Begeman}(1987)}]{Begeman87}
{Begeman}, K.~G. 1987, PhD thesis, , Kapteyn Institute, (1987)

\bibitem[{{Beifiori} {et~al.}(2012){Beifiori}, {Courteau}, {Corsini}, \&
  {Zhu}}]{Beifiori12}
{Beifiori}, A., {Courteau}, S., {Corsini}, E.~M., \& {Zhu}, Y. 2012, \mnras,
  419, 2497

\bibitem[{{Bell} {et~al.}(2003){Bell}, {McIntosh}, {Katz}, \&
  {Weinberg}}]{Bell03}
{Bell}, E.~F., {McIntosh}, D.~H., {Katz}, N., \& {Weinberg}, M.~D. 2003, \apjs,
  149, 289

\bibitem[{{Bigiel} {et~al.}(2008){Bigiel}, {Leroy}, {Walter}, {Brinks}, {de
  Blok}, {Madore}, \& {Thornley}}]{Bigiel08}
{Bigiel}, F., {Leroy}, A., {Walter}, F., {et~al.} 2008, \aj, 136, 2846

\bibitem[{{Bolatto} {et~al.}(2013){Bolatto}, {Warren}, {Leroy}, {Walter},
  {Veilleux}, {Ostriker}, {Ott}, {Zwaan}, {Fisher}, {Weiss}, {Rosolowsky}, \&
  {Hodge}}]{Bolatto13}
{Bolatto}, A.~D., {Warren}, S.~R., {Leroy}, A.~K., {et~al.} 2013, \nat, 499,
  450

\bibitem[{{Bonoli} {et~al.}(2014){Bonoli}, {Mayer}, \& {Callegari}}]{Bonoli14}
{Bonoli}, S., {Mayer}, L., \& {Callegari}, S. 2014, \mnras, 437, 1576

\bibitem[{{Boselli} {et~al.}(2015){Boselli}, {Fossati}, {Gavazzi}, {Ciesla},
  {Buat}, {Boissier}, \& {Hughes}}]{Boselli15}
{Boselli}, A., {Fossati}, M., {Gavazzi}, G., {et~al.} 2015, \aap, 579, A102

\bibitem[{{Buta} \& {Combes}(1996)}]{Buta96}
{Buta}, R., \& {Combes}, F. 1996, \fcp, 17, 95

\bibitem[{{Buta} \& {Crocker}(1992)}]{Buta92}
{Buta}, R., \& {Crocker}, D.~A. 1992, \aj, 103, 1804

\bibitem[{{Caplar} {et~al.}(2015){Caplar}, {Lilly}, \&
  {Trakhtenbrot}}]{Caplar15}
{Caplar}, N., {Lilly}, S.~J., \& {Trakhtenbrot}, B. 2015, \apj, 811, 148

\bibitem[{{Cappellari}(2002)}]{Cappellari02}
{Cappellari}, M. 2002, \mnras, 333, 400

\bibitem[{{Cappellari}(2008)}]{Cappellari08}
{Cappellari}. 2008, \mnras, 390, 71

\bibitem[{{Cardelli} {et~al.}(1989){Cardelli}, {Clayton}, \&
  {Mathis}}]{Cardelli89}
{Cardelli}, J.~A., {Clayton}, G.~C., \& {Mathis}, J.~S. 1989, \apj, 345, 245

\bibitem[{{Chilingarian} {et~al.}(2018){Chilingarian}, {Katkov}, {Zolotukhin},
  {Grishin}, {Beletsky}, {Boutsia}, \& {Osip}}]{Chilingarian18}
{Chilingarian}, I.~V., {Katkov}, I.~Y., {Zolotukhin}, I.~Y., {et~al.} 2018,
  \apj, 863, 1

\bibitem[{{Condon} {et~al.}(1998){Condon}, {Cotton}, {Greisen}, {Yin},
  {Perley}, {Taylor}, \& {Broderick}}]{Condon98}
{Condon}, J.~J., {Cotton}, W.~D., {Greisen}, E.~W., {et~al.} 1998, \aj, 115,
  1693

\bibitem[{{Davis}(2014)}]{Davis14}
{Davis}, T.~A. 2014, \mnras, 443, 911

\bibitem[{{Davis} {et~al.}(2013){Davis}, {Bureau}, {Cappellari}, {Sarzi}, \&
  {Blitz}}]{Davis13}
{Davis}, T.~A., {Bureau}, M., {Cappellari}, M., {Sarzi}, M., \& {Blitz}, L.
  2013, \nat, 494, 328

\bibitem[{{Davis} {et~al.}(2017){Davis}, {Bureau}, {Onishi}, {Cappellari},
  {Iguchi}, \& {Sarzi}}]{Davis17}
{Davis}, T.~A., {Bureau}, M., {Onishi}, K., {et~al.} 2017, \mnras, 468, 4675

\bibitem[{{Davis} {et~al.}(2018){Davis}, {Bureau}, {Onishi}, {van de Voort},
  {Cappellari}, {Iguchi}, {Liu}, {North}, {Sarzi}, \& {Smith}}]{Davis18}
{Davis} {et~al.}. 2018, \mnras, 473, 3818

\bibitem[{{de Vaucouleurs}(1975)}]{deVaucouleurs75}
{de Vaucouleurs}, G. 1975, \apj, 202, 319

\bibitem[{{Deller} \& {Middelberg}(2014)}]{Deller14}
{Deller}, A.~T., \& {Middelberg}, E. 2014, \aj, 147, 14

\bibitem[{{den Brok} {et~al.}(2015){den Brok}, {Seth}, {Barth}, {Carson},
  {Neumayer}, {Cappellari}, {Debattista}, {Ho}, {Hood}, \&
  {McDermid}}]{denbrok15}
{den Brok}, M., {Seth}, A.~C., {Barth}, A.~J., {et~al.} 2015, \apj, 809, 101

\bibitem[{{Desroches} {et~al.}(2009){Desroches}, {Greene}, \&
  {Ho}}]{Desroches09}
{Desroches}, L.-B., {Greene}, J.~E., \& {Ho}, L.~C. 2009, \apj, 698, 1515

\bibitem[{{Di Matteo} {et~al.}(2008){Di Matteo}, {Colberg}, {Springel},
  {Hernquist}, \& {Sijacki}}]{DiMatteo08}
{Di Matteo}, T., {Colberg}, J., {Springel}, V., {Hernquist}, L., \& {Sijacki},
  D. 2008, \apj, 676, 33

\bibitem[{{Dong} {et~al.}(2012){Dong}, {Ho}, {Yuan}, {Wang}, {Fan}, {Zhou}, \&
  {Jiang}}]{Dong12}
{Dong}, X.-B., {Ho}, L.~C., {Yuan}, W., {et~al.} 2012, \apj, 755, 167

\bibitem[{{Emsellem} {et~al.}(1994){Emsellem}, {Monnet}, \&
  {Bacon}}]{Emsellem94a}
{Emsellem}, E., {Monnet}, G., \& {Bacon}, R. 1994, \aap, 285, 723

\bibitem[{{Epinat} {et~al.}(2008){Epinat}, {Amram}, {Marcelin}, {Balkowski},
  {Daigle}, {Hernandez}, {Chemin}, {Carignan}, {Gach}, \& {Balard}}]{Epinat08}
{Epinat}, B., {Amram}, P., {Marcelin}, M., {et~al.} 2008, \mnras, 388, 500

\bibitem[{{Erroz-Ferrer} {et~al.}(2015){Erroz-Ferrer}, {Knapen}, {Leaman},
  {Cisternas}, {Font}, {Beckman}, {Sheth}, {Mu{\~n}oz-Mateos},
  {D{\'{\i}}az-Garc{\'{\i}}a}, {Bosma}, {Athanassoula}, {Elmegreen}, {Ho},
  {Kim}, {Laurikainen}, {Martinez-Valpuesta}, {Meidt}, \&
  {Salo}}]{Erroz-Ferrer15}
{Erroz-Ferrer}, S., {Knapen}, J.~H., {Leaman}, R., {et~al.} 2015, \mnras, 451,
  1004

\bibitem[{{Fabian}(2012)}]{Fabian12}
{Fabian}, A.~C. 2012, \araa, 50, 455

\bibitem[{{Ferrarese} \& {Merritt}(2000)}]{Ferrarese00}
{Ferrarese}, L., \& {Merritt}, D. 2000, \apjl, 539, L9

\bibitem[{{Fiacconi} \& {Rossi}(2016)}]{Fiacconi16}
{Fiacconi}, D., \& {Rossi}, E.~M. 2016, \mnras, 455, 2

\bibitem[{{Fiacconi} \& {Rossi}(2017)}]{Fiacconi17}
{Fiacconi} \& {Rossi}. 2017, \mnras, 464, 2259

\bibitem[{{Foreman-Mackey} {et~al.}(2013){Foreman-Mackey}, {Hogg}, {Lang}, \&
  {Goodman}}]{Foreman-Mackey13}
{Foreman-Mackey}, D., {Hogg}, D.~W., {Lang}, D., \& {Goodman}, J. 2013, \pasp,
  125, 306

\bibitem[{{Gallo} {et~al.}(2008){Gallo}, {Treu}, {Jacob}, {Woo}, {Marshall}, \&
  {Antonucci}}]{Gallo08}
{Gallo}, E., {Treu}, T., {Jacob}, J., {et~al.} 2008, \apj, 680, 154

\bibitem[{{Gallo} {et~al.}(2010){Gallo}, {Treu}, {Marshall}, {Woo}, {Leipski},
  \& {Antonucci}}]{Gallo10}
{Gallo}, E., {Treu}, T., {Marshall}, P.~J., {et~al.} 2010, \apj, 714, 25

\bibitem[{{Gebhardt} {et~al.}(2000){Gebhardt}, {Bender}, {Bower}, {Dressler},
  {Faber}, {Filippenko}, {Green}, {Grillmair}, {Ho}, {Kormendy}, {Lauer},
  {Magorrian}, {Pinkney}, {Richstone}, \& {Tremaine}}]{Gebhardt00}
{Gebhardt}, K., {Bender}, R., {Bower}, G., {et~al.} 2000, \apjl, 539, L13

\bibitem[{{Goodman} \& {Weare}(2010)}]{Goodman10}
{Goodman}, J., \& {Weare}, J. 2010, Communications in Applied Mathematics and
  Computational Science, Vol.~5, No.~1, p.~65-80, 2010, 5, 65

\bibitem[{{Graham} {et~al.}(2001){Graham}, {Erwin}, {Caon}, \&
  {Trujillo}}]{Graham01}
{Graham}, A.~W., {Erwin}, P., {Caon}, N., \& {Trujillo}, I. 2001, \apjl, 563,
  L11

\bibitem[{{Graham} \& {Scott}(2015)}]{Graham15}
{Graham}, A.~W., \& {Scott}, N. 2015, \apj, 798, 54

\bibitem[{{Greene} \& {Ho}(2007)}]{Greene07}
{Greene}, J.~E., \& {Ho}, L.~C. 2007, \apj, 670, 92

\bibitem[{{Greene} {et~al.}(2010){Greene}, {Peng}, {Kim}, {Kuo}, {Braatz},
  {Impellizzeri}, {Condon}, {Lo}, {Henkel}, \& {Reid}}]{Greene10}
{Greene}, J.~E., {Peng}, C.~Y., {Kim}, M., {et~al.} 2010, \apj, 721, 26

\bibitem[{{Greene} {et~al.}(2016){Greene}, {Seth}, {Kim}, {L{\"a}sker},
  {Goulding}, {Gao}, {Braatz}, {Henkel}, {Condon}, {Lo}, \& {Zhao}}]{Greene16}
{Greene}, J.~E., {Seth}, A., {Kim}, M., {et~al.} 2016, \apjl, 826, L32

\bibitem[{{G{\"u}ltekin} {et~al.}(2009){G{\"u}ltekin}, {Richstone}, {Gebhardt},
  {Lauer}, {Tremaine}, {Aller}, {Bender}, {Dressler}, {Faber}, {Filippenko},
  {Green}, {Ho}, {Kormendy}, {Magorrian}, {Pinkney}, \& {Siopis}}]{Gultekin09}
{G{\"u}ltekin}, K., {Richstone}, D.~O., {Gebhardt}, K., {et~al.} 2009, \apj,
  698, 198

\bibitem[{{H{\"a}ring} \& {Rix}(2004)}]{Haring04}
{H{\"a}ring}, N., \& {Rix}, H.-W. 2004, \apjl, 604, L89

\bibitem[{{Ho} \& {Filippenko}(1993)}]{Ho93}
{Ho}, L.~C., \& {Filippenko}, A.~V. 1993, \apss, 205, 19

\bibitem[{{Ho} {et~al.}(1997){Ho}, {Filippenko}, \& {Sargent}}]{Ho97}
{Ho}, L.~C., {Filippenko}, A.~V., \& {Sargent}, W.~L.~W. 1997, \apjs, 112, 315

\bibitem[{{Ho} {et~al.}(2009){Ho}, {Greene}, {Filippenko}, \& {Sargent}}]{Ho09}
{Ho}, L.~C., {Greene}, J.~E., {Filippenko}, A.~V., \& {Sargent}, W.~L.~W. 2009,
  \apjs, 183, 1

\bibitem[{{Ho} {et~al.}(1993){Ho}, {Shields}, \& {Filippenko}}]{Ho93b}
{Ho}, L.~C., {Shields}, J.~C., \& {Filippenko}, A.~V. 1993, \apj, 410, 567

\bibitem[{{Imanishi} {et~al.}(2018){Imanishi}, {Nakanishi}, {Izumi}, \&
  {Wada}}]{Imanishi18}
{Imanishi}, M., {Nakanishi}, K., {Izumi}, T., \& {Wada}, K. 2018, \apjl, 853,
  L25

\bibitem[{{Inayoshi} \& {Haiman}(2016)}]{Inayoshi16}
{Inayoshi}, K., \& {Haiman}, Z. 2016, \apj, 828, 110

\bibitem[{{Izumi} {et~al.}(2018){Izumi}, {Wada}, {Fukushige}, {Hamamura}, \&
  {Kohno}}]{Izumi18}
{Izumi}, T., {Wada}, K., {Fukushige}, R., {Hamamura}, S., \& {Kohno}, K. 2018,
  \apj, 867, 48

\bibitem[{{Jedrzejewski}(1987)}]{Jedrzejewski87a}
{Jedrzejewski}, R.~I. 1987, \mnras, 226, 747

\bibitem[{{Jedrzejewski} {et~al.}(1987){Jedrzejewski}, {Davies}, \&
  {Illingworth}}]{Jedrzejewski87b}
{Jedrzejewski}, R.~I., {Davies}, R.~L., \& {Illingworth}, G.~D. 1987, \aj, 94,
  1508

\bibitem[{{Keel}(1984)}]{Keel84}
{Keel}, W.~C. 1984, \apj, 282, 75

\bibitem[{{Kenney} {et~al.}(1993){Kenney}, {Carlstrom}, \& {Young}}]{Kenney93}
{Kenney}, J.~D.~P., {Carlstrom}, J.~E., \& {Young}, J.~S. 1993, \apj, 418, 687

\bibitem[{{Knapen} {et~al.}(2002){Knapen}, {P{\'e}rez-Ram{\'{\i}}rez}, \&
  {Laine}}]{Knapen02}
{Knapen}, J.~H., {P{\'e}rez-Ram{\'{\i}}rez}, D., \& {Laine}, S. 2002, \mnras,
  337, 808

\bibitem[{{Kormendy} \& {Bender}(2012)}]{Kormendy12}
{Kormendy}, J., \& {Bender}, R. 2012, \apjs, 198, 2

\bibitem[{{Kormendy} \& {Ho}(2013)}]{Kormendy13}
{Kormendy}, J., \& {Ho}, L.~C. 2013, \araa, 51, 511

\bibitem[{{Kormendy} \& {Richstone}(1995)}]{Kormendy95}
{Kormendy}, J., \& {Richstone}, D. 1995, \araa, 33, 581

\bibitem[{{Krajnovi{\'c}} {et~al.}(2006){Krajnovi{\'c}}, {Cappellari}, {de
  Zeeuw}, \& {Copin}}]{Krajnovic06}
{Krajnovi{\'c}}, D., {Cappellari}, M., {de Zeeuw}, P.~T., \& {Copin}, Y. 2006,
  \mnras, 366, 787

\bibitem[{{Krajnovi{\'c}} {et~al.}(2018){Krajnovi{\'c}}, {Cappellari},
  {McDermid}, {Thater}, {Nyland}, {de Zeeuw}, {Falc{\'o}n-Barroso}, {Khochfar},
  {Kuntschner}, {Sarzi}, \& {Young}}]{Krajnovic18}
{Krajnovi{\'c}}, D., {Cappellari}, M., {McDermid}, R.~M., {et~al.} 2018,
  \mnras, 477, 3030

\bibitem[{{Kuno} {et~al.}(2000){Kuno}, {Nishiyama}, {Nakai}, {Sorai},
  {Vila-Vilar{\'o}}, \& {Handa}}]{Kuno00}
{Kuno}, N., {Nishiyama}, K., {Nakai}, N., {et~al.} 2000, \pasj, 52, 775

\bibitem[{{Kuno} {et~al.}(2007){Kuno}, {Sato}, {Nakanishi}, {Hirota}, {Tosaki},
  {Shioya}, {Sorai}, {Nakai}, {Nishiyama}, \& {Vila-Vilar{\'o}}}]{Kuno07}
{Kuno}, N., {Sato}, N., {Nakanishi}, H., {et~al.} 2007, \pasj, 59, 117

\bibitem[{{Kuo} {et~al.}(2011){Kuo}, {Braatz}, {Condon}, {Impellizzeri}, {Lo},
  {Zaw}, {Schenker}, {Henkel}, {Reid}, \& {Greene}}]{Kuo11}
{Kuo}, C.~Y., {Braatz}, J.~A., {Condon}, J.~J., {et~al.} 2011, \apj, 727, 20

\bibitem[{{L{\"a}sker} {et~al.}(2016){L{\"a}sker}, {Greene}, {Seth}, {van de
  Ven}, {Braatz}, {Henkel}, \& {Lo}}]{Lasker16}
{L{\"a}sker}, R., {Greene}, J.~E., {Seth}, A., {et~al.} 2016, \apj, 825, 3

\bibitem[{{Laurikainen} {et~al.}(2004){Laurikainen}, {Salo}, {Buta}, \&
  {Vasylyev}}]{Laurikainen04}
{Laurikainen}, E., {Salo}, H., {Buta}, R., \& {Vasylyev}, S. 2004, \mnras, 355,
  1251

\bibitem[{{Lo}(2005)}]{Lo05}
{Lo}, K.~Y. 2005, \araa, 43, 625

\bibitem[{{Lodato} \& {Natarajan}(2006)}]{Lodato06}
{Lodato}, G., \& {Natarajan}, P. 2006, \mnras, 371, 1813

\bibitem[{{Magorrian} {et~al.}(1998){Magorrian}, {Tremaine}, {Richstone},
  {Bender}, {Bower}, {Dressler}, {Faber}, {Gebhardt}, {Green}, {Grillmair},
  {Kormendy}, \& {Lauer}}]{Magorrian98}
{Magorrian}, J., {Tremaine}, S., {Richstone}, D., {et~al.} 1998, \aj, 115, 2285

\bibitem[{{Maksym} {et~al.}(2013){Maksym}, {Ulmer}, {Eracleous}, {Guennou}, \&
  {Ho}}]{Maksym13}
{Maksym}, W.~P., {Ulmer}, M.~P., {Eracleous}, M.~C., {Guennou}, L., \& {Ho},
  L.~C. 2013, \mnras, 435, 1904

\bibitem[{{Maoz} {et~al.}(1998){Maoz}, {Koratkar}, {Shields}, {Ho},
  {Filippenko}, \& {Sternberg}}]{Maoz98}
{Maoz}, D., {Koratkar}, A., {Shields}, J.~C., {et~al.} 1998, \aj, 116, 55

\bibitem[{{Marconi} \& {Hunt}(2003)}]{Marconi03}
{Marconi}, A., \& {Hunt}, L.~K. 2003, \apjl, 589, L21

\bibitem[{{McConnell} {et~al.}(2013){McConnell}, {Chen}, {Ma}, {Greene},
  {Lauer}, \& {Gebhardt}}]{McConnell13}
{McConnell}, N.~J., {Chen}, S.-F.~S., {Ma}, C.-P., {et~al.} 2013, \apjl, 768,
  L21

\bibitem[{{Miller} {et~al.}(2015){Miller}, {Gallo}, {Greene}, {Kelly}, {Treu},
  {Woo}, \& {Baldassare}}]{Miller15}
{Miller}, B.~P., {Gallo}, E., {Greene}, J.~E., {et~al.} 2015, \apj, 799, 98

\bibitem[{{Mitzkus} {et~al.}(2017){Mitzkus}, {Cappellari}, \&
  {Walcher}}]{Mitzkus17}
{Mitzkus}, M., {Cappellari}, M., \& {Walcher}, C.~J. 2017, \mnras, 464, 4789

\bibitem[{{Miyoshi} {et~al.}(1995){Miyoshi}, {Moran}, {Herrnstein},
  {Greenhill}, {Nakai}, {Diamond}, \& {Inoue}}]{Miyoshi95}
{Miyoshi}, M., {Moran}, J., {Herrnstein}, J., {et~al.} 1995, \nat, 373, 127

\bibitem[{{Moran} {et~al.}(2014){Moran}, {Shahinyan}, {Sugarman}, {V{\'e}lez},
  \& {Eracleous}}]{Moran14}
{Moran}, E.~C., {Shahinyan}, K., {Sugarman}, H.~R., {V{\'e}lez}, D.~O., \&
  {Eracleous}, M. 2014, \aj, 148, 136

\bibitem[{{Netzer}(2015)}]{Netzer15}
{Netzer}, H. 2015, \araa, 53, 365

\bibitem[{{Neumayer} {et~al.}(2007){Neumayer}, {Cappellari}, {Reunanen}, {Rix},
  {van der Werf}, {de Zeeuw}, \& {Davies}}]{Neumayer07}
{Neumayer}, N., {Cappellari}, M., {Reunanen}, J., {et~al.} 2007, \apj, 671,
  1329

\bibitem[{{Nguyen} {et~al.}(2017){Nguyen}, {Seth}, {den Brok}, {Neumayer},
  {Cappellari}, {Barth}, {Caldwell}, {Williams}, \& {Binder}}]{Nguyen17}
{Nguyen}, D.~D., {Seth}, A.~C., {den Brok}, M., {et~al.} 2017, \apj, 836, 237

\bibitem[{{Nguyen} {et~al.}(2018){Nguyen}, {Seth}, {Neumayer}, {Kamann},
  {Voggel}, {Cappellari}, {Picotti}, {Nguyen}, {B{\"o}ker}, {Debattista},
  {Caldwell}, {McDermid}, {Bastian}, {Ahn}, \& {Pechetti}}]{Nguyen18}
{Nguyen}, D.~D., {Seth}, A.~C., {Neumayer}, N., {et~al.} 2018, \apj, 858, 118

\bibitem[{{Nguyen} {et~al.}(2019){Nguyen}, {Seth}, {Neumayer}, {Iguchi},
  {Cappellari}, {Strader}, {Chomiuk}, {Tremou}, {Pacucci}, {Nakanishi},
  {Bahramian}, {Nguyen}, {den Brok}, {Ahn}, {Voggel}, {Kacharov}, {Tsukui},
  {Ly}, {Dumont}, \& {Pechetti}}]{Nguyen19}
{Nguyen} {et~al.}. 2019, arXiv e-prints, arXiv:1901.05496

\bibitem[{{Nicholson} {et~al.}(1992){Nicholson}, {Bland-Hawthorn}, \&
  {Taylor}}]{Nicholson92}
{Nicholson}, R.~A., {Bland-Hawthorn}, J., \& {Taylor}, K. 1992, \apj, 387, 503

\bibitem[{{Onishi} {et~al.}(2017){Onishi}, {Iguchi}, {Davis}, {Bureau},
  {Cappellari}, {Sarzi}, \& {Blitz}}]{Onishi17}
{Onishi}, K., {Iguchi}, S., {Davis}, T.~A., {et~al.} 2017, ArXiv e-prints,
  arXiv:1703.05247

\bibitem[{{Onishi} {et~al.}(2015){Onishi}, {Iguchi}, {Sheth}, \&
  {Kohno}}]{Onishi15}
{Onishi}, K., {Iguchi}, S., {Sheth}, K., \& {Kohno}, K. 2015, \apj, 806, 39

\bibitem[{{Pacucci} \& {Loeb}(2018)}]{Pacucci18b}
{Pacucci}, F., \& {Loeb}, A. 2018, ArXiv e-prints, arXiv:1810.12302

\bibitem[{{Pacucci} {et~al.}(2018){Pacucci}, {Loeb}, {Mezcua}, \&
  {Mart{\'{\i}}n-Navarro}}]{Pacucci18}
{Pacucci}, F., {Loeb}, A., {Mezcua}, M., \& {Mart{\'{\i}}n-Navarro}, I. 2018,
  \apjl, 864, L6

\bibitem[{{Pacucci} {et~al.}(2017){Pacucci}, {Natarajan}, {Volonteri},
  {Cappelluti}, \& {Urry}}]{Pacucci17}
{Pacucci}, F., {Natarajan}, P., {Volonteri}, M., {Cappelluti}, N., \& {Urry},
  C.~M. 2017, \apjl, 850, L42

\bibitem[{{Pacucci} {et~al.}(2015){Pacucci}, {Volonteri}, \&
  {Ferrara}}]{Pacucci15}
{Pacucci}, F., {Volonteri}, M., \& {Ferrara}, A. 2015, \mnras, 452, 1922

\bibitem[{{Park} {et~al.}(2016){Park}, {Ricotti}, {Natarajan},
  {Bogdanovi{\'c}}, \& {Wise}}]{Park16}
{Park}, K., {Ricotti}, M., {Natarajan}, P., {Bogdanovi{\'c}}, T., \& {Wise},
  J.~H. 2016, \apj, 818, 184

\bibitem[{{Paturel} {et~al.}(2000){Paturel}, {Fang}, {Petit}, {Garnier}, \&
  {Rousseau}}]{Paturel00}
{Paturel}, G., {Fang}, Y., {Petit}, C., {Garnier}, R., \& {Rousseau}, J. 2000,
  \aaps, 146, 19

\bibitem[{{Peterson}(1982)}]{Peterson82}
{Peterson}, C.~J. 1982, \pasp, 94, 409

\bibitem[{{Quillen} {et~al.}(1992){Quillen}, {de Zeeuw}, {Phinney}, \&
  {Phillips}}]{Quillen92}
{Quillen}, A.~C., {de Zeeuw}, P.~T., {Phinney}, E.~S., \& {Phillips}, T.~G.
  1992, \apj, 391, 121

\bibitem[{{Randriamampandry} {et~al.}(2015){Randriamampandry}, {Combes},
  {Carignan}, \& {Deg}}]{Randriamampandry15}
{Randriamampandry}, T.~H., {Combes}, F., {Carignan}, C., \& {Deg}, N. 2015,
  \mnras, 454, 3743

\bibitem[{{Reines} {et~al.}(2013){Reines}, {Greene}, \& {Geha}}]{Reines13}
{Reines}, A.~E., {Greene}, J.~E., \& {Geha}, M. 2013, \apj, 775, 116

\bibitem[{{Reines} \& {Volonteri}(2015)}]{Reines15}
{Reines}, A.~E., \& {Volonteri}, M. 2015, \apj, 813, 82

\bibitem[{{Rieke} \& {Low}(1972)}]{Rieke72}
{Rieke}, G.~H., \& {Low}, F.~J. 1972, \apjl, 176, L95

\bibitem[{{Roediger} \& {Courteau}(2015)}]{Roediger15}
{Roediger}, J.~C., \& {Courteau}, S. 2015, \mnras, 452, 3209

\bibitem[{{Russell}(2002)}]{Russell02}
{Russell}, D.~G. 2002, \apj, 565, 681

\bibitem[{{Saglia} {et~al.}(2016){Saglia}, {Opitsch}, {Erwin}, {Thomas},
  {Beifiori}, {Fabricius}, {Mazzalay}, {Nowak}, {Rusli}, \&
  {Bender}}]{Saglia16}
{Saglia}, R.~P., {Opitsch}, M., {Erwin}, P., {et~al.} 2016, \apj, 818, 47

\bibitem[{{Salo} {et~al.}(2015){Salo}, {Laurikainen}, {Laine}, {Comer{\'o}n},
  {Gadotti}, {Buta}, {Sheth}, {Zaritsky}, {Ho}, {Knapen}, {Athanassoula},
  {Bosma}, {Laine}, {Cisternas}, {Kim}, {Mu{\~n}oz-Mateos}, {Regan}, {Hinz},
  {Gil de Paz}, {Menendez-Delmestre}, {Mizusawa}, {Erroz-Ferrer}, {Meidt}, \&
  {Querejeta}}]{Salo15}
{Salo}, H., {Laurikainen}, E., {Laine}, J., {et~al.} 2015, \apjs, 219, 4

\bibitem[{{Satyapal} {et~al.}(2009){Satyapal}, {B{\"o}ker}, {Mcalpine},
  {Gliozzi}, {Abel}, \& {Heckman}}]{Satyapal09}
{Satyapal}, S., {B{\"o}ker}, T., {Mcalpine}, W., {et~al.} 2009, \apj, 704, 439

\bibitem[{{Schawinski} {et~al.}(2007){Schawinski}, {Thomas}, {Sarzi},
  {Maraston}, {Kaviraj}, {Joo}, {Yi}, \& {Silk}}]{Schawinski07}
{Schawinski}, K., {Thomas}, D., {Sarzi}, M., {et~al.} 2007, \mnras, 382, 1415

\bibitem[{{Schlafly} \& {Finkbeiner}(2011)}]{Schlafly11}
{Schlafly}, E.~F., \& {Finkbeiner}, D.~P. 2011, \apj, 737, 103

\bibitem[{{Scott} {et~al.}(2013){Scott}, {Graham}, \& {Schombert}}]{Scott13}
{Scott}, N., {Graham}, A.~W., \& {Schombert}, J. 2013, \apj, 768, 76

\bibitem[{{Seth} {et~al.}(2010){Seth}, {Cappellari}, {Neumayer}, {Caldwell},
  {Bastian}, {Olsen}, {Blum}, {Debattista}, \citep{} {McDermid}, {Puzia}, \&
  {Stephens}}]{Seth10a}
{Seth}, A.~C., {Cappellari}, M., {Neumayer}, N., {et~al.} 2010, \apj, 714, 713

\bibitem[{{Seth} {et~al.}(2014){Seth}, {van den Bosch}, {Mieske}, {Baumgardt},
  {Brok}, {Strader}, {Neumayer}, {Chilingarian}, {Hilker}, {McDermid},
  {Spitler}, {Brodie}, {Frank}, \& {Walsh}}]{Seth14}
{Seth}, A.~C., {van den Bosch}, R., {Mieske}, S., {et~al.} 2014, \nat, 513, 398

\bibitem[{{She} {et~al.}(2017){She}, {Ho}, \& {Feng}}]{She17}
{She}, R., {Ho}, L.~C., \& {Feng}, H. 2017, \apj, 842, 131

\bibitem[{{Sheth} {et~al.}(2010){Sheth}, {Regan}, {Hinz}, {Gil de Paz},
  {Men{\'e}ndez-Delmestre}, {Mu{\~n}oz-Mateos}, {Seibert}, {Kim},
  {Laurikainen}, {Salo}, {Gadotti}, {Laine}, {Mizusawa}, {Armus},
  {Athanassoula}, {Bosma}, {Buta}, {Capak}, {Jarrett}, {Elmegreen},
  {Elmegreen}, {Knapen}, {Koda}, {Helou}, {Ho}, {Madore}, {Masters},
  {Mobasher}, {Ogle}, {Peng}, {Schinnerer}, {Surace}, {Zaritsky},
  {Comer{\'o}n}, {de Swardt}, {Meidt}, {Kasliwal}, \& {Aravena}}]{Sheth10}
{Sheth}, K., {Regan}, M., {Hinz}, J.~L., {et~al.} 2010, \pasp, 122, 1397

\bibitem[{{Silk} \& {Rees}(1998)}]{Silk98}
{Silk}, J., \& {Rees}, M.~J. 1998, \aap, 331, L1

\bibitem[{{Stone} {et~al.}(2017){Stone}, {K{\"u}pper}, \& {Ostriker}}]{Stone17}
{Stone}, N.~C., {K{\"u}pper}, A.~H.~W., \& {Ostriker}, J.~P. 2017, \mnras, 467,
  4180

\bibitem[{{Terrazas} {et~al.}(2017){Terrazas}, {Bell}, {Woo}, \&
  {Henriques}}]{Terrazas17}
{Terrazas}, B.~A., {Bell}, E.~F., {Woo}, J., \& {Henriques}, B.~M.~B. 2017,
  \apj, 844, 170

\bibitem[{{Thater} {et~al.}(2017){Thater}, {Krajnovi{\'c}}, {Bourne},
  {Cappellari}, {de Zeeuw}, {Emsellem}, {Magorrian}, {McDermid}, {Sarzi}, \&
  {van de Ven}}]{Thater17}
{Thater}, S., {Krajnovi{\'c}}, D., {Bourne}, M.~A., {et~al.} 2017, \aap, 597,
  A18

\bibitem[{{Thornton} {et~al.}(2008){Thornton}, {Barth}, {Ho}, {Rutledge}, \&
  {Greene}}]{Thornton08}
{Thornton}, C.~E., {Barth}, A.~J., {Ho}, L.~C., {Rutledge}, R.~E., \& {Greene},
  J.~E. 2008, \apj, 686, 892

\bibitem[{{Tristram} \& {Schartmann}(2011)}]{Tristram11}
{Tristram}, K.~R.~W., \& {Schartmann}, M. 2011, \aap, 531, A99

\bibitem[{{Valluri} {et~al.}(2005){Valluri}, {Ferrarese}, {Merritt}, \&
  {Joseph}}]{Valluri05}
{Valluri}, M., {Ferrarese}, L., {Merritt}, D., \& {Joseph}, C.~L. 2005, \apj,
  628, 137

\bibitem[{{van den Bosch} \& {de Zeeuw}(2010)}]{vandenBosch10}
{van den Bosch}, R.~C.~E., \& {de Zeeuw}, P.~T. 2010, \mnras, 401, 1770

\bibitem[{{van den Bosch} {et~al.}(2016){van den Bosch}, {Greene}, {Braatz},
  {Constantin}, \& {Kuo}}]{vandenBosch16}
{van den Bosch}, R.~C.~E., {Greene}, J.~E., {Braatz}, J.~A., {Constantin}, A.,
  \& {Kuo}, C.-Y. 2016, \apj, 819, 11

\bibitem[{{van Wassenhove} {et~al.}(2010){van Wassenhove}, {Volonteri},
  {Walker}, \& {Gair}}]{vanWassenhove10}
{van Wassenhove}, S., {Volonteri}, M., {Walker}, M.~G., \& {Gair}, J.~R. 2010,
  \mnras, 408, 1139

\bibitem[{{Verolme} {et~al.}(2002){Verolme}, {Cappellari}, {Copin}, {van der
  Marel}, {Bacon}, {Bureau}, {Davies}, {Miller}, \& {de Zeeuw}}]{Verolme02}
{Verolme}, E.~K., {Cappellari}, M., {Copin}, Y., {et~al.} 2002, \mnras, 335,
  517

\bibitem[{{Voggel} {et~al.}(2018){Voggel}, {Seth}, {Neumayer}, {Mieske},
  {Chilingarian}, {Ahn}, {Baumgardt}, {Hilker}, {Nguyen}, {Romanowsky},
  {Walsh}, {den Brok}, \& {Strader}}]{Voggel18}
{Voggel}, K.~T., {Seth}, A.~C., {Neumayer}, N., {et~al.} 2018, \apj, 858, 20

\bibitem[{{Volonteri}(2010)}]{Volonteri10}
{Volonteri}, M. 2010, \aapr, 18, 279

\bibitem[{{Volonteri}(2012)}]{Volonteri12b}
{Volonteri}, M. 2012, in American Institute of Physics Conference Series, Vol.
  1480, American Institute of Physics Conference Series, ed. M.~{Umemura} \&
  K.~{Omukai}, 289--296

\bibitem[{{Volonteri} \& {Bellovary}(2012)}]{Volonteri12a}
{Volonteri}, M., \& {Bellovary}, J. 2012, Reports on Progress in Physics, 75,
  124901

\bibitem[{{Volonteri} {et~al.}(2015){Volonteri}, {Capelo}, {Netzer},
  {Bellovary}, {Dotti}, \& {Governato}}]{Volonteri15}
{Volonteri}, M., {Capelo}, P.~R., {Netzer}, H., {et~al.} 2015, \mnras, 452, L6

\bibitem[{{Volonteri} {et~al.}(2008){Volonteri}, {Lodato}, \&
  {Natarajan}}]{Volonteri08}
{Volonteri}, M., {Lodato}, G., \& {Natarajan}, P. 2008, \mnras, 383, 1079

\end{thebibliography}
\end{document}